\documentclass[12pt]{article}

\usepackage{jheppub, bm, wrapfig,float,array}
\usepackage[utf8]{inputenc}
\numberwithin{equation}{section}
\renewcommand\afterTocRuleSpace{\bigskip}

\usepackage{amsfonts}
\usepackage{amsmath}
\usepackage{color}
\usepackage{graphicx}
\usepackage{float}
\usepackage[T1]{fontenc}
\usepackage[utf8]{inputenc}
\usepackage{alphabeta}
\usepackage{fancyvrb}
\usepackage[dvipsnames]{xcolor}
\usepackage{bold-extra}
\usepackage{lmodern}

\usepackage{tikz}
\usetikzlibrary{arrows}
\usetikzlibrary{shapes.geometric,calc,arrows, positioning,shapes.misc,decorations.markings}
\tikzset{
  big arrow/.style={
    decoration={markings,mark=at position 1 with {\arrow[scale=2,#1]{>}}},
    postaction={decorate},
    shorten >=0.4pt},
  big arrow/.default=black}

\newcommand{\bea}{\begin{eqnarray}}
\newcommand{\eea}{\end{eqnarray}}
\newcommand{\be}{\begin{equation}}
\newcommand{\ee}{\end{equation}}
\newcommand{\bit}{\begin{itemize}}
\newcommand{\eit}{\end{itemize}}
\newcommand{\ben}{\begin{enumerate}}
\newcommand{\een}{\end{enumerate}}

\renewcommand{\ni}{\noindent}

\newcommand{\wh}{\widehat}

\newcommand{\half}{\frac{1}{2}}

\newcommand{\Z}{{\mathbb Z}}

\newcommand{\C}{{\mathbb C}}

\newcommand{\cI}{\mathcal{I}}

\newcommand{\cN}{\mathcal{N}}

\newcommand{\F}{\mathsf{F}}
\renewcommand{\S}{\mathsf{S}}

\renewcommand{\C}{\mathsf{C}}

\renewcommand{\C}{\mathsf{C}}

\renewcommand{\L}{\mathsf{\Lambda}}

\newcommand{\fe}{\mathfrak{e}}
\newcommand{\ff}{\mathfrak{f}}
\newcommand{\fg}{\mathfrak{g}}

\newcommand{\su}{\mathfrak{su}}
\renewcommand{\sp}{\mathfrak{sp}}
\newcommand{\so}{\mathfrak{so}}

\newcommand{\ubf}[1]{\underline{\bf #1}}

\title{More $5d$ KK theories}

\author{Lakshya Bhardwaj}

\affiliation{Department of Physics, Harvard University\\17 Oxford St, Cambridge, MA 02138, USA}

\abstract{In this note, we discuss circle compactifications of $6d$ SCFTs for which a geometric M-theory construction is not known in previous literature.
}

\begin{document}

\maketitle

\section{Introduction} \label{I}
Recently, geometric constructions in M-theory have been very successful in progressing our understanding of $5d$ SCFTs and the $5d$ $\cN=1$ gauge theories arising on the extended\footnote{We define extended Coulomb branch to be the total space obtained by fibering the Coulomb branch over the space of supersymmetry preserving mass parameters.} Coulomb branch of $5d$ SCFTs \cite{DelZotto:2017pti,Jefferson:2018irk,Bhardwaj:2018yhy,Bhardwaj:2018vuu,Apruzzi:2018nre,Apruzzi:2019vpe,Apruzzi:2019opn,Bhardwaj:2019ngx,Apruzzi:2019enx,Bhardwaj:2019jtr,Bhardwaj:2019fzv,Bhardwaj:2019xeg,Apruzzi:2019syw,Bhardwaj:2020gyu}\footnote{See \cite{Jefferson:2017ahm,Hayashi:2016abm,Hayashi:2018bkd,Hayashi:2018lyv,Closset:2019mdz,Cabrera:2019hya,Kim:2019uqw,Kim:2019dqn,Fluder:2019szh,Uhlemann:2019ypp,Saxena:2019wuy,Hayashi:2019yxj,Closset:2018bjz,Hayashi:2015zka,Hayashi:2020sly,Eckhard:2020jyr,Bourget:2020gzi,Uhlemann:2019ors,Gu:2019pqj,Cota:2019cjx,Choi:2019miv,Chaney:2018gjc,Bastian:2018fba,Cheng:2018wll,Bah:2018lyv,Assel:2018rcw,Ashok:2017bld,Garozzo:2020pmz} for other recent related work on the subject of $5d$ $\cN=1$ QFTs.}. Thanks to this understanding, it has been possible to generate claims of obtaining a full classification of $5d$ SCFTs \cite{Jefferson:2018irk,Bhardwaj:2018yhy,Bhardwaj:2018vuu,Bhardwaj:2019fzv,Bhardwaj:2019jtr,Bhardwaj:2019xeg} which are backed by substantial evidence \cite{Jefferson:2018irk,Bhardwaj:2020gyu}. According to this classification proposal, all $5d$ SCFTs can be generated by performing special type of RG flows upon $5d$ theories obtained by compactifying $6d$ SCFTs on a circle of finite, non-zero radius. Such $5d$ theories are often referred to as $5d$ KK theories, and we will use this terminology throughout this paper. The special type of RG flows mentioned above can be understood as those processes that integrate out a set of BPS particles and strings \cite{Bhardwaj:2019xeg} from the extended Coulomb branch of a $5d$ KK theory.

Such RG flows obtain a clean characterization when the $5d$ theories under discussion are constructed by compactifying M-theory on a local Calabi-Yau threefold (CY3) with an isolated singularity. The extended Coulomb branch of the $5d$ theory is obtained by resolving the singularity, and the resulting resolved CY3 can be described by a collection of intersecting compact Kahler surfaces. The RG flows under discussion are then mapped to motions on the extended Kahler cone of the resolved CY3 which decompactify a set of compact complex curves and surfaces. This map is a consequence of the fact that M2/M5 branes compactified on compact complex curves/surfaces produce BPS particles/strings in the resulting $5d$ theory, and the volumes of these curves/surfaces can be identified as the masses/tensions of the corresponding BPS particles/strings.

Therefore, understanding the full set of $5d$ KK theories and the local CY3 associated to them is very important for the purposes of the classification program of $5d$ SCFTs based on the proposal described above. This task was undertaken by \cite{Bhardwaj:2018yhy,Bhardwaj:2018vuu,Bhardwaj:2019fzv}, but the work of \cite{Bhardwaj:2020gyu} featured some $5d$ KK theories which did not appear in \cite{Bhardwaj:2018yhy,Bhardwaj:2018vuu,Bhardwaj:2019fzv}. All such examples involve a twisted compactification of a $6d$ SCFT, which means that the observables in the $6d$ SCFT are acted upon by the action of a discrete global symmetry when transported around the circle. Twisted circle compactifications of $6d$ SCFTs and the CY3 associated to them were studied in \cite{Bhardwaj:2019fzv}. This article is devoted to a study of $5d$ KK theories (and the associated CY3) that are missing from their analysis.

The twists discussed in this paper fall into the following two different classes:
\ben
\item In the first class are the twists whose associated discrete global symmetry acts by outer automorphism of a gauge algebra appearing in the low energy\footnote{Let us recall that a $6d$ SCFT reduces to a $6d$ $\cN=(1,0)$ gauge theory interacting with tensor multiplets on its tensor branch.} $6d$ $\cN=(1,0)$ gauge theory, and its action on the hypermultiplets \emph{cannot} be represented as a permutation of the hypermultiplets. This is in contrast with the outer automorphism twists considered in \cite{Bhardwaj:2019fzv}, all of which \emph{could be} represented as a permutation of hypermultiplets. For example, let us contrast the two $6d$ SCFTs which have one dimensional tensor branch and carry $\su(5)$ gauge algebra at low energies on the tensor branch. One of the theories carries 10 fundamental hypers and the other carries 13 fundamental hypers plus a hyper in two-index antisymmetric representation. The outer automorphism of $\su(5)$ exchanges fields transforming in fundamental/antisymmetric representation with fields transforming in complex conjugate of fundamental/antisymmetric representation inside each hypermultiplet. In the case of first theory, this action is equivalent to organizing the 10 fundamental hypers into 5 pairs and exchanging the hypers in each pair, thus representing the action of outer automorphism as a permutation on hypermultiplets. In the case of second theory, the number of fundamental and antisymmetric hypers are odd, and hence the action cannot be represented fully as a permutation of hypermultiplets.
\item In the second class are the twists whose associated discrete global symmetry acts only on the hypermultiplets, but not on the vector and tensor multiplets. This is in contrast to the twists considered in \cite{Bhardwaj:2019fzv}, all of which acted either on vector or on tensor multiplets. Such twists can arise when we have $2n$ half-hypermultiplets transforming in a pseudoreal representation of some gauge algebra. These half-hypers are rotated by an $O(2n)$ global symmetry\footnote{In some cases, non-perturbative excitations force the global symmetry group to be $Spin(2n)$ instead of $O(2n)$, thus removing the possibility of such a twist.}, thus opening up the possibility of twisting the theory by a $\Z_2$ element of determinant $-1$ inside $O(2n)$. When we have $2n+1$ half-hypers, then a $\Z_2$ element of determinant $-1$ inside $O(2n+1)$ acts as a central element of the gauge group, thus reducing the global symmetry group to $SO(2n+1)$.
\een
These two kind of twists lead to new building blocks for $5d$ KK theories and new ways of gluing these building blocks to produce $5d$ KK theories. We will enlist these building blocks and their possible gluings. We will provide data of resolved CY3 associated to each new building block, and rules for gluing the associated resolved CY3s for each new gluing between the corresponding building blocks.

Our approach to the new resolved CY3s will differ from the approach employed in \cite{Bhardwaj:2019fzv}. This is because the approach used in \cite{Bhardwaj:2019fzv} relied on the knowledge of a low-energy $5d$ $\cN=1$ non-Abelian gauge theory description of the $5d$ KK theory. The data of this low-energy $5d$ gauge theory was obtained by modding out the data of the associated $6d$ gauge theory by the action of discrete symmetry. So, for the $6d$ SCFT carrying $\su(5)$ with 10 fundamental hypers compactified using an outer automorphism twist, the associated low energy $5d$ gauge theory carries $\sp(2)$ with 5 fundamental hypers. However, this procedure of projecting the data of $6d$ gauge theory does not work for the first kind of twists discussed above, since in those cases the associated discrete symmetry does not act by permutation of hypermultiplets.

Due to this reason, we will instead simply propose the resolved CY3s associated to the new building block $5d$ KK theories. The proposed resolved CY3 satisfy necessary geometric consistency conditions implying that they are consistent geometric backgrounds to compactify M-theory. Moreover, the proposed resolved CY3s will be presented in a special form which makes it manifest that the $5d$ $\cN=1$ theory resulting from M-theory compactification is actually a $5d$ KK theory. Such special presentations of the resolved CY3s associated to $5d$ KK theories were discussed at length in \cite{Bhardwaj:2019fzv,Bhardwaj:2020gyu} and will be reviewed briefly in this paper. According to the analysis of \cite{Bhardwaj:2020gyu}, one can easily read the data of the associated $6d$ SCFT and the type of twist from the data of the resolved CY3 presented in this special form. In this way, we will identify the $5d$ KK theory associated to each proposed resolved CY3.

This paper is organized as follows:\\
In Section \ref{BB}, we discuss the various new $5d$ KK theory building blocks that can arise by considering the two kinds of twists discussed above. These building blocks are $5d$ KK theories arising by compactifying $6d$ SCFTs having a one-dimensional tensor branch, or in other words carrying a single tensor multiplet. These building blocks are collected in Table \ref{KR1}.\\
In Section \ref{PCY}, we propose resolved CY3 that describe the extended Coulomb branches of these new building blocks.\\
In Section \ref{DCY}, we provide some checks of our proposal, where we review (following \cite{Bhardwaj:2020gyu}) how some data of the associated $6d$ SCFT and twist can be read from the data of resolved CY3.\\
In Section \ref{LE}, we provide further arguments in favor of our proposal put forward in Section \ref{PCY}. In this section, we discuss how one can compute, using the data of associated CY3, the various low-energy effective $5d$ $\cN=1$ non-abelian gauge theories arising upon compactifying a $6d$ SCFT (possibly with a twist) on a circle of finite, non-zero radius. These low-energy $5d$ gauge theories can be predicted by modding out the data of the $6d$ $\cN=(1,0)$ non-abelian gauge theory appearing on the tensor branch of the associated $6d$ SCFT by the action of the discrete symmetry generating the twist. We show that these predictions for new KK building blocks match the computations performed using the proposed associated resolved CY3s.\\
In Section \ref{CBB}, we discuss the various ways in which the new $5d$ KK theory building blocks can be combined with other new/old $5d$ KK theory building blocks to produce more general $5d$ KK theories whose associated $6d$ SCFTs have a tensor branch of dimension more than one. Such combinations are collected in Table \ref{KR2} where, due to reasons explained in Section \ref{CBB}, we have restricted our attention to new building blocks arising only from the twists of the first type discussed above.\\
In Section \ref{GR}, we propose rules for gluing the two CY3s associated to two building blocks, so that the combined CY3 describes the extended Coulomb branch of the KK theory produced by combining the two building blocks.\\
In Section \ref{RD}, we provide some checks of our proposal, where we review (following \cite{Bhardwaj:2020gyu}) how data of the discrete symmetry (used for twisting) permuting tensor multiplets in the associated $6d$ SCFT can be read from the data of gluing rules.\\
In Section \ref{LE2}, we describe how the gluing rules can be used to read the data of hypermultiplet content charged under multiple simple factors of the gauge algebra of $5d$ gauge theory appearing at low-energies. The discussion also supports some field-theoretic arguments made at the beginning of Section \ref{CBB} and used to compile Table \ref{KR2}.\\

\section{Building blocks}\label{BB}
In this paper, we are going to use the notation developed in \cite{Bhardwaj:2019fzv} to denote $5d$ KK theories. This captures the tensor branch data of the associated $6d$ SCFT and the action of discrete symmetry (used to twist the theory) on the tensor branch data. The notation used there for $5d$ KK theories arising from $6d$ SCFTs carrying a single tensor multiplet took the following form
\be
\begin{tikzpicture} [scale=1.9]
\node at (-4.6,0.8) {$k$};
\node at (-4.6,1.1) {$\fg^{(q)}$};
\end{tikzpicture}
\ee
where $\fg$ is a simple gauge algebra, $q$ denotes the order of outer automorphism acting on $\fg$, and $k$ (a positive integer) denotes the coefficient of Green-Schwarz term in the Lagrangian used for canceling 1-loop gauge anomaly. To incorporate twists of the second kind discussed in Section \ref{I}, we extend the above notation and use
\be
\begin{tikzpicture} [scale=1.9]
\node (v2) at (-4.6,0.8) {$k$};
\node at (-4.6,1.1) {$\fg^{(q)}$};
\node (v1) at (-5.5,0.8) {$\left[\Z^{(2)}_2\right]$};
\draw  (v1) edge (v2);
\end{tikzpicture}
\ee
to denote that we have an extra twist by a $\Z_2$ living inside an $O(2n)$ flavor symmetry.

\begin{table}[htbp]
\begin{center}
\begin{tabular}{|c|c|l|}
 \hline
 \raisebox{-.4\height}{ \begin{tikzpicture}
\node at (-0.5,0.4) {1};
\node at (-0.45,0.9) {$\su(n)^{(2)}$};
\end{tikzpicture}}&$n\ge5$
\\ \hline
\raisebox{-.4\height}{ \begin{tikzpicture}
\node at (-0.5,0.4) {1};
\node at (-0.45,0.9) {$\su(\wh n)^{(2)}$};
\end{tikzpicture}}&$n\ge8$
\\ \hline
\raisebox{-.4\height}{ \begin{tikzpicture}
\node at (-0.5,0.4) {$1$};
\node at (-0.45,0.9) {$\su(\tilde 6)^{(2)}$};
\end{tikzpicture}}&
\\ \hline
\raisebox{-.4\height}{ \begin{tikzpicture}
\node at (-0.5,0.4) {$k$};
\node at (-0.45,0.9) {$\so(10)^{(2)}$};
\end{tikzpicture}}&$k=1,3$
\\ \hline
\raisebox{-.4\height}{ \begin{tikzpicture}
\node at (-0.5,0.4) {$k$};
\node at (-0.45,0.9) {$\fe_6^{(2)}$};
\end{tikzpicture}}&$k=1,3,5$
\\ \hline
\raisebox{-.4\height}{ \begin{tikzpicture}
\node (v1) at (-2.3,0.4) {$\left[\Z^{(2)}_2\right]$};
\node (v2) at (-0.5,0.4) {2};
\node at (-0.45,0.9) {$\so(11)^{(1)}$};
\draw  (v1) edge (v2);
\end{tikzpicture}}&
\\ \hline
\raisebox{-.4\height}{ \begin{tikzpicture}
\node (v1) at (-2.3,0.4) {$\left[\Z^{(2)}_2\right]$};
\node (v2) at (-0.5,0.4) {2};
\node at (-0.45,0.9) {$\so(12)^{(1)}$};
\draw  (v1) edge (v2);
\end{tikzpicture}}&
\\ \hline
\raisebox{-.4\height}{ \begin{tikzpicture}
\node (v1) at (-2.3,0.4) {$\left[\Z^{(2)}_2\right]$};
\node (v2) at (-0.5,0.4) {1};
\node at (-0.45,0.9) {$\so(\wh{12})^{(1)}$};
\draw  (v1) edge (v2);
\end{tikzpicture}}&
\\ \hline
\raisebox{-.4\height}{\begin{tikzpicture}
\node (v1) at (-2.3,0.4) {$\left[\Z^{(2)}_2\right]$};
\node (v2) at (-0.5,0.4) {$k$};
\node at (-0.45,0.9) {$\fe_7^{(1)}$};
\draw  (v1) edge (v2);
\end{tikzpicture}}&$k=2,4,6$
\\ \hline
\end{tabular}
\end{center}
\caption{List of new building blocks for $5d$ KK theories. See text for more details.}	\label{KR1}
\end{table}

The new building block $5d$ KK theories arising from the two kinds of twists discussed in Section \ref{I} have been collected in Table \ref{KR1}, where they are expressed in the notation reviewed above. Let us discuss each of the entries in the table:
\bit
\item The $6d$ SCFT denoted by
\be\label{t1}
\begin{tikzpicture}
\node at (-0.5,0.4) {1};
\node at (-0.45,0.9) {$\su(n)$};
\end{tikzpicture}
\ee
carries $n+8$ hypers in fundamental and one hyper in two-index antisymmetric of $\su(n)$. Accordingly, a $\Z_2$ outer automorphism of $\su(n)$ is a symmetry of the theory since it complex conjugates all fields inside hypermultiplets. For $n=3,4$, the action of the outer automorphism can be represented as a permutation of hypermultiplets (see Section \ref{I}) and thus the corresponding $5d$ KK theories appeared already in \cite{Bhardwaj:2019fzv}.
\item The $6d$ SCFT denoted by
\be
\begin{tikzpicture}
\node at (-0.5,0.4) {1};
\node at (-0.45,0.9) {$\su(\wh n)$};
\end{tikzpicture}
\ee
carries $n-8$ hypers in fundamental and one hyper in two-index symmetric of $\su(n)$. The hat on top of $n$ in $\su(n)$ has been placed to distinguish this theory from the theory (\ref{t1}). A $\Z_2$ outer automorphism of $\su(n)$ is a symmetry of the theory since it complex conjugates all fields inside hypermultiplets.
\item For $\su(6)$ with Green-Schwarz coupling $k=1$, there is another possibility for matter content which we denote as
\be
\begin{tikzpicture}
\node at (-0.5,0.4) {1};
\node at (-0.45,0.9) {$\su(\tilde6)$};
\end{tikzpicture}
\ee
and it carries $15$ hypers in fundamental plus a half-hyper in three-index antisymmetric of $\su(6)$. A $\Z_2$ outer automorphism of $\su(6)$ is a symmetry of the theory since it complex conjugates all fields inside fundamental hypermultiplets and leaves the half-hyper in three-index antisymmetric invariant.
\item The $6d$ SCFT denoted by
\be
\begin{tikzpicture}
\node at (-0.5,0.4) {$k$};
\node at (-0.45,0.9) {$\so(10)$};
\end{tikzpicture}
\ee
carries $6-k$ hypers in fundamental representation and $4-k$ hypers in an irreducible spinor representation of $\so(10)$. A $\Z_2$ outer automorphism of $\so(10)$ leaves fundamental invariant but exchanges the spinor and cospinor representations. Since spinor and cospinor are complex conjugates for $\so(10)$, the outer automorphism acts as a symmetry of the theory. For $k=2,4$, the action of outer automorphism can be represented as a permutation of hypermultiplets and hence the corresponding $5d$ KK theories appeared already in \cite{Bhardwaj:2019fzv}.
\item The $6d$ SCFT denoted by
\be
\begin{tikzpicture}
\node at (-0.5,0.4) {$k$};
\node at (-0.45,0.9) {$\fe_6$};
\end{tikzpicture}
\ee
carries $6-k$ hypers in $\mathbf{27}$ dimensional representation of $\fe_6$. A $\Z_2$ outer automorphism of $\su(n)$ is a symmetry of the theory since it complex conjugates all fields inside hypermultiplets. For $k=2,4,6$, the action of outer automorphism can be represented as a permutation of hypermultiplets and hence the corresponding $5d$ KK theories appeared already in \cite{Bhardwaj:2019fzv}.
\item The $6d$ SCFT denoted by
\be
\begin{tikzpicture}
\node at (-0.5,0.4) {2};
\node at (-0.45,0.9) {$\so(11)$};
\end{tikzpicture}
\ee
carries five hypers in fundamental and two half-hypers in spinor of $\so(11)$. The half-hypers in spinor are thus rotated by an $O(2)$ global symmetry and we can twist by a $\Z_2$ element of determinant $-1$ in $O(2)$. The superscript $(1)$ on $\so(11)$ denotes that there is no outer automorphism twist involved.
\item The $6d$ SCFT denoted by
\be
\begin{tikzpicture}
\node at (-0.5,0.4) {2};
\node at (-0.45,0.9) {$\so(12)$};
\end{tikzpicture}
\ee
carries six hypers in fundamental representation and two half-hypers in irreducible spinor representation of $\so(12)$. Thus, we can twist by a $\Z_2$ element of determinant $-1$ in $O(2)$ rotating the two half-hypers in spinor representation.
\item The $6d$ SCFT denoted by
\be
\begin{tikzpicture}
\node at (-0.5,0.4) {1};
\node at (-0.45,0.9) {$\so(\wh{12})$};
\end{tikzpicture}
\ee
carries seven hypers in fundamental, two half-hypers in spinor and one half-hyper in cospinor of $\so(12)$. Thus, we can twist by a $\Z_2$ element of determinant $-1$ in $O(2)$ rotating the two half-hypers in spinor representation. The hat on $12$ in $\so(12)$ has been placed to distinguish it from the $6d$ SCFT denoted as
\be
\begin{tikzpicture}
\node at (-0.5,0.4) {1};
\node at (-0.45,0.9) {$\so(12)$};
\end{tikzpicture}
\ee
which carries seven hypers in fundamental and three half-hypers in spinor of $\so(12)$.
\item The $6d$ SCFT denoted by
\be
\begin{tikzpicture}
\node at (-0.5,0.4) {$k$};
\node at (-0.45,0.9) {$\fe_7$};
\end{tikzpicture}
\ee
carries $8-k$ half-hypers in $\mathbf{56}$ dimensional representation of $\fe_7$. Thus, we can twist by a $\Z_2$ element of determinant $-1$ in $O(8-k)$ rotating the $8-k$ half-hypers if $k$ is even (see Section \ref{I} for explanation).
\eit
The last four entries in Table \ref{KR1} involve twists of the second type discussed in Section \ref{I}. Such twists require the presence of matter in pseudo-real representations of the gauge algebra. For $6d$ SCFTs, along with the cases discussed above, there is another case carrying pseudo-real representations. This $6d$ SCFT is denoted as
\be\label{t2}
\begin{tikzpicture}
\node at (-0.5,0.4) {1};
\node at (-0.45,0.9) {$\sp(n)$};
\end{tikzpicture}
\ee
and carries $4n+16$ half-hypers in fundamental of $\sp(n)$. However, the global symmetry associated to these half-hypers is $Spin(4n+16)$ rather than $O(4n+16)$ since an instanton string tranforms in irreducible spinor representation of the $\so(4n+16)$ global symmetry algebra. This obstructs the existence of a $\Z_2$ element of determinant $-1$ in the associated global symmetry group since it exchanges spinor and cospinor representations of $\so(4n+16)$, and thus is not a symmetry of the $6d$ SCFT. Consequently (\ref{t2}) does not lead to any new $5d$ KK theory building blocks.

\subsection{Associated CY3}\label{PCY}
In this subsection we will propose the resolved CY3 associated to new $5d$ KK theory building blocks appearing in Table \ref{KR1}. The data for resolved CY3 will be presented in terms of a graph. The vertices of the graph denote different irreducible compact Kahler surfaces and the edges indicate intersections between these surfaces. An intersection between two surfaces $S_1$ and $S_2$ can be described as a gluing of a curve $C_1$ in $S_1$ to some curve $C_2$ in $S_2$. We indicate the data of $C_1$ and $C_2$ at the two ends of the corresponding edge. We refer the reader to Sections 5.1, 5.2 and Appendix A of \cite{Bhardwaj:2019fzv} for further geometric details used throughout the rest of this paper. The proposed resolved CY3 are described below. We will use an integer $\nu$ to parametrize different CY3s associated to a single KK theory building block, with the CY3s for different values of $\nu$ related by flop transitions. This parameter $\nu$ will be helpful for us when we discuss the gluing rules in Section \ref{GR}.

\be\label{PCY1}
\begin{tikzpicture} [scale=1.9]
\node (v2) at (0.6,1) {$\mathbf{(n-1)_{n-9+\nu}}$};
\node (v3) at (2.4,1) {$\mathbf{(n-2)_{n-11+\nu}}$};
\node (v10) at (0.6,-0.6) {$\mathbf{n_0^{(n-1)+1+\nu}}$};
\node (v4) at (4.3,1) {$\mathbf{(n-3)_{n-13+\nu}}$};
\node (v5) at (5.5,1) {$\cdots$};
\node (v7) at (6.4,-0.6) {$\mathbf{0^{(n+4-\nu)+(n+4-\nu)}_{6}}$};
\node (v6) at (6.4,1) {$\mathbf{1_{n+5-\nu}}$};
\draw  (v2) edge (v3);
\draw  (v3) edge (v4);
\draw  (v4) edge (v5);
\draw  (v5) edge (v6);
\draw  (v6) edge (v7);
\node at (3.5,1.1) {\scriptsize{$h$}};
\node[rotate=0] at (0.3,0.6) {\scriptsize{$h$+$f$}};
\node at (1.3,1.1) {\scriptsize{$e$}};
\node at (1.65,1.1) {\scriptsize{$h$}};
\node at (3.2,1.1) {\scriptsize{$e$}};
\node at (5.1,1.1) {\scriptsize{$e$}};
\node at (5.9,1.1) {\scriptsize{$e$}};
\node[rotate=0] at (6.6,0.7) {\scriptsize{$2h$}};
\node at (6.9,-0.3) {\scriptsize{$e$-$\sum x_i$-$\sum y_i$}};
\node at (1.55,-0.05) {\scriptsize{$x_2$-$x_1$}};
\node at (0,-0.2) {\scriptsize{$2e$+$f$-$\sum x_i$-$\sum z_i$}};
\node at (5.8,0.7) {\scriptsize{$f$}};
\draw  (v2) edge (v10);
\draw  (v3) edge (v10);
\node at (1.9,0.7) {\scriptsize{$f$}};
\node at (5.4,-0.75) {\scriptsize{$f,f$}};
\node at (3.6,0.8) {\scriptsize{$f$}};
\node at (1.95,-0.75) {\scriptsize{$f$-$x_{n\text{-}1}$-$x_{n\text{-}2},x_{n\text{-}1}$-$x_{n\text{-}2}$}};
\begin{scope}[shift={(4.1,1.1)}]
\node (v1_1) at (-0.5,0.7) {1};
\node at (-0.45,1) {$\su(2n+1)^{(2)}$};
\draw  (-0.5,0.9) ellipse (0.75 and 0.6);
\end{scope}
\draw  (v10) edge (v4);
\node[rotate=0] at (1,0) {\scriptsize{$x_{1}$-$y$}};
\node at (2.2,-0.2) {$\cdots$};
\draw (v10) .. controls (4.9,-0.3) and (5.5,0.2) .. (v6);
\node[rotate=5] at (2.6,-0.35) {\scriptsize{$x_{n\text{-}2}$-$x_{n\text{-}3}$}};
\node (v1) at (3.9,-0.6) {\scriptsize{2}};
\draw  (v10) edge (v1);
\draw  (v1) edge (v7);
\node at (5.9,-1.1) {\scriptsize{$x_i$}};
\node at (6.95,-1.1) {\scriptsize{$y_i$}};
\node (v) at (6.4,-1.5) {\scriptsize{$n+4-\nu$}};
\draw (v7) .. controls (6.3,-0.8) and (5.8,-1.2) .. (v);
\draw (v7) .. controls (6.5,-0.8) and (7,-1.2) .. (v);
\end{tikzpicture}
\ee
where $n\ge3$, and let us clarify that there is an edge between $S_i$ and $S_n$ gluing $x_{n-1-i}-x_{n-2-i}$ to $f_i$ for $1\le i\le n-3$. For $n=2$, we have
\be\label{su5T}
\begin{tikzpicture} [scale=1.9]
\node (v2) at (0.5,1) {$\mathbf{1_{5-\nu}}$};
\node (v3) at (2.9,1) {$\mathbf{0^{(6-\nu)+(6-\nu)}_{6}}$};
\node (v10) at (0.5,-0.6) {$\mathbf{2_0^{1+1+\nu}}$};
\draw  (v2) edge (v3);
\node[rotate=0] at (0.3,0.7) {\scriptsize{$e$}};
\node at (1,1.1) {\scriptsize{$2h$+$2f$}};
\node at (1.9,1.1) {\scriptsize{$e$-$\sum x_i$-$\sum y_i$}};
\node at (0,-0.2) {\scriptsize{$2e$+$f$-$x$-$\sum z_i$}};
\draw  (v2) edge (v10);
\node at (2.6,0.6) {\scriptsize{$f,f$}};
\begin{scope}[shift={(2.3,1.3)}]
\node (v1_1) at (-0.5,0.7) {1};
\node at (-0.45,1) {$\su(5)^{(2)}$};
\draw  (-0.5,0.9) ellipse (0.5 and 0.5);
\end{scope}
\node[rotate=0] at (1.3,-0.4) {\scriptsize{$f$-$x$-$y$,$x$-$y$}};
\node (v1) at (1.75,0.2) {\scriptsize{2}};
\draw  (v10) edge (v1);
\draw  (v1) edge (v3);
\node at (3.4,1.5) {\scriptsize{$x_i$}};
\node at (3.4,0.5) {\scriptsize{$y_i$}};
\node (v) at (3.9,1) {\scriptsize{$6-\nu$}};
\draw (v3) .. controls (3.2,1.4) and (3.8,1.6) .. (v);
\draw (v3) .. controls (3.2,0.6) and (3.8,0.4) .. (v);
\end{tikzpicture}
\ee

\be\label{ACY1}
\begin{tikzpicture} [scale=1.9]
\node (v2) at (0.6,1) {$\mathbf{(n-1)_{n-10+\nu}}$};
\node (v3) at (2.5,1) {$\mathbf{(n-2)_{n-12+\nu}}$};
\node (v10) at (0.6,-0.6) {$\mathbf{n_0^{(n-2)+1+\nu}}$};
\node (v4) at (4.4,1) {$\mathbf{(n-3)_{n-14+\nu}}$};
\node (v5) at (5.6,1) {$\cdots$};
\node (v7) at (6.5,-0.6) {$\mathbf{0^{(n+4-\nu)+(n+4-\nu)}_{n+6-\nu}}$};
\node (v6) at (6.5,1) {$\mathbf{2_{n+4-\nu}}$};
\draw  (v2) edge (v3);
\draw  (v3) edge (v4);
\draw  (v4) edge (v5);
\draw  (v5) edge (v6);
\draw  (v6) edge (v7);
\node at (3.6,1.1) {\scriptsize{$h$}};
\node[rotate=0] at (0.3,0.6) {\scriptsize{$h$+$f$}};
\node at (1.4,1.1) {\scriptsize{$e$}};
\node at (1.7,1.1) {\scriptsize{$h$}};
\node at (3.3,1.1) {\scriptsize{$e$}};
\node at (5.2,1.1) {\scriptsize{$e$}};
\node at (6,1.1) {\scriptsize{$e$}};
\node[rotate=0] at (6.4,0.6) {\scriptsize{$h$}};
\node at (6.4,-0.2) {\scriptsize{$e$}};
\node at (1.6,-0.05) {\scriptsize{$x_2$-$x_1$}};
\node at (0,-0.2) {\scriptsize{$2e$+$f$-$\sum x_i$-$\sum z_i$}};
\node at (5.9,0.7) {\scriptsize{$f$}};
\draw  (v2) edge (v10);
\draw  (v3) edge (v10);
\node at (2,0.7) {\scriptsize{$f$}};
\node at (5.4,-0.75) {\scriptsize{$f$}};
\node at (3.7,0.8) {\scriptsize{$f$}};
\node at (2,-0.75) {\scriptsize{$f$-$x_{n\text{-}2}$-$x_{n\text{-}3}$}};
\begin{scope}[shift={(4.8,1.1)}]
\node (v1_1) at (-0.5,0.7) {1};
\node at (-0.45,1) {$\su(2n)^{(2)}$};
\draw  (-0.5,0.9) ellipse (0.55 and 0.5);
\end{scope}
\draw  (v10) edge (v4);
\node[rotate=0] at (1,0) {\scriptsize{$x_{1}$-$y$}};
\node at (2.3,-0.2) {$\cdots$};
\draw (v10) .. controls (5,-0.3) and (5.5,0.2) .. (v6);
\node[rotate=5] at (2.7,-0.35) {\scriptsize{$x_{n\text{-}3}$-$x_{n\text{-}4}$}};
\draw  (v10) edge (v7);
\node (v1) at (8.3,1) {$\mathbf{1_{n+6-\nu}}$};
\draw  (v6) edge (v1);
\node at (7,1.1) {\scriptsize{$h$}};
\node at (7.8,1.1) {\scriptsize{$e$}};
\draw (v10) .. controls (2.5,-1.3) and (9.3,-2.4) .. (v1);
\node[rotate=0] at (1.6,-1.1) {\scriptsize{$x_{n\text{-}2}$-$x_{n\text{-}3}$}};
\node at (8.5,0.7) {\scriptsize{$f$}};
\node (v8) at (7.4,0.2) {\scriptsize{$n+4-\nu$}};
\draw  (v7) edge (v8);
\draw  (v8) edge (v1);
\node at (8.05,0.6) {\scriptsize{$f$}};
\node at (7.2,-0.3) {\scriptsize{$f$-$x_i$-$y_i$}};
\end{tikzpicture}
\ee
for $n\ge4$. For $n=3$, we have
\be\label{RCY1}
\begin{tikzpicture} [scale=1.9]
\node (v2) at (0.5,1) {$\mathbf{2_{5-\nu}}$};
\node (v3) at (2.3,1) {$\mathbf{0^{(7-\nu)+(7-\nu)}_{9-\nu}}$};
\node (v10) at (0.5,-0.6) {$\mathbf{1_{9-\nu}}$};
\draw  (v2) edge (v3);
\node[rotate=0] at (0.4,-0.2) {\scriptsize{$e$}};
\node at (1,1.1) {\scriptsize{$h$+$f$}};
\node at (1.6,1.1) {\scriptsize{$e$}};
\node at (-0.8,1.1) {\scriptsize{$2e$+$f$-$x$-$\sum z_i$}};
\draw  (v2) edge (v10);
\node at (0.9,-0.4) {\scriptsize{$f$}};
\begin{scope}[shift={(1,1.5)}]
\node (v1_1) at (-0.5,0.7) {1};
\node at (-0.45,1) {$\su(6)^{(2)}$};
\draw  (-0.5,0.9) ellipse (0.5 and 0.5);
\end{scope}
\node[rotate=0] at (-1.7,1.4) {\scriptsize{$f$-$x$-$y$}};
\node (v1) at (-1.7,1) {$\mathbf{3^{1+1+\nu}_{0}}$};
\draw  (v1) edge (v2);
\node at (0.3,0.7) {\scriptsize{$h$+$f$}};
\node at (0.1,1.1) {\scriptsize{$e$}};
\node[rotate=0] at (-1.5,0.6) {\scriptsize{$x$-$y$}};
\node at (0,-0.4) {\scriptsize{$f$}};
\draw  (v1) edge (v10);
\node at (2.3,1.4) {\scriptsize{$f$}};
\node at (2.2,0.6) {\scriptsize{$f$-$x_i$-$y_i$}};
\node (v4) at (1.4,0.2) {\scriptsize{$7-\nu$}};
\draw  (v10) edge (v4);
\draw  (v4) edge (v3);
\draw (v1) .. controls (-1.4,1.8) and (2,1.8) .. (v3);
\end{tikzpicture}
\ee

\be\label{PCY2}
\begin{tikzpicture} [scale=1.9]
\node (v2) at (0.6,1) {$\mathbf{(n-1)_{n-1+\nu}}$};
\node (v3) at (2.4,1) {$\mathbf{(n-2)_{n-3+\nu}}$};
\node (v10) at (0.6,-0.6) {$\mathbf{n_0^{1+1+n+\nu}}$};
\node (v4) at (4.2,1) {$\mathbf{(n-3)_{n-5+\nu}}$};
\node (v5) at (5.4,1) {$\cdots$};
\node (v7) at (6.3,-0.6) {$\mathbf{0^{(n-4-\nu)+(n-4-\nu)}_{6}}$};
\node (v6) at (6.3,1) {$\mathbf{1_{n-3-\nu}}$};
\draw  (v2) edge (v3);
\draw  (v3) edge (v4);
\draw  (v4) edge (v5);
\draw  (v5) edge (v6);
\draw  (v6) edge (v7);
\node at (3.5,1.1) {\scriptsize{$h$}};
\node[rotate=0] at (0.3,0.6) {\scriptsize{$h,f$}};
\node at (1.3,1.1) {\scriptsize{$e$}};
\node at (1.7,1.1) {\scriptsize{$h$}};
\node at (3.1,1.1) {\scriptsize{$e$}};
\node at (4.9,1.1) {\scriptsize{$e$}};
\node at (5.8,1.1) {\scriptsize{$e$}};
\node[rotate=0] at (6.5,0.7) {\scriptsize{$2h$}};
\node at (6.8,-0.3) {\scriptsize{$e$-$\sum x_i$-$\sum y_i$}};
\node at (1.55,-0.05) {\scriptsize{$z_3$-$z_2$}};
\node at (-0.4,-0.2) {\scriptsize{$2e$+$f$-$\sum z_i$-$x$-$2y$-$\sum w_i,z_1$-$x$}};
\node at (5.7,0.7) {\scriptsize{$f$}};
\draw  (v3) edge (v10);
\node at (1.9,0.7) {\scriptsize{$f$}};
\node at (5.3,-0.75) {\scriptsize{$f,f$}};
\node at (3.5,0.8) {\scriptsize{$f$}};
\node at (1.9,-0.75) {\scriptsize{$f$-$z_{n\text{-}1}$-$z_n,z_n$-$z_{n\text{-}1}$}};
\begin{scope}[shift={(4.1,1.1)}]
\node (v1_1) at (-0.5,0.7) {1};
\node at (-0.45,1) {$\su(\wh{2n+1})^{(2)}$};
\draw  (-0.5,0.9) ellipse (0.75 and 0.6);
\end{scope}
\draw  (v10) edge (v4);
\node[rotate=0] at (1,0) {\scriptsize{$z_{2}$-$z_1$}};
\node at (2.2,-0.2) {$\cdots$};
\draw (v10) .. controls (5,-0.3) and (5.4,0.2) .. (v6);
\node[rotate=5] at (2.6,-0.35) {\scriptsize{$z_{n\text{-}1}$-$z_{n\text{-}2}$}};
\draw (v10) .. controls (0,-1.2) and (1.2,-1.2) .. (v10);
\node at (0.2,-1) {\scriptsize{$x$}};
\node at (1,-1) {\scriptsize{$y$}};
\node (v1) at (0.6,0.2) {\scriptsize{2}};
\draw  (v2) edge (v1);
\draw  (v1) edge (v10);
\node (v8) at (3.8,-0.6) {\scriptsize{2}};
\draw  (v10) edge (v8);
\draw  (v8) edge (v7);
\node at (5.8,-1.1) {\scriptsize{$x_i$}};
\node at (6.85,-1.1) {\scriptsize{$y_i$}};
\node (v) at (6.3,-1.5) {\scriptsize{$n-4-\nu$}};
\draw (v7) .. controls (6.2,-0.8) and (5.7,-1.2) .. (v);
\draw (v7) .. controls (6.4,-0.8) and (7,-1.2) .. (v);
\end{tikzpicture}
\ee

\be\label{ACY2}
\begin{tikzpicture} [scale=1.9]
\node (v2) at (0.6,1) {$\mathbf{(n-1)_{n-2+\nu}}$};
\node (v3) at (2.3,1) {$\mathbf{(n-2)_{n-4+\nu}}$};
\node (v10) at (0.6,-0.6) {$\mathbf{n_0^{1+1+(n-1)+\nu}}$};
\node (v4) at (4,1) {$\mathbf{(n-3)_{n-6+\nu}}$};
\node (v5) at (5.05,1) {$\cdots$};
\node (v7) at (5.9,-0.6) {$\mathbf{0^{(n-4-\nu)+(n-4-\nu)}_{n-2-\nu}}$};
\node (v6) at (5.9,1) {$\mathbf{2_{n-4-\nu}}$};
\draw  (v2) edge (v3);
\draw  (v3) edge (v4);
\draw  (v4) edge (v5);
\draw  (v5) edge (v6);
\draw  (v6) edge (v7);
\node at (3.3,1.1) {\scriptsize{$h$}};
\node[rotate=0] at (0.3,0.6) {\scriptsize{$h,f$}};
\node at (1.3,1.1) {\scriptsize{$e$}};
\node at (1.6,1.1) {\scriptsize{$h$}};
\node at (3,1.1) {\scriptsize{$e$}};
\node at (4.7,1.1) {\scriptsize{$e$}};
\node at (5.4,1.1) {\scriptsize{$e$}};
\node[rotate=0] at (5.8,0.6) {\scriptsize{$h$}};
\node at (5.8,-0.2) {\scriptsize{$e$}};
\node at (1.45,-0.05) {\scriptsize{$z_3$-$z_2$}};
\node at (-0.4,-0.2) {\scriptsize{$2e$+$f$-$\sum z_i$-$x$-$2y$-$\sum w_i,z_1$-$x$}};
\node at (5.3,0.7) {\scriptsize{$f$}};
\draw  (v3) edge (v10);
\node at (1.8,0.7) {\scriptsize{$f$}};
\node at (4.9,-0.75) {\scriptsize{$f$}};
\node at (3.3,0.8) {\scriptsize{$f$}};
\node at (1.8,-0.75) {\scriptsize{$f$-$z_{n\text{-}1}$-$z_{n\text{-}2}$}};
\begin{scope}[shift={(4.7,1.1)}]
\node (v1_1) at (-0.5,0.7) {1};
\node at (-0.45,1) {$\su(\wh{2n})^{(2)}$};
\draw  (-0.5,0.9) ellipse (0.55 and 0.5);
\end{scope}
\draw  (v10) edge (v4);
\node[rotate=0] at (1,0) {\scriptsize{$z_{2}$-$z_1$}};
\node at (2.1,-0.2) {$\cdots$};
\draw (v10) .. controls (4.6,-0.3) and (5,0.2) .. (v6);
\node[rotate=5] at (2.5,-0.35) {\scriptsize{$z_{n\text{-}2}$-$z_{n\text{-}3}$}};
\draw (v10) .. controls (0,-1.3) and (1.2,-1.3) .. (v10);
\node at (0.2,-1) {\scriptsize{$x$}};
\node at (1,-1) {\scriptsize{$y$}};
\node (v1) at (0.6,0.2) {\scriptsize{2}};
\draw  (v2) edge (v1);
\draw  (v1) edge (v10);
\node (v9) at (7.7,1) {$\mathbf{1_{n-2-\nu}}$};
\draw  (v6) edge (v9);
\node[rotate=0] at (6.4,1.1) {\scriptsize{$h$}};
\node[rotate=0] at (7.2,1.1) {\scriptsize{$e$}};
\draw (v10) .. controls (2.3,-1.3) and (7.9,-2.6) .. (v9);
\node at (1.7,-1.2) {\scriptsize{$z_{n\text{-}1}$-$z_{n\text{-}2}$}};
\draw  (v10) edge (v7);
\node at (7.9,0.7) {\scriptsize{$f$}};
\node (v8) at (6.8,0.2) {\scriptsize{$n-4-\nu$}};
\draw  (v7) edge (v8);
\draw  (v8) edge (v9);
\node at (7.45,0.6) {\scriptsize{$f$}};
\node at (6.6,-0.3) {\scriptsize{$f$-$x_i$-$y_i$}};
\end{tikzpicture}
\ee

\be\label{RCY2}
\begin{tikzpicture} [scale=1.9]
\node (v2) at (0.6,-0.2) {$\mathbf{2_{5-\nu}}$};
\node (v3) at (2.3,-0.2) {$\mathbf{1_{11-\nu}}$};
\node (v10) at (0.6,-1.7) {$\mathbf{3_0^{1+1+\nu}}$};
\draw  (v2) edge (v3);
\node[rotate=0] at (0.5,0.1) {\scriptsize{$h$}};
\node at (1.1,-0.1) {\scriptsize{$h$+$2f$}};
\node at (1.8,-0.1) {\scriptsize{$e$}};
\node at (0.1,-1.3) {\scriptsize{$2e$+$f$-$x$-$\sum z_i$}};
\draw  (v2) edge (v10);
\node at (2.2,-0.5) {\scriptsize{$f,f$}};
\begin{scope}[shift={(1.9,1)}]
\node (v1_1) at (-0.5,0.7) {1};
\node at (-0.45,1) {$\su(\tilde6)^{(2)}$};
\draw  (-0.5,0.9) ellipse (0.5 and 0.5);
\end{scope}
\node[rotate=0] at (1.35,-1.4) {\scriptsize{$f$-$x$-$y,x$-$y$}};
\node (v1) at (0.6,1.1) {$\mathbf{0^{(7-\nu)+(7-\nu)}_{7-\nu}}$};
\draw  (v1) edge (v2);
\node at (0.4,-0.5) {\scriptsize{$e$}};
\node at (0.5,0.8) {\scriptsize{$e$}};
\node (v4) at (1.5,0.4) {\scriptsize{$7-\nu$}};
\draw  (v1) edge (v4);
\draw  (v4) edge (v3);
\node at (1.3,0.9) {\scriptsize{$f$-$x_i$-$y_i$}};
\node at (2.1,0.15) {\scriptsize{$f$}};
\node (v5) at (1.5,-0.9) {\scriptsize{2}};
\draw  (v10) edge (v5);
\draw  (v5) edge (v3);
\end{tikzpicture}
\ee

\be
\begin{tikzpicture} [scale=1.9]
\node (v2) at (0.6,1) {$\mathbf{3_{3-\nu}}$};
\node (v3) at (2.3,1) {$\mathbf{2_{1-\nu}}$};
\node (v10) at (0.6,-0.6) {$\mathbf{4_6^{1+1+(3-\nu)+(3-\nu)}}$};
\draw  (v2) edge (v3);
\node[rotate=0] at (0.1,-0.3) {\scriptsize{$e$-$\sum z_i$-$\sum w_i$}};
\node at (1,1.1) {\scriptsize{$e$}};
\node at (2.7,1.1) {\scriptsize{$e$}};
\node at (0.3,0.7) {\scriptsize{$2h$+$f$}};
\draw  (v2) edge (v10);
\draw  (v3) edge (v10);
\node at (1.9,0.8) {\scriptsize{$f$}};
\begin{scope}[shift={(2.8,1)}]
\node (v1_1) at (-0.5,0.7) {3};
\node at (-0.45,1) {$\so(10)^{(2)}$};
\draw  (-0.5,0.9) ellipse (0.5 and 0.5);
\end{scope}
\node[rotate=0] at (0.95,0) {\scriptsize{$f$-$x$-$y$}};
\node (v1) at (4,1) {$\mathbf{1^{1}_{1+\nu}}$};
\draw  (v3) edge (v1);
\node at (1.9,1.1) {\scriptsize{$h$}};
\node at (3.6,1.1) {\scriptsize{$e$}};
\draw  (v10) edge (v1);
\node[rotate=0] at (1.4,-0.1) {\scriptsize{$y$}};
\node (v4) at (4,-0.6) {$\mathbf{0^{\nu+\nu}_{5}}$};
\draw  (v1) edge (v4);
\node at (4.3,0.7) {\scriptsize{$2h$-$x$}};
\node[rotate=0] at (4.5,-0.3) {\scriptsize{$e$-$\sum x_i$-$\sum y_i$}};
\draw  (v10) edge (v4);
\node[rotate=0] at (1.6,-0.5) {\scriptsize{$x$-$y$}};
\node at (3.6,-0.5) {\scriptsize{$f$}};
\node[rotate=0] at (3.3,0.8) {\scriptsize{$f$-$x$}};
\node (v) at (0.6,-1.3) {\scriptsize{$3-\nu$}};
\draw (v10) .. controls (0.5,-0.8) and (0,-1.2) .. (v);
\draw (v10) .. controls (0.7,-0.8) and (1.2,-1.2) .. (v);
\node at (0.1,-1.1) {\scriptsize{$z_i$}};
\node at (1.1,-1.1) {\scriptsize{$w_i$}};
\begin{scope}[shift={(3.4,0)}]
\node (w) at (0.6,-1.3) {\scriptsize{$\nu$}};
\draw (v4) .. controls (0.5,-0.8) and (0,-1.2) .. (w);
\draw (v4) .. controls (0.7,-0.8) and (1.2,-1.2) .. (w);
\node at (0.1,-1.1) {\scriptsize{$x_i$}};
\node at (1.1,-1.1) {\scriptsize{$y_i$}};
\end{scope}
\end{tikzpicture}
\ee

\be
\begin{tikzpicture} [scale=1.9]
\node (v2) at (0.6,1) {$\mathbf{3_{3}}$};
\node (v3) at (2.3,1) {$\mathbf{2_{1}}$};
\node (v10) at (0.6,-0.6) {$\mathbf{4_6^{3+3+5+5}}$};
\draw  (v2) edge (v3);
\node[rotate=0] at (0.1,-0.3) {\scriptsize{$e$-$\sum z_i$-$\sum w_i$}};
\node at (0.9,1.1) {\scriptsize{$e$}};
\node at (2.6,1.1) {\scriptsize{$e$}};
\node at (0.3,0.7) {\scriptsize{$2h$+$3f$}};
\draw  (v2) edge (v10);
\node at (1.9,0.8) {\scriptsize{$f$}};
\begin{scope}[shift={(3.1,1)}]
\node (v1_1) at (-0.5,0.7) {1};
\node at (-0.45,1) {$\so(10)^{(2)}$};
\draw  (-0.5,0.9) ellipse (0.5 and 0.5);
\end{scope}
\node[rotate=0] at (0.95,0) {\scriptsize{$f$-$x_i$-$y_i$}};
\node (v1) at (4.3,1) {$\mathbf{1^{3}_{1}}$};
\draw  (v3) edge (v1);
\node at (2,1.1) {\scriptsize{$h$}};
\node at (4,1.1) {\scriptsize{$e$}};
\node[rotate=0] at (1.5,-0.1) {\scriptsize{$y_i$}};
\node (v4) at (4.3,-0.6) {$\mathbf{0_{3}}$};
\draw  (v1) edge (v4);
\node at (4.7,0.7) {\scriptsize{$2h$-$\sum x_i$}};
\node[rotate=0] at (4.5,-0.3) {\scriptsize{$e$}};
\node[rotate=0] at (1.8,-0.5) {\scriptsize{$x_1$-$y_2,x_2$-$y_1,x_3$-$y_3$}};
\node at (3.9,-0.5) {\scriptsize{$f,f,f$}};
\node[rotate=0] at (3.5,0.8) {\scriptsize{$f$-$x_i$}};
\node (v5) at (1.6,0.3) {\scriptsize{3}};
\draw  (v10) edge (v5);
\draw  (v5) edge (v3);
\node (v6) at (2.5,0.2) {\scriptsize{3}};
\draw  (v10) edge (v6);
\draw  (v6) edge (v1);
\node (v7) at (3,-0.6) {\scriptsize{3}};
\draw  (v10) edge (v7);
\draw  (v7) edge (v4);
\node (v) at (0.6,-1.3) {\scriptsize{5}};
\draw (v10) .. controls (0.5,-0.8) and (0,-1.2) .. (v);
\draw (v10) .. controls (0.7,-0.8) and (1.2,-1.2) .. (v);
\node at (0.1,-1.1) {\scriptsize{$z_i$}};
\node at (1.1,-1.1) {\scriptsize{$w_i$}};
\end{tikzpicture}
\ee

\be
\begin{tikzpicture} [scale=1.9]
\node (v2) at (0.6,1) {$\mathbf{3_{0}}$};
\node (v3) at (2.3,1) {$\mathbf{2_{6}}$};
\node (v10) at (0.6,-0.6) {$\mathbf{4_2^{1+1}}$};
\draw  (v2) edge (v3);
\node[rotate=0] at (0.4,-0.3) {\scriptsize{$e$}};
\node at (1,1.1) {\scriptsize{$2e$+$f$}};
\node at (2,1.1) {\scriptsize{$e$}};
\node at (0.4,0.7) {\scriptsize{$e$}};
\draw  (v2) edge (v10);
\draw  (v3) edge (v10);
\node at (1.9,0.8) {\scriptsize{$f$}};
\begin{scope}[shift={(2.8,1)}]
\node (v1_1) at (-0.5,0.7) {5};
\node at (-0.45,1) {$\fe_6^{(2)}$};
\draw  (-0.5,0.9) ellipse (0.5 and 0.5);
\end{scope}
\node[rotate=0] at (0.95,0) {\scriptsize{$f$-$x$-$y$}};
\node (v1) at (3.8,1) {$\mathbf{1^{1}_{8}}$};
\draw  (v3) edge (v1);
\node at (2.6,1.1) {\scriptsize{$h$}};
\node at (3.5,1.1) {\scriptsize{$e$}};
\draw  (v10) edge (v1);
\node[rotate=0] at (1.4,-0.1) {\scriptsize{$y$}};
\node (v4) at (3.8,-0.6) {$\mathbf{0_{10}}$};
\draw  (v1) edge (v4);
\node at (4,0.7) {\scriptsize{$h$}};
\node[rotate=0] at (4,-0.3) {\scriptsize{$e$}};
\draw  (v10) edge (v4);
\node[rotate=0] at (1.3,-0.5) {\scriptsize{$x$-$y$}};
\node at (3.4,-0.5) {\scriptsize{$f$}};
\node[rotate=0] at (3.1,0.8) {\scriptsize{$f$-$x$}};
\end{tikzpicture}
\ee

\be
\begin{tikzpicture} [scale=1.9]
\node (v2) at (0.6,1) {$\mathbf{3_{0}}$};
\node (v3) at (2.3,1) {$\mathbf{2_{6}}$};
\node (v10) at (0.6,-0.6) {$\mathbf{4_4^{3+3}}$};
\draw  (v2) edge (v3);
\node[rotate=0] at (0.4,-0.3) {\scriptsize{$e$}};
\node at (1,1.1) {\scriptsize{$2e$+$f$}};
\node at (2,1.1) {\scriptsize{$e$}};
\node at (0.4,0.7) {\scriptsize{$e$+$f$}};
\draw  (v2) edge (v10);
\node at (1.9,0.8) {\scriptsize{$f$}};
\begin{scope}[shift={(3.1,1)}]
\node (v1_1) at (-0.5,0.7) {3};
\node at (-0.45,1) {$\fe_6^{(2)}$};
\draw  (-0.5,0.9) ellipse (0.5 and 0.5);
\end{scope}
\node[rotate=0] at (0.95,0) {\scriptsize{$f$-$x_i$-$y_i$}};
\node (v1) at (4.3,1) {$\mathbf{1^{3}_{8}}$};
\draw  (v3) edge (v1);
\node at (2.6,1.1) {\scriptsize{$h$}};
\node at (4,1.1) {\scriptsize{$e$}};
\node[rotate=0] at (1.5,-0.1) {\scriptsize{$y_i$}};
\node (v4) at (4.3,-0.6) {$\mathbf{0_{10}}$};
\draw  (v1) edge (v4);
\node at (4.5,0.7) {\scriptsize{$h$}};
\node[rotate=0] at (4.5,-0.3) {\scriptsize{$e$}};
\node[rotate=0] at (1.7,-0.5) {\scriptsize{$x_1$-$y_2,x_2$-$y_1,x_3$-$y_3$}};
\node at (3.8,-0.5) {\scriptsize{$f,f,f$}};
\node[rotate=0] at (3.5,0.8) {\scriptsize{$f$-$x_i$}};
\node (v5) at (1.6,0.3) {\scriptsize{3}};
\draw  (v10) edge (v5);
\draw  (v5) edge (v3);
\node (v6) at (2.5,0.2) {\scriptsize{3}};
\draw  (v10) edge (v6);
\draw  (v6) edge (v1);
\node (v7) at (3,-0.6) {\scriptsize{3}};
\draw  (v10) edge (v7);
\draw  (v7) edge (v4);
\end{tikzpicture}
\ee

\be
\begin{tikzpicture} [scale=1.9]
\node (v2) at (0.6,1) {$\mathbf{3_{0}}$};
\node (v3) at (2.3,1) {$\mathbf{2_{6}}$};
\node (v10) at (0.6,-0.6) {$\mathbf{4_6^{5+5}}$};
\draw  (v2) edge (v3);
\node[rotate=0] at (0.4,-0.3) {\scriptsize{$e$}};
\node at (1,1.1) {\scriptsize{$2e$+$f$}};
\node at (2,1.1) {\scriptsize{$e$}};
\node at (0.3,0.7) {\scriptsize{$e$+$2f$}};
\draw  (v2) edge (v10);
\node at (1.9,0.8) {\scriptsize{$f$}};
\begin{scope}[shift={(3.1,1)}]
\node (v1_1) at (-0.5,0.7) {1};
\node at (-0.45,1) {$\fe_6^{(2)}$};
\draw  (-0.5,0.9) ellipse (0.5 and 0.5);
\end{scope}
\node[rotate=0] at (0.95,0) {\scriptsize{$f$-$x_i$-$y_i$}};
\node (v1) at (4.3,1) {$\mathbf{1^{5}_{8}}$};
\draw  (v3) edge (v1);
\node at (2.6,1.1) {\scriptsize{$h$}};
\node at (4,1.1) {\scriptsize{$e$}};
\node[rotate=0] at (1.5,-0.1) {\scriptsize{$y_i$}};
\node (v4) at (4.3,-0.6) {$\mathbf{0_{10}}$};
\draw  (v1) edge (v4);
\node at (4.5,0.7) {\scriptsize{$h$}};
\node[rotate=0] at (4.5,-0.3) {\scriptsize{$e$}};
\node[rotate=0] at (1.9,-0.5) {\scriptsize{$x_3$-$y_4,x_4$-$y_3,x_5$-$y_5$}};
\node at (3.7,-0.5) {\scriptsize{$f,f,f,f,f$}};
\node[rotate=0] at (3.5,0.8) {\scriptsize{$f$-$x_i$}};
\node (v5) at (1.6,0.3) {\scriptsize{5}};
\draw  (v10) edge (v5);
\draw  (v5) edge (v3);
\node (v6) at (2.5,0.2) {\scriptsize{5}};
\draw  (v10) edge (v6);
\draw  (v6) edge (v1);
\node (v7) at (2.9,-0.6) {\scriptsize{5}};
\draw  (v10) edge (v7);
\draw  (v7) edge (v4);
\node[rotate=0] at (2,-0.35) {\scriptsize{$x_1$-$y_2,x_2$-$y_1,$}};
\end{tikzpicture}
\ee

\be\label{Int}
\begin{tikzpicture} [scale=1.9]
\node (v2) at (0.6,1) {$\mathbf{4_{4}}$};
\node (v3) at (2.3,1) {$\mathbf{3_{2}}$};
\node (v10) at (0.6,-0.6) {$\mathbf{5_6^{2+2+5+5}}$};
\draw  (v2) edge (v3);
\node[rotate=0] at (0.1,-0.3) {\scriptsize{$e$-$\sum z_i$-$\sum w_i$}};
\node at (0.9,1.1) {\scriptsize{$e$}};
\node at (2.6,1.1) {\scriptsize{$e$}};
\node at (0.3,0.7) {\scriptsize{$2h$+$2f$}};
\draw  (v2) edge (v10);
\node at (1.9,0.8) {\scriptsize{$f$}};
\begin{scope}[shift={(4.4,1.3)}]
\node (v1_1) at (-0.5,0.7) {2};
\node at (-0.45,1) {$\so(11)^{(1)}$};
\draw  (-1.1,0.8) ellipse (1.3 and 0.7);
\end{scope}
\node[rotate=0] at (0.95,0) {\scriptsize{$f$-$x_i$-$y_i$}};
\node (v1) at (4.3,1) {$\mathbf{2^{2}_{0}}$};
\draw  (v3) edge (v1);
\node at (2,1.1) {\scriptsize{$h$}};
\node at (4,1.1) {\scriptsize{$e$}};
\node[rotate=0] at (1.5,-0.1) {\scriptsize{$y_i$}};
\node (v4) at (4.3,-0.6) {$\mathbf{1_{1}}$};
\draw  (v1) edge (v4);
\node at (4.5,0.7) {\scriptsize{$e$-$x_2$}};
\node[rotate=0] at (4.5,-0.3) {\scriptsize{$e$}};
\node at (3.9,-0.5) {\scriptsize{$f$}};
\node[rotate=0] at (3.5,0.8) {\scriptsize{$f$-$x_i$}};
\node (v5) at (1.6,0.3) {\scriptsize{2}};
\draw  (v10) edge (v5);
\draw  (v5) edge (v3);
\node (v6) at (2.5,0.2) {\scriptsize{2}};
\draw  (v10) edge (v6);
\draw  (v6) edge (v1);
\node[rotate=0] at (1.4,-0.5) {\scriptsize{$x_1$-$y_1$}};
\node (v8) at (2.6,2) {$\left[\Z_2^{(2)}\right]$};
\draw  (v8) edge (v1_1);
\draw  (v10) edge (v4);
\node (v7) at (5.9,1) {$\mathbf{0_{1}}$};
\node (v) at (0.6,-1.3) {\scriptsize{5}};
\draw (v10) .. controls (0.5,-0.8) and (0,-1.2) .. (v);
\draw (v10) .. controls (0.7,-0.8) and (1.2,-1.2) .. (v);
\node at (0.1,-1.1) {\scriptsize{$z_i$}};
\node at (1.1,-1.1) {\scriptsize{$w_i$}};
\draw  (v1) edge (v7);
\node at (4.7,1.1) {\scriptsize{$e$-$x_1$}};
\node[rotate=0] at (5.6,1.1) {\scriptsize{$e$}};
\draw (v10) .. controls (2.3,-1.3) and (5.9,-2.6) .. (v7);
\node at (5.7,0.7) {\scriptsize{$f$}};
\node[rotate=0] at (1.7,-0.9) {\scriptsize{$x_2$-$y_2$}};
\end{tikzpicture}
\ee

\be
\begin{tikzpicture} [scale=1.9]
\node (v2) at (0.6,1) {$\mathbf{4_6^2}$};
\node (v3) at (2.3,1) {$\mathbf{3_{4}}$};
\node (v10) at (-1.1,1) {$\mathbf{5_7}$};
\draw  (v2) edge (v3);
\node at (1.2,1.1) {\scriptsize{$e$}};
\node at (2.7,1.1) {\scriptsize{$e$}};
\node at (0.4,1.3) {\scriptsize{$h$-$x_2$}};
\draw  (v2) edge (v10);
\node at (-0.8,0.8) {\scriptsize{$f$}};
\begin{scope}[shift={(3.3,2.5)}]
\node (v1_1) at (-0.5,0.7) {2};
\node at (-0.45,1) {$\so(12)^{(1)}$};
\draw  (-1.1,0.8) ellipse (1.3 and 0.7);
\end{scope}
\node[rotate=0] at (0,2.1) {\scriptsize{$f$-$x_i$-$y_i$}};
\node (v1) at (4.3,1) {$\mathbf{2_{0}}$};
\draw  (v3) edge (v1);
\node at (2,1.1) {\scriptsize{$h$}};
\node at (3.9,1.1) {\scriptsize{$e$+$f$}};
\node (v4) at (4.3,-0.6) {$\mathbf{1_2^{1+1}}$};
\draw  (v1) edge (v4);
\node at (4.4,0.7) {\scriptsize{$e$}};
\node[rotate=0] at (4.4,-0.3) {\scriptsize{$e$}};
\node at (-0.9,1.3) {\scriptsize{$f$}};
\node[rotate=0] at (3.8,-0.6) {\scriptsize{$x$-$y$}};
\node (v8) at (1.5,3.2) {$\left[\Z_2^{(2)}\right]$};
\draw  (v8) edge (v1_1);
\draw  (v10) edge (v4);
\node (v7) at (4.3,2.3) {$\mathbf{0_2^{1+1}}$};
\node[rotate=0] at (0.5,1.9) {\scriptsize{$e$}};
\draw  (v1) edge (v7);
\node at (4.4,1.3) {\scriptsize{$e$}};
\node[rotate=0] at (3.8,2.4) {\scriptsize{$x$-$y$}};
\node (v9) at (0.6,2.3) {$\mathbf{6^{6+6}_{7}}$};
\draw  (v9) edge (v2);
\node at (0.2,1.1) {\scriptsize{$h$-$x_1$}};
\node[rotate=0] at (-0.7,1.1) {\scriptsize{$e$}};
\node (v11) at (-0.3,1.6) {\scriptsize{6}};
\draw  (v10) edge (v11);
\draw  (v11) edge (v9);
\draw  (v9) edge (v7);
\node at (1,2.4) {\scriptsize{$f$}};
\node at (4.4,2) {\scriptsize{$e$}};
\draw  (v3) edge (v7);
\node at (2.95,1.3) {\scriptsize{$f$}};
\node at (4,1.9) {\scriptsize{$f$-$x$-$y$}};
\draw  (v2) edge (v7);
\node at (3.4,2.1) {\scriptsize{$y$}};
\node at (1,1.3) {\scriptsize{$f$-$x_1$}};
\draw  (v3) edge (v4);
\node at (3.95,-0.1) {\scriptsize{$f$-$x$-$y$}};
\node at (2.9,0.7) {\scriptsize{$f$}};
\draw  (v2) edge (v4);
\node at (3.4,-0.1) {\scriptsize{$y$}};
\node at (1.5,0.8) {\scriptsize{$f$-$x_2$}};
\end{tikzpicture}
\ee

\be\label{int}
\begin{tikzpicture} [scale=1.9]
\node (v2) at (0.6,1) {$\mathbf{4_7^{2+1}}$};
\node (v3) at (2.3,1) {$\mathbf{3_{5}}$};
\node (v10) at (-1.1,1) {$\mathbf{5_7}$};
\draw  (v2) edge (v3);
\node at (1.2,1.1) {\scriptsize{$e$}};
\node at (2.7,1.1) {\scriptsize{$e$}};
\node at (0.4,1.3) {\scriptsize{$h$-$x_1$}};
\draw  (v2) edge (v10);
\node at (-0.8,0.8) {\scriptsize{$f$}};
\begin{scope}[shift={(3.5,3.5)}]
\node (v1_1) at (-0.5,0.7) {1};
\node at (-0.45,1) {$\so(\wh{12})^{(1)}$};
\draw  (-1.1,0.8) ellipse (1.3 and 0.7);
\end{scope}
\node[rotate=0] at (0,2.1) {\scriptsize{$f$-$x_i$-$y_i$}};
\node (v1) at (4.3,1) {$\mathbf{2_{1}}$};
\draw  (v3) edge (v1);
\node at (2,1.1) {\scriptsize{$h$}};
\node at (3.9,1.1) {\scriptsize{$h$+$f$}};
\node (v4) at (4.3,-0.6) {$\mathbf{1_3^{2+2}}$};
\draw  (v1) edge (v4);
\node at (4.4,0.7) {\scriptsize{$h$}};
\node[rotate=0] at (4.4,-0.3) {\scriptsize{$e$}};
\node at (-0.9,1.3) {\scriptsize{$f$}};
\node[rotate=0] at (3.55,-0.5) {\scriptsize{$x_1$-$y_1$}};
\node (v8) at (1.7,4.2) {$\left[\Z_2^{(2)}\right]$};
\draw  (v8) edge (v1_1);
\draw  (v10) edge (v4);
\node (v7) at (4.3,2.3) {$\mathbf{0_1^{1+1}}$};
\node[rotate=0] at (0.5,1.9) {\scriptsize{$e$}};
\draw  (v1) edge (v7);
\node at (4.4,1.3) {\scriptsize{$e$}};
\node[rotate=0] at (3.8,2.4) {\scriptsize{$x$-$y$}};
\node (v9) at (0.6,2.3) {$\mathbf{6^{7+7}_{8}}$};
\draw  (v9) edge (v2);
\node at (0.1,1.1) {\scriptsize{$h$-$x_2$-$y$}};
\node[rotate=0] at (-0.7,1.1) {\scriptsize{$e$}};
\node (v11) at (-0.3,1.6) {\scriptsize{7}};
\draw  (v10) edge (v11);
\draw  (v11) edge (v9);
\draw  (v9) edge (v7);
\node at (1.1,2.4) {\scriptsize{$f$}};
\node at (4.4,2) {\scriptsize{$e$}};
\draw  (v3) edge (v7);
\node at (2.95,1.3) {\scriptsize{$f$}};
\node at (4,1.9) {\scriptsize{$f$-$x$-$y$}};
\draw  (v2) edge (v7);
\node at (3.4,2.1) {\scriptsize{$y$}};
\node at (1,1.3) {\scriptsize{$f$-$y$}};
\node at (4,-0.05) {\scriptsize{$f$-$x_i$-$y_i$}};
\node at (2.9,0.7) {\scriptsize{$f$}};
\node at (3.4,-0.1) {\scriptsize{$y_i$}};
\node at (1.5,0.8) {\scriptsize{$f$-$x_i$}};
\node at (0.9,2.7) {\scriptsize{$f$}};
\node[rotate=0] at (5,-0.4) {\scriptsize{$x_2$-$y_2$}};
\node (v5) at (3.3,0.2) {\scriptsize{2}};
\draw  (v3) edge (v5);
\draw  (v5) edge (v4);
\node (v6) at (2.2,0.3) {\scriptsize{2}};
\draw  (v2) edge (v6);
\draw  (v6) edge (v4);
\draw (v4) .. controls (5.3,0) and (6.4,1.8) .. (5.2,2.7);
\draw (v9) .. controls (1.5,3) and (4.1,3.5) .. (5.2,2.7);
\end{tikzpicture}
\ee

\be
\begin{tikzpicture} [scale=1.9]
\node (v2) at (0.6,1) {$\mathbf{4_0}$};
\node (v3) at (2.2,1) {$\mathbf{5^{1+1}_{1}}$};
\node (v10) at (-1.1,1) {$\mathbf{3^{1+1}_1}$};
\draw  (v2) edge (v3);
\node at (0.9,1.1) {\scriptsize{$e$}};
\node at (2.6,1.1) {\scriptsize{$h$}};
\node at (0.4,1.4) {\scriptsize{$e$+$f$}};
\draw  (v2) edge (v10);
\begin{scope}[shift={(1.7,2.6)}]
\node (v1_1) at (-0.5,0.7) {6};
\node at (-0.45,1) {$\fe_7^{(1)}$};
\draw  (-1.1,0.8) ellipse (1.3 and 0.7);
\end{scope}
\node (v1) at (4.1,1) {$\mathbf{6_{3}}$};
\draw  (v3) edge (v1);
\node at (1.7,1.1) {\scriptsize{$e$-$x$}};
\node at (4.3,0.7) {\scriptsize{$h$+$f$}};
\node (v4) at (4.1,-0.6) {$\mathbf{0_7}$};
\draw  (v1) edge (v4);
\node at (3.8,1.1) {\scriptsize{$e$}};
\node[rotate=0] at (4.2,-0.3) {\scriptsize{$e$}};
\node at (0.1,2.2) {\scriptsize{$f$}};
\node (v8) at (-0.1,3.3) {$\left[\Z_2^{(2)}\right]$};
\draw  (v8) edge (v1_1);
\node (v7) at (-2.7,1) {$\mathbf{2_3}$};
\node[rotate=0] at (0.5,2) {\scriptsize{$e$}};
\node at (-2.8,-0.3) {\scriptsize{$e$}};
\node[rotate=0] at (-0.9,1.4) {\scriptsize{$x$-$y$}};
\node (v9) at (0.6,2.4) {$\mathbf{7_{4}}$};
\draw  (v9) edge (v2);
\node at (0.3,1.1) {\scriptsize{$e$}};
\node[rotate=0] at (-0.6,1.1) {\scriptsize{$e$-$x$}};
\node at (-2.3,-0.4) {\scriptsize{$f$}};
\node at (-2.4,1.1) {\scriptsize{$e$}};
\node at (1.1,2.2) {\scriptsize{$f$}};
\node at (-1.2,0.6) {\scriptsize{$f$-$x$-$y$}};
\draw  (v3) edge (v4);
\node at (2.3,0.6) {\scriptsize{$f$-$x$-$y$}};
\node at (3.6,-0.4) {\scriptsize{$f$}};
\draw  (v7) edge (v10);
\node (v5) at (-2.7,-0.6) {$\mathbf{1_7}$};
\draw  (v7) edge (v5);
\draw  (v10) edge (v5);
\draw  (v10) edge (v9);
\draw  (v9) edge (v3);
\node[rotate=0] at (2,1.4) {\scriptsize{$x$-$y$}};
\node at (-1.5,1.1) {\scriptsize{$h$}};
\node at (-2.9,0.7) {\scriptsize{$h$+$f$}};
\end{tikzpicture}
\ee

\be
\begin{tikzpicture} [scale=1.9]
\node (v2) at (0.6,1) {$\mathbf{4_2}$};
\node (v3) at (2.2,1) {$\mathbf{5^{3+3}_{1}}$};
\node (v10) at (-1.1,1) {$\mathbf{3^{1+1}_1}$};
\draw  (v2) edge (v3);
\node at (0.9,1.1) {\scriptsize{$h$}};
\node at (2.6,1.1) {\scriptsize{$h$}};
\node at (0.4,1.4) {\scriptsize{$h$+$f$}};
\draw  (v2) edge (v10);
\begin{scope}[shift={(1.7,2.6)}]
\node (v1_1) at (-0.5,0.7) {4};
\node at (-0.45,1) {$\fe_7^{(1)}$};
\draw  (-1.1,0.8) ellipse (1.3 and 0.7);
\end{scope}
\node (v1) at (4,1) {$\mathbf{6_{3}}$};
\draw  (v3) edge (v1);
\node at (1.6,1.1) {\scriptsize{$e$-$\sum x_i$}};
\node at (4.3,0.7) {\scriptsize{$h$+$3f$}};
\node (v4) at (4,-0.6) {$\mathbf{0_{11}}$};
\draw  (v1) edge (v4);
\node at (3.7,1.1) {\scriptsize{$e$}};
\node[rotate=0] at (4.2,-0.3) {\scriptsize{$e$}};
\node at (0.1,2.2) {\scriptsize{$f$}};
\node (v8) at (-0.1,3.3) {$\left[\Z_2^{(2)}\right]$};
\draw  (v8) edge (v1_1);
\node (v7) at (-2.7,1) {$\mathbf{2_1}$};
\node[rotate=0] at (0.5,2) {\scriptsize{$e$}};
\node at (-2.9,-0.3) {\scriptsize{$e$}};
\node[rotate=0] at (-0.9,1.4) {\scriptsize{$x$-$y$}};
\node (v9) at (0.6,2.4) {$\mathbf{7_{6}}$};
\draw  (v9) edge (v2);
\node at (0.3,1.1) {\scriptsize{$e$}};
\node[rotate=0] at (-0.6,1.1) {\scriptsize{$h$-$x$}};
\node at (-2.3,-0.4) {\scriptsize{$f$}};
\node at (-2.4,1.1) {\scriptsize{$e$}};
\node at (1.2,2.2) {\scriptsize{$f,f,f$}};
\node at (-1.2,0.6) {\scriptsize{$f$-$x$-$y$}};
\node at (2.3,0.6) {\scriptsize{$f$-$x_i$-$y_i$}};
\node at (3.5,-0.4) {\scriptsize{$f$}};
\draw  (v7) edge (v10);
\node (v5) at (-2.7,-0.6) {$\mathbf{1_5}$};
\draw  (v7) edge (v5);
\draw  (v10) edge (v5);
\draw  (v10) edge (v9);
\node[rotate=0] at (2.5,1.4) {\scriptsize{$x_1$-$y_2,x_2$-$y_1,x_3$-$y_3$}};
\node at (-1.5,1.1) {\scriptsize{$e$}};
\node at (-2.9,0.7) {\scriptsize{$h$+$f$}};
\node (v6) at (3.1,0.2) {\scriptsize{3}};
\draw  (v3) edge (v6);
\draw  (v6) edge (v4);
\node (v11) at (1.4,1.7) {\scriptsize{3}};
\draw  (v9) edge (v11);
\draw  (v11) edge (v3);
\end{tikzpicture}
\ee

\be
\begin{tikzpicture} [scale=1.9]
\node (v2) at (0.6,1) {$\mathbf{4_4}$};
\node (v3) at (2.2,1) {$\mathbf{5^{5+5}_{1}}$};
\node (v10) at (-1.1,1) {$\mathbf{3^{1+1}_3}$};
\draw  (v2) edge (v3);
\node at (0.9,1.1) {\scriptsize{$h$}};
\node at (2.6,1.1) {\scriptsize{$h$}};
\node at (0.4,1.4) {\scriptsize{$h$+$f$}};
\draw  (v2) edge (v10);
\begin{scope}[shift={(1.7,2.6)}]
\node (v1_1) at (-0.5,0.7) {2};
\node at (-0.45,1) {$\fe_7^{(1)}$};
\draw  (-1.1,0.8) ellipse (1.3 and 0.7);
\end{scope}
\node (v1) at (4,1) {$\mathbf{6_{3}}$};
\draw  (v3) edge (v1);
\node at (1.6,1.1) {\scriptsize{$e$-$\sum x_i$}};
\node at (4.3,0.7) {\scriptsize{$h$+$5f$}};
\node (v4) at (4,-0.6) {$\mathbf{0_{15}}$};
\draw  (v1) edge (v4);
\node at (3.7,1.1) {\scriptsize{$e$}};
\node[rotate=0] at (4.2,-0.3) {\scriptsize{$e$}};
\node at (0.1,2.2) {\scriptsize{$f$}};
\node (v8) at (-0.1,3.3) {$\left[\Z_2^{(2)}\right]$};
\draw  (v8) edge (v1_1);
\node (v7) at (-2.7,1) {$\mathbf{2_1}$};
\node[rotate=0] at (0.5,2) {\scriptsize{$e$}};
\node at (-2.9,-0.3) {\scriptsize{$e$}};
\node[rotate=0] at (-0.9,1.4) {\scriptsize{$x$-$y$}};
\node (v9) at (0.6,2.4) {$\mathbf{7_{8}}$};
\draw  (v9) edge (v2);
\node at (0.3,1.1) {\scriptsize{$e$}};
\node[rotate=0] at (-0.6,1.1) {\scriptsize{$h$-$x$}};
\node at (-2.3,-0.4) {\scriptsize{$f$}};
\node at (-2.4,1.1) {\scriptsize{$h$}};
\node at (1.3,2.2) {\scriptsize{$f,f,f,f,f$}};
\node at (-1.2,0.6) {\scriptsize{$f$-$x$-$y$}};
\node at (2.3,0.6) {\scriptsize{$f$-$x_i$-$y_i$}};
\node at (3.5,-0.4) {\scriptsize{$f$}};
\draw  (v7) edge (v10);
\node (v5) at (-2.7,-0.6) {$\mathbf{1_3}$};
\draw  (v7) edge (v5);
\draw  (v10) edge (v5);
\draw  (v10) edge (v9);
\node[rotate=0] at (2.9,1.4) {\scriptsize{$x_1$-$y_2,x_2$-$y_1,x_3$-$y_4,x_4$-$y_3,x_5$-$y_5$}};
\node at (-1.5,1.1) {\scriptsize{$e$}};
\node at (-2.9,0.7) {\scriptsize{$h$}};
\node (v6) at (3.1,0.2) {\scriptsize{5}};
\draw  (v3) edge (v6);
\draw  (v6) edge (v4);
\node (v11) at (1.4,1.7) {\scriptsize{5}};
\draw  (v9) edge (v11);
\draw  (v11) edge (v3);
\end{tikzpicture}
\ee

Let us also collect CY3s associated to some KK theory building blocks already discussed in \cite{Bhardwaj:2019fzv}. We are doing so because we will need some flop frames of these CY3s (in Section \ref{GR}) which were not described in \cite{Bhardwaj:2019fzv}.

\be
\begin{tikzpicture}[scale=1.9]
\node (v2) at (-2.9,-0.5) {$\mathbf{n^\nu_1}$};
\node (v3) at (-1,-0.5) {$\mathbf{(n-1)_{6-\nu}}$};
\node (v4) at (0.4,-0.5) {$\mathbf{\cdots}$};
\node (v5) at (1.65,-0.5) {$\mathbf{2_{2n-\nu}}$};
\node (v6) at (3.5,-1.55) {$\mathbf{0^{(2n-\nu)+(2n-\nu)}_{2n+2-\nu}}$};
\draw  (v2) edge (v3);
\draw  (v3) edge (v4);
\draw  (v4) edge (v5);
\draw  (v5) edge (v6);
\node at (3.7,-0.2) {\scriptsize{$f$}};
\node at (3.8,-1.2) {\scriptsize{$f$-$ x_i$}-$ y_i$};
\node at (-2.35,-0.4) {\scriptsize{$2h$-$\sum x_i$}};
\node at (-1.7,-0.4) {\scriptsize{$e$}};
\node at (-0.3,-0.4) {\scriptsize{$h$}};
\node at (2,-0.9) {\scriptsize{$h$}};
\node at (2.8,-1.3) {\scriptsize{$e$}};
\node at (1.15,-0.4) {\scriptsize{$e$}};
\node (v7) at (3.5,-0.65) {\scriptsize{$2n-\nu$}};
\node (v1) at (3.5,0.2) {$\mathbf{1_{2n+2-\nu}}$};
\draw  (v5) edge (v1);
\draw  (v1) edge (v7);
\draw  (v6) edge (v7);
\node at (2,-0.2) {\scriptsize{$h$}};
\node at (2.9,0.1) {\scriptsize{$e$}};
\begin{scope}[shift={(0.7,-0.2)}]
\node at (-0.5,0.7) {$2$};
\node at (-0.45,1) {$\su(2n)^{(2)}$};
\draw  (-0.5,0.9) ellipse (0.6 and 0.6);
\end{scope}
\end{tikzpicture}
\ee
for $n\ge3$. For $n=2$, we have
\be
\begin{tikzpicture} [scale=1.9]
\node (v1) at (-4.3,-0.5) {$\mathbf{0^{(4-\nu)+(4-\nu)}_{6-\nu}}$};
\node (v2) at (-2.5,-0.5) {$\mathbf{2^\nu_1}$};
\node (v3) at (-0.9,-0.5) {$\mathbf{1_{6-\nu}}$};
\draw  (v1) edge (v2);
\draw  (v2) edge (v3);
\node at (-3.6,-0.4) {\scriptsize{$e$}};
\node at (-3,-0.4) {\scriptsize{$2h$-$\sum x_i$}};
\node at (-2,-0.4) {\scriptsize{$2h$-$\sum x_i$}};
\node at (-1.3,-0.4) {\scriptsize{$e$}};
\begin{scope}[]
\node (v4) at (-2.5,-1.3) {\scriptsize{$4-\nu$}};
\end{scope}
\draw (v1) .. controls (-4.35,-1.1) and (-3.75,-1.3) .. (v4);
\draw (v4) .. controls (-1.4,-1.3) and (-0.9,-1.1) .. (v3);
\begin{scope}[shift={(-2,-0.6)}]
\node at (-0.5,0.75) {$2$};
\node at (-0.45,1) {$\su(4)^{(2)}$};
\draw  (-0.5,0.9) ellipse (0.5 and 0.5);
\end{scope}
\node at (-4.55,-0.9) {\scriptsize{$f$-$x_i$-$y_i$}};
\node at (-0.8,-0.8) {\scriptsize{$f$}};
\end{tikzpicture}
\ee

\be
\begin{tikzpicture} [scale=1.9]
\node (v1) at (-4.6,-0.5) {$\mathbf{0^{(6-\nu)+(6-\nu)+1+1}_{6-\nu}}$};
\node (v2) at (-2.1,-0.5) {$\mathbf{2^\nu_1}$};
\node (v3) at (-0.5,-0.5) {$\mathbf{1_{6-\nu}}$};
\draw  (v1) edge (v2);
\draw  (v2) edge (v3);
\node at (-3.6,-0.4) {\scriptsize{$e$-$z$-$w$}};
\node at (-2.7,-0.4) {\scriptsize{$2h$+$f$-$\sum x_i$}};
\node at (-1.6,-0.4) {\scriptsize{$2h$-$\sum x_i$}};
\node at (-0.9,-0.4) {\scriptsize{$e$}};
\begin{scope}[shift={(0.1,-0.1)}]
\node (v4) at (-2.5,-1.3) {\scriptsize{$6-\nu$}};
\end{scope}
\draw (v1) .. controls (-4.65,-1.1) and (-4.15,-1.4) .. (v4);
\draw (v4) .. controls (-0.9,-1.4) and (-0.5,-1.1) .. (v3);
\begin{scope}[shift={(-2,-0.5)}]
\node at (-0.5,0.75) {$1$};
\node at (-0.45,1) {$\su(4)^{(2)}$};
\draw  (-0.5,0.9) ellipse (0.5 and 0.5);
\end{scope}
\node at (-4.75,-1.1) {\scriptsize{$f$-$x_i$-$y_i$}};
\node at (-0.4,-0.8) {\scriptsize{$f$}};
\draw (v1) .. controls (-5.95,0.3) and (-5.95,-1.3) .. (v1);
\node at (-5.55,-0.1) {\scriptsize{$z$}};
\node at (-5.55,-0.9) {\scriptsize{$w$}};
\end{tikzpicture}
\ee

\be
\begin{tikzpicture} [scale=1.9]
\node (v2) at (-2.2,-0.5) {$\mathbf{n^\nu_1}$};
\node (v3) at (-0.5,-0.5) {$\mathbf{(n-1)_{6-\nu}}$};
\node (v4) at (0.6,-0.5) {$\cdots$};
\node (v5) at (1.6,-0.5) {$\mathbf{ 1_{2n+2-\nu}}$};
\node (v6) at (4.3,-0.5) {$\mathbf{0^{(2n+1-\nu)+(2n+1-\nu)}_{6}}$};
\draw  (v2) edge (v3);
\draw  (v3) edge (v4);
\draw  (v4) edge (v5);
\draw  (v5) edge (v6);
\node at (4.7,-1.1) {\scriptsize{$y_i$}};
\node[rotate=0] at (4.8,0.2) {\scriptsize{$x_i$}};
\node at (-1.7,-0.4) {\scriptsize{$2h$-$\sum x_i$}};
\node at (-1.1,-0.4) {\scriptsize{$e$}};
\node at (0.1,-0.4) {\scriptsize{$h$}};
\node at (2.15,-0.4) {\scriptsize{$2h$}};
\node at (3,-0.4) {\scriptsize{$e$-$\sum x_i$-$\sum y_i$}};
\node at (1,-0.4) {\scriptsize{$e$}};
\node (v7) at (5.9,-0.5) {\scriptsize{$2n+1-\nu$}};
\draw (v6) .. controls (4.4,0.1) and (5.7,0.5) .. (v7);
\draw (v6) .. controls (4.4,-1) and (5.7,-1.5) .. (v7);
\begin{scope}[shift={(2,-0.1)}]
\node at (-0.5,0.7) {$2$};
\node at (-0.45,1) {$\su(2n+1)^{(2)}$};
\draw  (-0.5,0.9) ellipse (0.7 and 0.7);
\end{scope}
\end{tikzpicture}
\ee
for $n\ge2$. For $n=1$, we have
\be
\begin{tikzpicture} [scale=1.9]
\node (v1) at (-4.25,-0.5) {$\mathbf{0^{(3-\nu)+(3-\nu)}_{6}}$};
\node (v2) at (-1.3,-0.5) {$\mathbf{1^\nu_0}$};
\draw  (v1) edge (v2);
\node at (-3.1,-0.4) {\scriptsize{$e$-$\sum x_i$-$\sum y_i$}};
\node at (-1.95,-0.4) {\scriptsize{$4 e$+$2f$-$2\sum x_i$}};
\node at (-4.3,-0.1) {\scriptsize{$x_i$}};
\node at (-4.3,-0.9) {\scriptsize{$y_i$}};
\begin{scope}[shift={(-2.6,0)}]
\node at (-0.5,0.75) {$2$};
\node at (-0.45,1) {$\su(3)^{(2)}$};
\draw  (-0.5,0.9) ellipse (0.5 and 0.5);
\end{scope}
\node (v3) at (-5.7,-0.5) {\scriptsize{$3-\nu$}};
\draw (v1) .. controls (-4.5,0.2) and (-5.6,0.2) .. (v3);
\draw (v1) .. controls (-4.5,-1.2) and (-5.6,-1.2) .. (v3);
\end{tikzpicture}
\ee

\be
\begin{tikzpicture} [scale=1.9]
\node (v1) at (-4.25,-0.5) {$\mathbf{0^{(6-\nu)+(6-\nu)}_{2}}$};
\node (v2) at (-1.3,-0.5) {$\mathbf{1^\nu_0}$};
\draw  (v1) edge (v2);
\node at (-3.1,-0.4) {\scriptsize{$e$-$\sum x_i$-$\sum y_i$}};
\node at (-1.95,-0.4) {\scriptsize{$4e$+$3f$-$2\sum x_i$}};
\node at (-4.3,-0.1) {\scriptsize{$x_i$}};
\node at (-4.3,-0.9) {\scriptsize{$y_i$}};
\begin{scope}[shift={(-2.6,0)}]
\node at (-0.5,0.75) {$1$};
\node at (-0.45,1) {$\su(3)^{(2)}$};
\draw  (-0.5,0.9) ellipse (0.5 and 0.5);
\end{scope}
\node (v3) at (-5.7,-0.5) {\scriptsize{$6-\nu$}};
\draw (v1) .. controls (-4.5,0.2) and (-5.6,0.2) .. (v3);
\draw (v1) .. controls (-4.5,-1.2) and (-5.6,-1.2) .. (v3);
\end{tikzpicture}
\ee

\subsection{Reading the data of $5d$ KK theory from CY3}\label{DCY}
In this subsection, we review the method of \cite{Bhardwaj:2020gyu} which allows one to identify the $5d$ KK theory if the associated CY3 is presented in a specific form. In this form, all the compact surfaces $S_i$ are presented as Hirzebruch surfaces, with fibers $f_i$, such that the intersection matrix
\be\label{IM}
\cI_{ij}:=-f_i\cdot S_j
\ee
takes the form of Cartan matrix of an (untwisted or twisted) affine Lie algebra $\fg^{(q)}$. Then, $\fg$ is identified as the gauge algebra appearing on the tensor branch of the associated $6d$ SCFT and $q$ captures the order of outer automorphism acting on $\fg$ while compactifying the $6d$ SCFT on circle. $q=1$ indicates that there is no outer automorphism twist. All of the proposed CY3 appearing in Section \ref{PCY} have been presented in this specific form and the reader can check that the intersection matrix reproduces the associated affine Lie algebra displayed there.

Moreover, every blowup $x$ must satisfy
\be
\left(\sum_i d^\vee_i S_i\right)\cdot x = 0
\ee
where $d^\vee_i$ are dual Coxeter numbers associated to $\fg^{(q)}$. This condition captures and generalizes the ``shifting of prepotential'' proposal of \cite{Bhardwaj:2019fzv}. The reader can check that every blowup appearing in every CY3 proposed in Section \ref{PCY} satisfies this condition.

The GS coupling of the associated $6d$ SCFT is captured by
\be
k=-\left(\sum_i d^\vee_i S_i\right)\cdot \tilde e
\ee
where $\tilde e$ denotes the $e$ curve of a specific Hirzebruch surface. This can be taken to be the surface denoted by $S_n$ for $\su(2n+1)^{(2)},\su(2n)^{(2)}$; the surface denoted by $S_3$ for $\so(10)^{(2)},\fe_6^{(2)}$; and the surface denoted by $S_1$ for $\so(11)^{(1)},\so(12)^{(1)},\fe_7^{(1)}$. The reader can now verify that the GS coupling for each proposed CY3 in Section \ref{PCY} matches the GS coupling displayed there. For more details, we refer the reader to Sections 3.3 and 3.4 of \cite{Bhardwaj:2020gyu}.

Using the above information, one can determine a lot of information about the $5d$ KK theory associated to a resolved CY3. This includes, as we have seen, the gauge algebra arising on the tensor branch of the associated $6d$ SCFT and the GS coupling associated to the $6d$ SCFT. In many cases, these two data are sufficient to uniquely determine the matter content arising on the tensor branch of the $6d$ SCFT. However, in some cases, there exist multiple $6d$ SCFTs with the same associated GS coupling and gauge algebra, but having different hypermultiplet content. To distinguish between such $6d$ SCFTs, we need further analysis to which we turn in the next subsection.

\subsection{Low energy effective $5d$ gauge theories from CY3}\label{LE}
At certain locations in the extended Coulomb branch, the $5d$ KK theories under discussion reduce at low-energies to non-Abelian $5d$ $\cN=1$ gauge theories. The hypermultiplet content of a particular such low-energy $5d$ gauge theory encodes the hypermultiplet content of the parent $6d$ SCFT. This particular low energy description is obtained by contracting a maximal set of fibers and blowups to zero size when the CY3 associated to a KK theory is presented in the form described in Section \ref{DCY}.

The (inverse of the) radius of compactification captures the volume of the curve associated to the KK mode of the $5d$ KK theory. This curve $f$ is a genus-one fiber that can be written in terms of the fibers $f_i$ of $S_i$ as \cite{Bhardwaj:2019fzv}
\be
f=\sum_i d_i f_i
\ee
where $d_i$ are the Coxeter numbers associated to $\fg^{(q)}$. For a finite value of the radius, the volume of $f$ must be strictly positive, which implies that not all $f_i$ can be shrunk to zero volume simultaneously.

Thus, the maximum number of fibers that can be contracted to zero size must be one less than the total number of fibers in the CY3. Consequently, the Dynkin diagram of the gauge theory obtained after contracting a maximal set of fibers can be obtained by deleting one node from the Dynkin diagram of the affine algebra $\fg^{(q)}$. Any such deletion leads to the Dynkin diagram of a finite semi-simple Lie algebra, thus guaranteeing that the gauge algebra for the low-energy $5d$ gauge theory must be finite and cannot be affine. This is a crucial consistency check since a gauge theory with an affine gauge algebra would have a troublesome positive semi-definite kinetic matrix\footnote{The kinetic matrix is captured by the Killing form.}.

Let us now study an example of $5d$ KK theory which was mentioned in Section \ref{I}. Consider the $6d$ SCFT carrying $\su(5)$ with $10$ fundamental hypers compactified with an outer automorphism twist. According to \cite{Bhardwaj:2019fzv}, the associated resolved CY3 can be expressed as
\be\label{BF}
\begin{tikzpicture} [scale=1.9]
\node (v2) at (-1.6,-0.5) {$\mathbf{2_1}$};
\node (v3) at (-0.3,-0.5) {$\mathbf{1_6}$};
\node (v6) at (1.9,-0.5) {$\mathbf{0^{5+5}_{6}}$};
\draw  (v2) edge (v3);
\node at (2.5,-1) {\scriptsize{$y_i$}};
\node[rotate=0] at (2.5,0) {\scriptsize{$x_i$}};
\node at (-1.3,-0.4) {\scriptsize{$2h$}};
\node at (-0.6,-0.4) {\scriptsize{$e$}};
\node at (0,-0.4) {\scriptsize{$2h$}};
\node at (1.1,-0.4) {\scriptsize{$e$-$\sum x_i$-$\sum y_i$}};
\node (v7) at (2.6,-0.5) {\scriptsize{$5$}};
\draw (v6) .. controls (2,-0.1) and (2.5,0.2) .. (v7);
\draw (v6) .. controls (2,-0.9) and (2.5,-1.2) .. (v7);
\draw  (v3) edge (v6);
\end{tikzpicture}
\ee
which let us rewrite into the following form for our convenience
\be\label{AF}
\begin{tikzpicture} [scale=1.9]
\node (v2) at (-1.6,-0.5) {$\mathbf{2_1}$};
\node (v3) at (0,-0.5) {$\mathbf{1^5_1}$};
\node (v6) at (1.4,-0.5) {$\mathbf{0_{6}}$};
\draw  (v2) edge (v3);
\node at (-1.3,-0.4) {\scriptsize{$2h$}};
\node at (-0.4,-0.4) {\scriptsize{$e$-$\sum x_i$}};
\node at (0.3,-0.4) {\scriptsize{$2h$}};
\node at (1.1,-0.4) {\scriptsize{$e$}};
\draw  (v3) edge (v6);
\end{tikzpicture}
\ee
This form is achieved by flopping all the blowups in (\ref{BF}) and performing an isomorphism on $S_1$. Let us choose to contract $f_1$ and $f_2$ to zero volume while keeping the volume of $f_0$ non-zero. This gives rise to an $\sp(2)$ gauge algebra at low energies. Moreover, we can choose all $x_i$ to have zero volume as well, without any obstruction. Doing this we obtain 5 massless hypermultiplets transforming in fundamental representation of $\sp(2)$ \cite{Bhardwaj:2019fzv}. That is, making a maximal number of BPS particles massless leads to a location in the extended Coulomb branch of the $5d$ KK theory
\be\label{T3}
\begin{tikzpicture}
\node at (-0.5,0.4) {2};
\node at (-0.45,0.9) {$\su(5)^{(2)}$};
\end{tikzpicture}
\ee
where the low-energy effective theory is a $5d$ $\cN=1$ gauge theory with gauge algebra $\sp(2)$ and 5 fundamental hypers. The volumes of the $e$ curves of $S_1$ and $S_2$ capture the gauge coupling of the low-energy $\sp(2)$ gauge theory.

Notice that $\sp(2)$ is the gauge algebra left invariant by the action of an outer automorphism of $\su(5)$, and projecting out by the action such an outer automorphism on 10 fundamental hypers of $\su(5)$ in the $6d$ theory, we indeed are left with 5 fundamental hypers (see Section \ref{I}) of $\su(5)$ which descend to 5 fundamental hypers of $\sp(2)$ under the projection. Thus, the low-energy theory associated to (\ref{T3}) can be determined by projecting out the tensor branch data of the associated $6d$ SCFT by the action of the discrete symmetry used for twisting. In fact, this was true for all KK theories studied in \cite{Bhardwaj:2019fzv}, where this fact was used to obtain the prepotential for the $5d$ KK theory starting from the prepotential of the associated low-energy $5d$ theory.

However, such a projection is not neatly defined (upon the hypermultiplet spectrum) for the twists of first type discussed in Section \ref{I}. An example for such a twist is provided by the KK theory
\be\label{T4}
\begin{tikzpicture}
\node at (-0.5,0.4) {1};
\node at (-0.45,0.9) {$\su(5)^{(2)}$};
\end{tikzpicture}
\ee
as discussed in Section \ref{I}. Let us start from the CY3 (\ref{su5T}) proposed for this KK theory in Section \ref{PCY}, which we reproduce below after a flop and an isomorphism,
\be\label{F1}
\begin{tikzpicture} [scale=1.9]
\node (v2) at (0.6,1) {$\mathbf{1^6_1}$};
\node (v3) at (2.9,1) {$\mathbf{0_{6}}$};
\node (v10) at (0.6,-0.6) {$\mathbf{2_0^{1+1}}$};
\draw  (v2) edge (v3);
\node[rotate=0] at (0.3,0.7) {\scriptsize{$h$-$\sum x_i$}};
\node at (0.9,1.1) {\scriptsize{$2h$}};
\node at (2.6,1.1) {\scriptsize{$e$}};
\node at (0.3,-0.2) {\scriptsize{$2e$+$f$-$x$}};
\draw  (v2) edge (v10);
\node at (2.7,0.7) {\scriptsize{$f,f$}};
\node[rotate=0] at (1.3,-0.4) {\scriptsize{$f$-$x$-$y$,$x$-$y$}};
\node (v1) at (1.9,0.3) {\scriptsize{2}};
\draw  (v10) edge (v1);
\draw  (v1) edge (v3);
\end{tikzpicture}
\ee
and determine the associated low-energy $5d$ gauge theory. Again, we would like to contract $f_1$ and $f_2$ while keeping $f_0$ at non-zero size. However, this is not possible when CY3 is presented in the form (\ref{F1}) since, according to one of the gluings
\be
f_0\sim f_2-x-y
\ee
which implies that $f_2$ must remain at a non-zero size as well. However, this problem can be alleviated if we flop the blowup $y$ living in $S_2$ to obtain the following form of the associated CY3
\be
\begin{tikzpicture} [scale=1.9]
\node (v2) at (0.6,1) {$\mathbf{1^6_1}$};
\node (v3) at (2.9,1) {$\mathbf{0^{1+1}_{6}}$};
\node (v10) at (0.6,-0.6) {$\mathbf{2_0^{1}}$};
\draw  (v2) edge (v3);
\node[rotate=0] at (0.3,0.7) {\scriptsize{$h$-$\sum x_i$}};
\node at (0.9,1.1) {\scriptsize{$2h$}};
\node at (2.5,1.1) {\scriptsize{$e$}};
\node at (0.3,-0.2) {\scriptsize{$2e$+$f$-$x$}};
\draw  (v2) edge (v10);
\node at (2.7,0.55) {\scriptsize{$f$-$x,f$-$y$}};
\node[rotate=0] at (1.3,-0.4) {\scriptsize{$f$-$x$,$x$}};
\node (v1) at (1.7993,0.2258) {\scriptsize{2}};
\draw  (v10) edge (v1);
\draw  (v1) edge (v3);
\draw (v3) .. controls (3.9,1.7) and (3.9,0.3) .. (v3);
\node at (3.4,1.4) {\scriptsize{$x$}};
\node at (3.4,0.6) {\scriptsize{$y$}};
\end{tikzpicture}
\ee
Now, we can provide a volume to $x,y$ living in $S_0$ which is equal to the volume of $f_0$, thus contracting the curves $f_0-x, f_0-y$ living in $S_0$. According to the gluings, this implies that the curves $f_2-x$ and $x$ living in $S_2$ must have zero volume, which can be consistently achieved if both the curves $f_2$ and $x$ living in $S_2$ have zero volume. We can also contract all the blowups $x_i$ living in $S_1$ without any obstruction. Thus, the low-energy $5d$ gauge theory associated to the KK theory (\ref{T4}) is
\be\label{5D}
\sp(2)+7\F
\ee
that is, $\sp(2)$ with 7 fundamental hypers. Six of the hypers arise from the six blowups living in $S_1$ and one hyper arises from the blowup $x$ living in $S_2$ \cite{Bhardwaj:2019ngx,Bhardwaj:2020gyu}.

Comparing (\ref{5D}) with the data of the $6d$ theory
\be\label{6D}
\su(5)+\L^2+13\F
\ee
(where $\L^2$ denotes the irreducible two-index antisymmetric representation) we can see that (\ref{5D}) is not a neat projection of (\ref{6D}). However, it is still possible to understand (\ref{5D}) morally as a projection of (\ref{6D}). The action of outer automorphism exchanges fields (in pairs) living inside hypermultiplets valued in $\L^2+13\F$, thus projecting it to ``$\half(\L^2+13\F)$'', which descends to\footnote{Upto singlets, $\F$ of $\su(5)$ descends to $\F$ of $\sp(2)$ and $\L^2$ of $\su(5)$ descends to $\L^2+\F$ of $\sp(2)$.} ``$\half\L^2$'' plus $7\F$ of $\sp(2)$. Since $\L^2$ is a strictly real representation for $\sp(n)$, the degrees of freedom in ``$\half\L^2$'' should be completely projected out, leaving us only with the matter content shown in (\ref{5D}). The finite volume blowups $x,y$ living in $S_0$ can be thought of as the remnant of ``$\half\L^2$'' since they lead to massive BPS particles transforming in $\L^2$ of $\sp(2)$.

Similarly, we would expect the following low energy descriptions for those KK theories in Table \ref{KR1} that involve twist of the first type (see intoduction):
\be
\begin{tikzpicture}
\node at (-0.5,0.4) {1};
\node at (-0.45,0.9) {$\su(2n+1)^{(2)}$};
\draw (1,0.4) -- (2.4,0.4) -- (2.3,0.5);
\draw (2.3,0.3) -- (2.4,0.4);
\node at (4.4,0.4) {$\sp(n)+(n+5)\F$};
\end{tikzpicture}
\ee
\be
\begin{tikzpicture}
\node at (-0.5,0.4) {1};
\node at (-0.45,0.9) {$\su(2n)^{(2)}$};
\draw (1,0.4) -- (2.4,0.4) -- (2.3,0.5);
\draw (2.3,0.3) -- (2.4,0.4);
\node at (4.4,0.4) {$\sp(n)+(n+4)\F$};
\end{tikzpicture}
\ee
\be
\begin{tikzpicture}
\node at (-0.5,0.4) {1};
\node at (-0.45,0.9) {$\su(\wh{2n+1})^{(2)}$};
\draw (1,0.4) -- (2.4,0.4) -- (2.3,0.5);
\draw (2.3,0.3) -- (2.4,0.4);
\node at (4.4,0.4) {$\sp(n)+(n-3)\F$};
\end{tikzpicture}
\ee
\be
\begin{tikzpicture}
\node at (-0.5,0.4) {1};
\node at (-0.45,0.9) {$\su(\wh{2n})^{(2)}$};
\draw (1,0.4) -- (2.4,0.4) -- (2.3,0.5);
\draw (2.3,0.3) -- (2.4,0.4);
\node at (4.4,0.4) {$\sp(n)+(n-4)\F$};
\end{tikzpicture}
\ee
\be\label{P1}
\begin{tikzpicture}
\node at (-0.5,0.4) {1};
\node at (-0.45,0.9) {$\su(\tilde 6)^{(2)}$};
\draw (1,0.4) -- (2.4,0.4) -- (2.3,0.5);
\draw (2.3,0.3) -- (2.4,0.4);
\node at (4.4,0.4) {$\sp(3)+\half\L^3+8\F$};
\end{tikzpicture}
\ee
\be
\begin{tikzpicture}
\node at (-0.5,0.4) {3};
\node at (-0.45,0.9) {$\so(10)^{(2)}$};
\draw (1,0.4) -- (2.4,0.4) -- (2.3,0.5);
\draw (2.3,0.3) -- (2.4,0.4);
\node at (4.4,0.4) {$\so(9)+3\F$};
\end{tikzpicture}
\ee
\be
\begin{tikzpicture}
\node at (-0.5,0.4) {1};
\node at (-0.45,0.9) {$\so(10)^{(2)}$};
\draw (1,0.4) -- (2.4,0.4) -- (2.3,0.5);
\draw (2.3,0.3) -- (2.4,0.4);
\node at (4.4,0.4) {$\so(9)+5\F+\S$};
\end{tikzpicture}
\ee
\be
\begin{tikzpicture}
\node at (-0.5,0.4) {5};
\node at (-0.45,0.9) {$\fe_6^{(2)}$};
\draw (1,0.4) -- (2.4,0.4) -- (2.3,0.5);
\draw (2.3,0.3) -- (2.4,0.4);
\node at (4.4,0.4) {$\ff_4$};
\end{tikzpicture}
\ee
\be
\begin{tikzpicture}
\node at (-0.5,0.4) {3};
\node at (-0.45,0.9) {$\fe_6^{(2)}$};
\draw (1,0.4) -- (2.4,0.4) -- (2.3,0.5);
\draw (2.3,0.3) -- (2.4,0.4);
\node at (4.4,0.4) {$\ff_4+\F$};
\end{tikzpicture}
\ee
\be
\begin{tikzpicture}
\node at (-0.5,0.4) {1};
\node at (-0.45,0.9) {$\fe_6^{(2)}$};
\draw (1,0.4) -- (2.4,0.4) -- (2.3,0.5);
\draw (2.3,0.3) -- (2.4,0.4);
\node at (4.4,0.4) {$\ff_4+2\F$};
\end{tikzpicture}
\ee
where $\S$ and $\L^3$ denote the irreducible spinor and 3-index antisymmetric representations respectively, and $\F$ of $\ff_4$ refers to the $\mathbf{26}$ dimensional irreduicble representation of $\ff_4$. Here, we emphasize that we have made a choice of outer automorphisms so that the gauge algebras left invariant by them coincide with the algebras appearing on the right hand side of the above equations.

We can also compute the low-energy theories associated to these KK theories according to their respective resolved CY3 proposed in Section \ref{PCY}. The results are collected in Table \ref{LG1}. Notice that the results match the above expectations except for the case of $\su(\tilde 6)$. This mismatch can be explained if we recall that an $\sp(3)$ $\cN=1$ gauge theory in $5d$ cannot contain $\half\L^3+n\F$, but can carry $\half\L^3+\frac{2n+1}{2}\F$. Thus an extra $\half\F$ should be projected out from the expectation (\ref{P1}).

\begin{table}[htbp]
\begin{center}
\begin{tabular}{|c|c|c|}
\hline
KK theory&Conditions&Low-energy theory\\
\hline
\hline
\raisebox{-.4\height}{ \begin{tikzpicture}
\node at (-0.5,0.4) {1};
\node at (-0.45,0.9) {$\su(2n+1)^{(2)}$};
\end{tikzpicture}}&$n\ge2$&$\sp(n)+(n+5)\F$
\\ \hline
 \raisebox{-.4\height}{ \begin{tikzpicture}
\node at (-0.5,0.4) {1};
\node at (-0.45,0.9) {$\su(2n)^{(2)}$};
\end{tikzpicture}}&$n\ge3$&$\sp(n)+(n+4)\F$
\\ \hline
\raisebox{-.4\height}{ \begin{tikzpicture}
\node at (-0.5,0.4) {1};
\node at (-0.45,0.9) {$\su(\wh{2n+1})^{(2)}$};
\end{tikzpicture}}&$n\ge4$&$\sp(n)+(n-3)\F$
\\ \hline
\raisebox{-.4\height}{ \begin{tikzpicture}
\node at (-0.5,0.4) {1};
\node at (-0.45,0.9) {$\su(\wh{2n})^{(2)}$};
\end{tikzpicture}}&$n\ge4$&$\sp(n)+(n-4)\F$
\\ \hline
\raisebox{-.4\height}{ \begin{tikzpicture}
\node at (-0.5,0.4) {$1$};
\node at (-0.45,0.9) {$\su(\tilde 6)^{(2)}$};
\end{tikzpicture}}&&$\sp(3)+\half\L^3+\frac{15}2\F$
\\ \hline
\raisebox{-.4\height}{ \begin{tikzpicture}
\node at (-0.5,0.4) {3};
\node at (-0.45,0.9) {$\so(10)^{(2)}$};
\end{tikzpicture}}&&$\so(9)+3\F$
\\ \hline
\raisebox{-.4\height}{ \begin{tikzpicture}
\node at (-0.5,0.4) {1};
\node at (-0.45,0.9) {$\so(10)^{(2)}$};
\end{tikzpicture}}&&$\so(9)+5\F+\S$
\\ \hline
\raisebox{-.4\height}{ \begin{tikzpicture}
\node at (-0.5,0.4) {5};
\node at (-0.45,0.9) {$\fe_6^{(2)}$};
\end{tikzpicture}}&&$\ff_4$
\\ \hline
\raisebox{-.4\height}{ \begin{tikzpicture}
\node at (-0.5,0.4) {3};
\node at (-0.45,0.9) {$\fe_6^{(2)}$};
\end{tikzpicture}}&&$\ff_4+\F$
\\ \hline
\raisebox{-.4\height}{ \begin{tikzpicture}
\node at (-0.5,0.4) {1};
\node at (-0.45,0.9) {$\fe_6^{(2)}$};
\end{tikzpicture}}&&$\ff_4+2\F$
\\ \hline
\end{tabular}
\end{center}
\caption{Low-energy $5d$ non-abelian gauge theories associated to $5d$ KK theory building blocks involving twists of first type.}	\label{LG1}
\end{table}

Now it can be understood how we associated the proposed CY3 (\ref{PCY1}) and (\ref{PCY2}) to 
\be\label{TT1}
\begin{tikzpicture}
\node at (-0.5,0.4) {1};
\node at (-0.45,0.9) {$\su(2n+1)^{(2)}$};
\end{tikzpicture}
\ee
and
\be\label{TT2}
\begin{tikzpicture}
\node at (-0.5,0.4) {1};
\node at (-0.45,0.9) {$\su(\wh{2n+1})^{(2)}$};
\end{tikzpicture}
\ee
respectively, despite the two CY3 having the same associated affine Lie algebra and same associated GS coupling. The low-energy theory associated to CY3 (\ref{PCY1}) is the low-energy theory expected for (\ref{TT1}), and low-energy theory associated to CY3 (\ref{PCY2}) is the low-energy theory expected for (\ref{TT2}). The same discussion holds true for the CY3 (\ref{ACY1}) and (\ref{ACY2}) associated to
\be
\begin{tikzpicture}
\node at (-0.5,0.4) {1};
\node at (-0.45,0.9) {$\su(2n)^{(2)}$};
\end{tikzpicture}
\ee
and
\be
\begin{tikzpicture}
\node at (-0.5,0.4) {1};
\node at (-0.45,0.9) {$\su(\wh{2n})^{(2)}$};
\end{tikzpicture}
\ee
respectively, and for (\ref{RCY1}) and (\ref{RCY2}) associated to
\be
\begin{tikzpicture}
\node at (-0.5,0.4) {1};
\node at (-0.45,0.9) {$\su(6)^{(2)}$};
\end{tikzpicture}
\ee
and
\be
\begin{tikzpicture}
\node at (-0.5,0.4) {1};
\node at (-0.45,0.9) {$\su(\tilde 6)^{(2)}$};
\end{tikzpicture}
\ee
respectively.

For KK theories containing twists of second type, we expect that the $\Z_2$ holonomy projects out a half-hyper, which can indeed be verified from the proposed CY3s. We collect the low-energy theories associated to these KK theories in Table \ref{LG2}.

\begin{table}[htbp]
\begin{center}
\begin{tabular}{|c|c|}
\hline
KK theory&Low-energy theory\\
\hline
\hline
\raisebox{-.4\height}{ \begin{tikzpicture}
\node (v1) at (-2.3,0.4) {$\left[\Z^{(2)}_2\right]$};
\node (v2) at (-0.5,0.4) {2};
\node at (-0.45,0.9) {$\so(11)^{(1)}$};
\draw  (v1) edge (v2);
\end{tikzpicture}}&$\so(11)+5\F+\half\S$
\\ \hline
\raisebox{-.4\height}{ \begin{tikzpicture}
\node (v1) at (-2.3,0.4) {$\left[\Z^{(2)}_2\right]$};
\node (v2) at (-0.5,0.4) {2};
\node at (-0.45,0.9) {$\so(12)^{(1)}$};
\draw  (v1) edge (v2);
\end{tikzpicture}}&$\so(12)+6\F+\half\S$
\\ \hline
\raisebox{-.4\height}{ \begin{tikzpicture}
\node (v1) at (-2.3,0.4) {$\left[\Z^{(2)}_2\right]$};
\node (v2) at (-0.5,0.4) {1};
\node at (-0.45,0.9) {$\so(\wh{12})^{(1)}$};
\draw  (v1) edge (v2);
\end{tikzpicture}}&$\so(12)+7\F+\half\S+\half\C$
\\ \hline
\raisebox{-.4\height}{ \begin{tikzpicture}
\node (v1) at (-2.3,0.4) {$\left[\Z^{(2)}_2\right]$};
\node (v2) at (-0.5,0.4) {6};
\node at (-0.45,0.9) {$\fe_7^{(1)}$};
\draw  (v1) edge (v2);
\end{tikzpicture}}&$\fe_7+\half\F$
\\ \hline
\raisebox{-.4\height}{ \begin{tikzpicture}
\node (v1) at (-2.3,0.4) {$\left[\Z^{(2)}_2\right]$};
\node (v2) at (-0.5,0.4) {4};
\node at (-0.45,0.9) {$\fe_7^{(1)}$};
\draw  (v1) edge (v2);
\end{tikzpicture}}&$\fe_7+\frac32\F$
\\ \hline
\raisebox{-.4\height}{ \begin{tikzpicture}
\node (v1) at (-2.3,0.4) {$\left[\Z^{(2)}_2\right]$};
\node (v2) at (-0.5,0.4) {2};
\node at (-0.45,0.9) {$\fe_7^{(1)}$};
\draw  (v1) edge (v2);
\end{tikzpicture}}&$\fe_7+\frac52\F$
\\ \hline
\end{tabular}
\end{center}
\caption{Low-energy $5d$ non-abelian gauge theories associated to $5d$ KK theory building blocks involving twists of second type. $\C$ denotes the conjugate spinor representation of $\so(2n)$ and $\F$ for $\fe_7$ denotes the $\mathbf{56}$ dimensional irreducible representation of $\fe_7$.}	\label{LG2}
\end{table}

Out of these cases, the case of
\be\label{tint}
\begin{tikzpicture}
\node (v1) at (-2.3,0.4) {$\left[\Z^{(2)}_2\right]$};
\node (v2) at (-0.5,0.4) {1};
\node at (-0.45,0.9) {$\so(\wh{12})^{(1)}$};
\draw  (v1) edge (v2);
\end{tikzpicture}
\ee
is interesting since it allows one to obtain either
\be\label{TI1}
\so(12)+7\F+\half\S+\half\C
\ee
or
\be\label{TI2}
\so(12)+7\F+\half\S
\ee
after contracting a maximal set of fibers and blowups, depending on the choice of the maximal set of fibers one decides to contract. As one can see from the CY3 (\ref{int}) associated to (\ref{tint}), there are multiple sets of fibers that can be contracted to yield an $\so(12)$ gauge theory. One could choose to contract all fibers except $f_0$, or all fibers except $f_1$, or all fibers except $f_5$ or all fibers except $f_6$. 

Let us first choose to contract all fibers except $f_0$. This will require that both $x$ and $y$ in $S_0$ remain at non-zero volume, which in turn implies that $f_4$ remains at non-zero volume. So, it is not possible to perform this contraction in the frame (\ref{int}). However, we can perform some flops and write (\ref{int}) in the following form
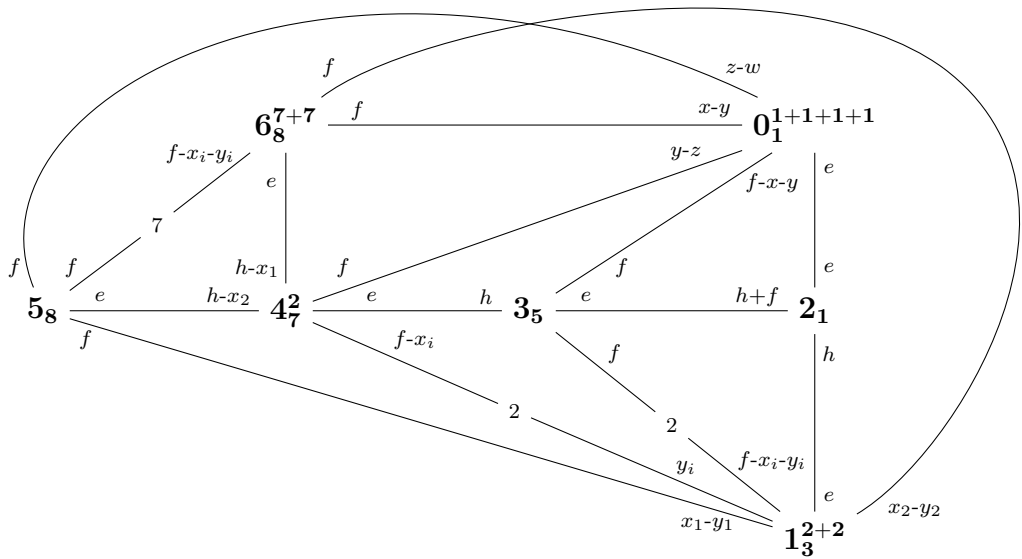
\be
\begin{tikzpicture} [scale=1.9]
\node (v2) at (0.6,1) {$\mathbf{4_7^{2}}$};
\node (v3) at (2.3,1) {$\mathbf{3_{5}}$};
\node (v10) at (-1.1,1) {$\mathbf{5_8}$};
\draw  (v2) edge (v3);
\node at (1.2,1.1) {\scriptsize{$e$}};
\node at (2.7,1.1) {\scriptsize{$e$}};
\node at (0.4,1.3) {\scriptsize{$h$-$x_1$}};
\draw  (v2) edge (v10);
\node at (-0.8,0.8) {\scriptsize{$f$}};
\node[rotate=0] at (0,2.1) {\scriptsize{$f$-$x_i$-$y_i$}};
\node (v1) at (4.3,1) {$\mathbf{2_{1}}$};
\draw  (v3) edge (v1);
\node at (2,1.1) {\scriptsize{$h$}};
\node at (3.9,1.1) {\scriptsize{$h$+$f$}};
\node (v4) at (4.3,-0.6) {$\mathbf{1_3^{2+2}}$};
\draw  (v1) edge (v4);
\node at (4.4,0.7) {\scriptsize{$h$}};
\node[rotate=0] at (4.4,-0.3) {\scriptsize{$e$}};
\node at (-0.9,1.3) {\scriptsize{$f$}};
\node[rotate=0] at (3.55,-0.5) {\scriptsize{$x_1$-$y_1$}};
\draw  (v10) edge (v4);
\node (v7) at (4.3,2.3) {$\mathbf{0_1^{1+1+1+1}}$};
\node[rotate=0] at (0.5,1.9) {\scriptsize{$e$}};
\draw  (v1) edge (v7);
\node at (4.4,1.3) {\scriptsize{$e$}};
\node[rotate=0] at (3.6,2.4) {\scriptsize{$x$-$y$}};
\node (v9) at (0.6,2.3) {$\mathbf{6^{7+7}_{8}}$};
\draw  (v9) edge (v2);
\node at (0.2,1.1) {\scriptsize{$h$-$x_2$}};
\node[rotate=0] at (-0.7,1.1) {\scriptsize{$e$}};
\node (v11) at (-0.3,1.6) {\scriptsize{7}};
\draw  (v10) edge (v11);
\draw  (v11) edge (v9);
\draw  (v9) edge (v7);
\node at (1.1,2.4) {\scriptsize{$f$}};
\node at (4.4,2) {\scriptsize{$e$}};
\draw  (v3) edge (v7);
\node at (2.95,1.3) {\scriptsize{$f$}};
\node at (4,1.9) {\scriptsize{$f$-$x$-$y$}};
\draw  (v2) edge (v7);
\node at (3.4,2.1) {\scriptsize{$y$-$z$}};
\node at (1,1.3) {\scriptsize{$f$}};
\node at (4,-0.05) {\scriptsize{$f$-$x_i$-$y_i$}};
\node at (2.9,0.7) {\scriptsize{$f$}};
\node at (3.4,-0.1) {\scriptsize{$y_i$}};
\node at (1.5,0.8) {\scriptsize{$f$-$x_i$}};
\node at (0.9,2.7) {\scriptsize{$f$}};
\node[rotate=0] at (5,-0.4) {\scriptsize{$x_2$-$y_2$}};
\draw (v10) .. controls (-1.7,2.5) and (1,3.9) .. (v7);
\node at (-1.3,1.3) {\scriptsize{$f$}};
\node[rotate=0] at (3.8,2.7) {\scriptsize{$z$-$w$}};
\node (v5) at (3.3,0.2) {\scriptsize{2}};
\draw  (v3) edge (v5);
\draw  (v5) edge (v4);
\node (v6) at (2.2,0.3) {\scriptsize{2}};
\draw  (v2) edge (v6);
\draw  (v6) edge (v4);
\draw (v4) .. controls (5.3,0) and (6.4,1.8) .. (5.2,2.7);
\draw (v9) .. controls (1.5,3) and (4.1,3.5) .. (5.2,2.7);
\end{tikzpicture}
\ee
in which it is possible to perform this contraction. We can contract all the blowups except for the four blowups $x,y,z,w$ living in $S_0$. This limit leads to (\ref{TI1}) as the low-energy description for (\ref{tint}).

Now, let us choose to contract all fibers except $f_1$. For similar reason as above, it is not possible to do so when the CY3 is expressed in the form (\ref{int}). But, after doing some flops, and representing it as
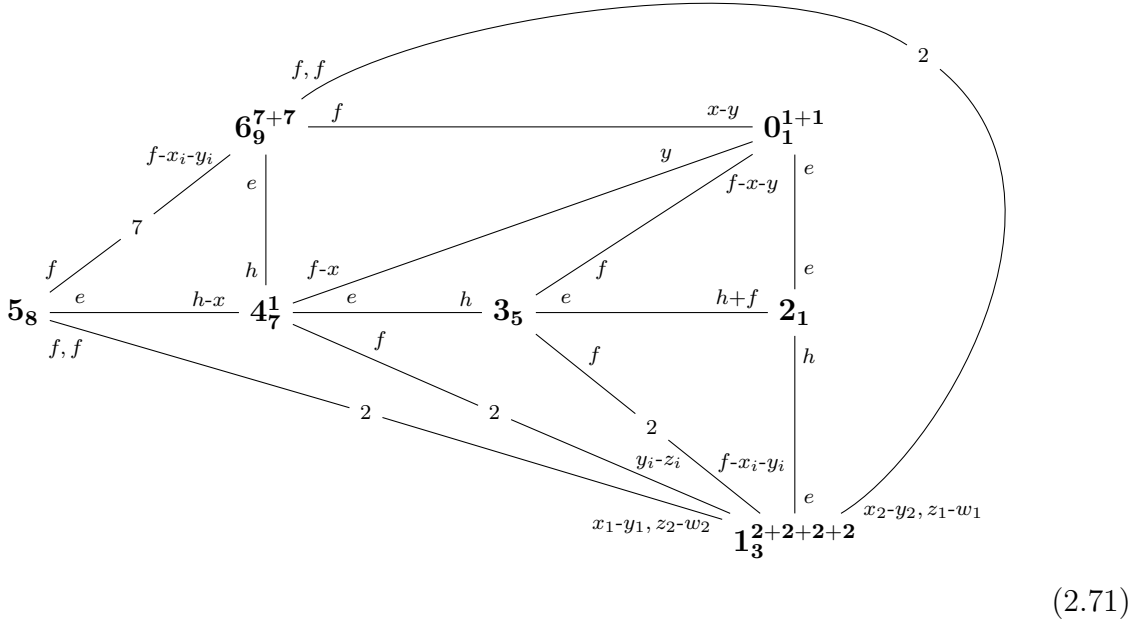
\be
\begin{tikzpicture} [scale=1.9]
\node (v2) at (0.6,1) {$\mathbf{4_7^{1}}$};
\node (v3) at (2.3,1) {$\mathbf{3_{5}}$};
\node (v10) at (-1.1,1) {$\mathbf{5_8}$};
\draw  (v2) edge (v3);
\node at (1.2,1.1) {\scriptsize{$e$}};
\node at (2.7,1.1) {\scriptsize{$e$}};
\node at (0.5,1.3) {\scriptsize{$h$}};
\draw  (v2) edge (v10);
\node at (-0.8,0.75) {\scriptsize{$f,f$}};
\node[rotate=0] at (0,2.1) {\scriptsize{$f$-$x_i$-$y_i$}};
\node (v1) at (4.3,1) {$\mathbf{2_{1}}$};
\draw  (v3) edge (v1);
\node at (2,1.1) {\scriptsize{$h$}};
\node at (3.9,1.1) {\scriptsize{$h$+$f$}};
\node (v4) at (4.3,-0.6) {$\mathbf{1_3^{2+2+2+2}}$};
\draw  (v1) edge (v4);
\node at (4.4,0.7) {\scriptsize{$h$}};
\node[rotate=0] at (4.4,-0.3) {\scriptsize{$e$}};
\node at (-0.9,1.3) {\scriptsize{$f$}};
\node[rotate=0] at (3.3,-0.5) {\scriptsize{$x_1$-$y_1,z_2$-$w_2$}};
\node (v7) at (4.3,2.3) {$\mathbf{0_1^{1+1}}$};
\node[rotate=0] at (0.5,1.9) {\scriptsize{$e$}};
\draw  (v1) edge (v7);
\node at (4.4,1.3) {\scriptsize{$e$}};
\node[rotate=0] at (3.8,2.4) {\scriptsize{$x$-$y$}};
\node (v9) at (0.6,2.3) {$\mathbf{6^{7+7}_{9}}$};
\draw  (v9) edge (v2);
\node at (0.2,1.1) {\scriptsize{$h$-$x$}};
\node[rotate=0] at (-0.7,1.1) {\scriptsize{$e$}};
\node (v11) at (-0.3,1.6) {\scriptsize{7}};
\draw  (v10) edge (v11);
\draw  (v11) edge (v9);
\draw  (v9) edge (v7);
\node at (1.1,2.4) {\scriptsize{$f$}};
\node at (4.4,2) {\scriptsize{$e$}};
\draw  (v3) edge (v7);
\node at (2.95,1.3) {\scriptsize{$f$}};
\node at (4,1.9) {\scriptsize{$f$-$x$-$y$}};
\draw  (v2) edge (v7);
\node at (3.4,2.1) {\scriptsize{$y$}};
\node at (1,1.3) {\scriptsize{$f$-$x$}};
\node at (4,-0.05) {\scriptsize{$f$-$x_i$-$y_i$}};
\node at (2.9,0.7) {\scriptsize{$f$}};
\node at (3.35,-0.05) {\scriptsize{$y_i$-$z_i$}};
\node at (1.4,0.8) {\scriptsize{$f$}};
\node at (0.9,2.7) {\scriptsize{$f,f$}};
\node[rotate=0] at (5.2,-0.4) {\scriptsize{$x_2$-$y_2,z_1$-$w_1$}};
\node (v5) at (3.3,0.2) {\scriptsize{2}};
\draw  (v3) edge (v5);
\draw  (v5) edge (v4);
\node (v6) at (2.2,0.3) {\scriptsize{2}};
\draw  (v2) edge (v6);
\draw  (v6) edge (v4);
\node (v8) at (1.3,0.3) {\scriptsize{2}};
\draw  (v10) edge (v8);
\draw  (v8) edge (v4);
\node (v) at (5.2,2.8) {\scriptsize{2}};
\draw (v4) .. controls (5.3,0) and (6.4,1.8) .. (v);
\draw (v9) .. controls (1.5,3) and (4.1,3.5) .. (v);
\end{tikzpicture}
\ee
allows us to take this limit. We are also able to contract all the blowups except for the eight blowups $x_i,y_i,z_i,w_i$ living in $S_1$. This limit leads to (\ref{TI2}) as the low-energy description for (\ref{tint}).

Similarly, choosing to contract all fibers except $f_6$ leads to the low-energy theory (\ref{TI1}), and choosing to contract all fibers except $f_5$ leads to the low-energy theory (\ref{TI2}), as the reader can check. We propose that this existence of multiple low-energy limits is simply a reflection of the fact that while twisting the $6d$ SCFT
\be
\begin{tikzpicture}
\node at (-0.5,0.4) {1};
\node at (-0.45,0.9) {$\so(\wh{12})$};
\end{tikzpicture}
\ee
carrying $7\F+\S+\half\C$ (in $6d$) by the $\Z_2$ element of determinant $-1$ in $O(2)$ symmetry rotating $\S$, one can also include a holonomy in the $O(1)\simeq\Z_2$ symmetry rotating $\half\C$. If this holonomy is included, we expect the low-energy description to be (\ref{TI2}); while, if this holonomy is not including, we expect the low-energy description to be (\ref{TI1}). The fact that these two low-energy theories are continuously connected inside the extended Coulomb branch of the $5d$ KK theory
\be
\begin{tikzpicture}
\node (v1) at (-2.3,0.4) {$\left[\Z^{(2)}_2\right]$};
\node (v2) at (-0.5,0.4) {1};
\node at (-0.45,0.9) {$\so(\wh{12})^{(1)}$};
\draw  (v1) edge (v2);
\end{tikzpicture}
\ee
means that the inclusion of this holonomy does not lead to a physically distinguishable twist, as we already argued in Section \ref{I}.

We can test this proposal in other similar cases. For example, consider the CY3 for
\be
\begin{tikzpicture}
\node at (-0.5,0.4) {3};
\node at (-0.45,0.9) {$\so(11)^{(1)}$};
\end{tikzpicture}
\ee
which is \cite{Bhardwaj:2018yhy,Bhardwaj:2018vuu,Bhardwaj:2019fzv}
\be
\begin{tikzpicture} [scale=1.9]
\node (v2) at (0.6,1) {$\mathbf{4_{4}}$};
\node (v3) at (2.3,1) {$\mathbf{3_{2}}$};
\node (v10) at (0.6,-0.6) {$\mathbf{5_6^{1+1+4+4}}$};
\draw  (v2) edge (v3);
\node[rotate=0] at (0.1,-0.3) {\scriptsize{$e$-$\sum z_i$-$\sum w_i$}};
\node at (0.9,1.1) {\scriptsize{$e$}};
\node at (2.6,1.1) {\scriptsize{$e$}};
\node at (0.3,0.7) {\scriptsize{$2h$+$f$}};
\draw  (v2) edge (v10);
\node at (1.9,0.8) {\scriptsize{$f$}};
\node[rotate=0] at (0.95,0) {\scriptsize{$f$-$x$-$y$}};
\node (v1) at (4.3,1) {$\mathbf{2^{1}_{0}}$};
\draw  (v3) edge (v1);
\node at (2,1.1) {\scriptsize{$h$}};
\node at (4,1.1) {\scriptsize{$e$}};
\node[rotate=0] at (1.5,-0.1) {\scriptsize{$y$}};
\node (v4) at (4.3,-0.6) {$\mathbf{1_{2}}$};
\draw  (v1) edge (v4);
\node at (4.5,0.7) {\scriptsize{$e$}};
\node[rotate=0] at (4.5,-0.3) {\scriptsize{$e$}};
\node at (3.9,-0.5) {\scriptsize{$f$}};
\node[rotate=0] at (3.5,0.8) {\scriptsize{$f$-$x$}};
\node[rotate=0] at (1.4,-0.5) {\scriptsize{$x$-$y$}};
\draw  (v10) edge (v4);
\node (v7) at (5.9,1) {$\mathbf{0_{1}}$};
\draw  (v1) edge (v7);
\node at (4.7,1.1) {\scriptsize{$e$-$x$}};
\node[rotate=0] at (5.6,1.1) {\scriptsize{$e$}};
\draw  (v10) edge (v3);
\draw  (v10) edge (v1);
\node (v) at (0.6,-1.3) {\scriptsize{4}};
\draw (v10) .. controls (0.5,-0.8) and (0,-1.2) .. (v);
\draw (v10) .. controls (0.7,-0.8) and (1.2,-1.2) .. (v);
\node at (0.1,-1.1) {\scriptsize{$z_i$}};
\node at (1.1,-1.1) {\scriptsize{$w_i$}};
\end{tikzpicture}
\ee
Contracting all fibers except $f_0$ leads to the low-energy $5d$ theory
\be
\so(11)+4\F+\half\S
\ee
while contracting all fibers except $f_1$ leads to the low-energy $5d$ theory
\be
\so(11)+4\F
\ee
which can be explained as the absence or presence of a holonomy in $\Z_2$ symmetry acting by reflection on $\half\S$ in the associated $6d$ $\cN=(1,0)$ theory $\so(11)+4\F+\half\S$.

Now consider the CY3 for
\be\label{E1}
\begin{tikzpicture}
\node at (-0.5,0.4) {2};
\node at (-0.45,0.9) {$\so(11)^{(1)}$};
\end{tikzpicture}
\ee
which is \cite{Bhardwaj:2018yhy,Bhardwaj:2018vuu,Bhardwaj:2019fzv}
\be
\begin{tikzpicture} [scale=1.9]
\node (v2) at (0.6,1) {$\mathbf{4_{4}}$};
\node (v3) at (2.3,1) {$\mathbf{3_{2}}$};
\node (v10) at (0.6,-0.6) {$\mathbf{5_6^{2+2+5+5}}$};
\draw  (v2) edge (v3);
\node[rotate=0] at (0.1,-0.3) {\scriptsize{$e$-$\sum z_i$-$\sum w_i$}};
\node at (0.9,1.1) {\scriptsize{$e$}};
\node at (2.6,1.1) {\scriptsize{$e$}};
\node at (0.3,0.7) {\scriptsize{$2h$+$2f$}};
\draw  (v2) edge (v10);
\node at (1.9,0.8) {\scriptsize{$f$}};
\node[rotate=0] at (0.95,0) {\scriptsize{$f$-$x_i$-$y_i$}};
\node (v1) at (4.3,1) {$\mathbf{2^{2}_{0}}$};
\draw  (v3) edge (v1);
\node at (2,1.1) {\scriptsize{$h$}};
\node at (4,1.1) {\scriptsize{$e$}};
\node[rotate=0] at (1.5,-0.1) {\scriptsize{$y_i$}};
\node (v4) at (4.3,-0.6) {$\mathbf{1_{2}}$};
\draw  (v1) edge (v4);
\node at (4.5,0.7) {\scriptsize{$e$}};
\node[rotate=0] at (4.5,-0.3) {\scriptsize{$e$}};
\node at (3.9,-0.5) {\scriptsize{$f,f$}};
\node[rotate=0] at (3.5,0.8) {\scriptsize{$f$-$x_i$}};
\node (v5) at (1.6,0.3) {\scriptsize{2}};
\draw  (v10) edge (v5);
\draw  (v5) edge (v3);
\node (v6) at (2.5,0.2) {\scriptsize{2}};
\draw  (v10) edge (v6);
\draw  (v6) edge (v1);
\node[rotate=0] at (1.6,-0.5) {\scriptsize{$x_1$-$y_2,x_2$-$y_1$}};
\node (v7) at (5.9,1) {$\mathbf{0_{0}}$};
\draw  (v1) edge (v7);
\node at (4.8,1.1) {\scriptsize{$e$-$\sum x_i$}};
\node[rotate=0] at (5.6,1.1) {\scriptsize{$e$}};
\node (v8) at (2.7,-0.6) {\scriptsize{2}};
\draw  (v10) edge (v8);
\draw  (v8) edge (v4);
\node (v) at (0.6,-1.3) {\scriptsize{5}};
\draw (v10) .. controls (0.5,-0.8) and (0,-1.2) .. (v);
\draw (v10) .. controls (0.7,-0.8) and (1.2,-1.2) .. (v);
\node at (0.1,-1.1) {\scriptsize{$z_i$}};
\node at (1.1,-1.1) {\scriptsize{$w_i$}};
\end{tikzpicture}
\ee
The reader can check that no matter whether one contracts all fibers except $f_0$ or contracts all fibers except $f_1$, one lands on the following low-energy $5d$ theory
\be\label{LE1}
\so(11)+5\F+\S
\ee
On the other hand, no matter whether one contracts all fibers except $f_0$ or contracts all fibers except $f_1$ in (\ref{Int}), one lands on the following low-energy $5d$ theory
\be
\so(11)+5\F+\half\S
\ee
Thus, the absence or presence of holonomy in $\Z_2$ element of determinant $-1$ in $O(2)$ rotating $\S$ in the $6d$ gauge theory $\so(11)+5\F+\S$ associated to the $6d$ SCFT
\be
\begin{tikzpicture}
\node at (-0.5,0.4) {2};
\node at (-0.45,0.9) {$\so(11)$};
\end{tikzpicture}
\ee
leads to two distinct low-energy $5d$ theories which are not connected to each other. The low-energy theory (\ref{LE1}) is found in the extended Coulomb branch of the KK theory denoted by (\ref{E1}), while the low-energy theory (\ref{LE1}) is found in the extended Coulomb branch of the KK theory denoted by
\be
\begin{tikzpicture}
\node (v1) at (-2.3,0.4) {$\left[\Z^{(2)}_2\right]$};
\node (v2) at (-0.5,0.4) {2};
\node at (-0.45,0.9) {$\so(11)^{(1)}$};
\draw  (v1) edge (v2);
\end{tikzpicture}
\ee
This verifies the arguments presented in Section \ref{I} in relation to the twists of second type.

\section{Combining the building blocks}\label{CBB}
In this section, we describe how the new building blocks can be combined with new/old building blocks to produce more general $5d$ KK theories. Let us first consider a $6d$ SCFT of the form
\be\label{S}
\begin{tikzpicture}
\node (v1) at (-2.3,0.4) {$k$};
\node (v2) at (-0.5,0.4) {$l$};
\node at (-0.45,0.9) {$\su(2m)$};
\draw  (v1) edge (v2);
\node at (-2.25,0.9) {$\su(2n)$};
\end{tikzpicture}
\ee
which makes sense for $k,l\in\{1,2\}$ and $k+l\ge3$. The GS coupling is a $2\times 2$ matrix whose diagonal entries are captrued by $k$ and $l$, while a single edge in (\ref{S}) denotes that both off-diagonal entries are $-1$. The mixed hypermultiplet content is a bifundamental of $\su(2n)\oplus\su(2m)$. We would like to perform an outer automorphism twist on $\su(2n)$, which should project the degrees of freedom living inside the bifundamental by a factor of half. If there is no outer automorphism twist acting on the $\su(2m)$ factor, then we would expect a half-hyper in bifundamental of $\sp(n)\oplus\su(2m)$ in the associated low-energy effective $5d$ gauge theory. However, a half-hyper in bifundamental is not allowed for this set of gauge algebras. 

We can achieve a consistent projection if we also act by an outer automorphism on $\su(2m)$, but we choose this outer automorphism such that it projects $\su(2m)$ to $\so(2m)$ rather than $\sp(m)$. Then, we would expect the associated low-energy effective $5d$ gauge theory to contain a half-hyper in bifundamental of $\sp(n)\oplus\so(2m)$, which is indeed an allowed matter content. This expectation is verified geometrically where we observe that there is a consistent resolved CY3 associated to
\be\label{S2}
\begin{tikzpicture}
\node (v1) at (-2.3,0.4) {$k$};
\node (v2) at (0.1,0.4) {$l$};
\node at (0.15,0.9) {$\su(2m)^{(2)}$};
\draw  (v1) edge (v2);
\node at (-2.25,0.9) {$\su(2n)^{(2)}$};
\end{tikzpicture}
\ee
but no consistent resolved CY3 associated to
\be
\begin{tikzpicture}
\node (v1) at (-2.3,0.4) {$k$};
\node (v2) at (0.1,0.4) {$l$};
\node at (0.15,0.9) {$\su(2m)^{(1)}$};
\draw  (v1) edge (v2);
\node at (-2.25,0.9) {$\su(2n)^{(2)}$};
\end{tikzpicture}
\ee
We can compute from the CY3 associated to (\ref{S2}), presented later in this section, that the associated low-energy theory indeed contains a half-hyper in bifundamental of $\sp(n)\oplus\so(2m)$.

Let us now consider the $6d$ SCFT
\be\label{G1}
\begin{tikzpicture}
\node (v1) at (-2.3,0.4) {$2$};
\node (v2) at (-0.5,0.4) {$2$};
\node at (-0.45,0.9) {$\su(2m)$};
\draw  (v1) edge (v2);
\node at (-2.25,0.9) {$\su(2n)$};
\begin{scope}[shift={(1.8,0)}]
\node (v3) at (-0.5,0.4) {$2$};
\node at (-0.45,0.9) {$\su(2n)$};
\end{scope}
\draw  (v2) edge (v3);
\end{tikzpicture}
\ee
How many twists are possible for this $6d$ SCFT? First of all, we can simultaneously act by outer automorphism on all three algebras to obtain
\be\label{G2}
\begin{tikzpicture}
\node (v1) at (-2.3,0.4) {$2$};
\node (v2) at (0.1,0.4) {$2$};
\node at (0.15,0.9) {$\su(2m)^{(2)}$};
\draw  (v1) edge (v2);
\node at (-2.25,0.9) {$\su(2n)^{(2)}$};
\begin{scope}[shift={(2.4,0)}]
\node (v3) at (0.1,0.4) {$2$};
\node at (0.15,0.9) {$\su(2n)^{(2)}$};
\end{scope}
\draw  (v2) edge (v3);
\end{tikzpicture}
\ee
Second, we can exchange the two $\su(n)$ (along with the two bifundamentals). The corresponding KK theory was denoted in \cite{Bhardwaj:2019fzv} as
\be
\begin{tikzpicture}
\node (v1) at (-2.3,0.4) {$2$};
\node (v2) at (0.1,0.4) {$2$};
\node at (0.15,0.9) {$\su(2n)^{(1)}$};
\node at (-2.25,0.9) {$\su(2m)^{(1)}$};
\node (v3) at (-1.1,0.4) {\tiny{2}};
\draw  (v1) edge (v3);
\draw [->] (v3) -- (v2);
\end{tikzpicture}
\ee
which represents a folding of the graph (\ref{G1}), and describes that no outer automorphism is acting on either of the two algebras (the projection by the exchange operation identifies the two $\su(2n)$).\\
Another possibility is to perform the outer automorphism on $\su(2m)$ alone, while representing the action of outer automorphism as an exchange of the two bifundamentals. Since the action of outer automorphism is a permutation on the hypermultiplet spectrum, this twist was already considered in \cite{Bhardwaj:2019fzv} where the corresponding KK theory was denoted as
\be
\begin{tikzpicture}
\node (v1) at (-2.3,0.4) {$2$};
\node (v2) at (0.1,0.4) {$2$};
\node at (0.15,0.9) {$\su(2n)^{(1)}$};
\node at (-2.25,0.9) {$\su(2m)^{(2)}$};
\node (v3) at (-1.1,0.4) {\tiny{2}};
\draw  (v1) edge (v3);
\draw [->] (v3) -- (v2);
\end{tikzpicture}
\ee
which represents a folding of the graph (\ref{G1}), and describes that an outer automorphism is acting upon the $\su(2m)$ gauge algebra.\\
We claim that there is yet another possible twist which can be represented as
\be
\begin{tikzpicture}
\node (v1) at (-2.3,0.4) {$2$};
\node (v2) at (0.1,0.4) {$2$};
\node at (0.15,0.9) {$\su(2n)^{(2)}$};
\node at (-2.25,0.9) {$\su(2m)^{(2)}$};
\node (v3) at (-1.1,0.4) {\tiny{2}};
\draw  (v1) edge (v3);
\draw [->] (v3) -- (v2);
\end{tikzpicture}
\ee
where along with an exchange, both the $\su(2m)$ and $\su(2n)$ algebras have an outer automorphism acting upon them. This can be thought of as first reducing to (\ref{G2}) whose associated low-energy theory contains
\be
\begin{tikzpicture}
\node (v1) at (-2.3,0.4) {$\sp(n)$};
\node (v2) at (0.1,0.4) {$\so(2m)$};
\draw  (v1) edge (v2);
\begin{scope}[shift={(2.4,0)}]
\node (v3) at (0.1,0.4) {$\sp(n)$};
\end{scope}
\draw  (v2) edge (v3);
\end{tikzpicture}
\ee
where each edge denotes a half-hyper in bifundamental. We can now add a further exchange symmetry to the twist, whose action on the above low-energy theory exchanges the two half-bifundamentals and the two $\sp(n)$.

In light of the above discussion, let us revisit the case of (\ref{S}) when $m=n$ and $k=l$. The $6d$ SCFT can then be represented as
\be
\begin{tikzpicture}
\node (v1) at (-2.3,0.4) {$2$};
\node (v2) at (-0.5,0.4) {$2$};
\node at (-0.45,0.9) {$\su(2n)$};
\draw  (v1) edge (v2);
\node at (-2.25,0.9) {$\su(2n)$};
\end{tikzpicture}
\ee
Are there any twists which involve the exchange of two $\su(2n)$? If there is no outer automorphism involved, then this twist was already discussed in \cite{Bhardwaj:2019fzv} and the corresponding KK theory was denoted as
\be
\begin{tikzpicture}
\node (v1) at (-2.3,0.4) {$2$};
\node at (-2.25,0.9) {$\su(2n)^{(1)}$};
\draw (v1) .. controls (-3.1,-0.6) and (-1.5,-0.6) .. (v1);
\end{tikzpicture}
\ee
However, if there is an outer automorphism involved, we don't expect that the exchange is a symmetry since the low-energy theory associated to
\be
\begin{tikzpicture}
\node (v1) at (-2.3,0.4) {$2$};
\node (v2) at (0.1,0.4) {$2$};
\node at (0.15,0.9) {$\su(2n)^{(2)}$};
\draw  (v1) edge (v2);
\node at (-2.25,0.9) {$\su(2n)^{(2)}$};
\end{tikzpicture}
\ee
contains
\be
\begin{tikzpicture}
\node (v1) at (-2.3,0.4) {$\sp(n)$};
\node (v2) at (0.1,0.4) {$\so(2n)$};
\draw  (v1) edge (v2);
\end{tikzpicture}
\ee
which does not admit any exchange symmetry. That is, we do not expect the existence of a KK theory of the form
\be
\begin{tikzpicture}
\node (v1) at (-2.3,0.4) {$2$};
\node at (-2.25,0.9) {$\su(2n)^{(2)}$};
\draw (v1) .. controls (-3.1,-0.6) and (-1.5,-0.6) .. (v1);
\end{tikzpicture}
\ee
Correspondingly, we do not find any consistent CY3 that could be associated to such a KK theory.

\begin{table}[htbp]
\begin{center}
\begin{tabular}{|c|c|l|} \hline
 \raisebox{-.4\height}{\begin{tikzpicture}
\node (v1) at (-0.5,0.4) {1};
\node at (-0.45,0.9) {$\sp(n_\alpha)^{(1)}$};
\begin{scope}[shift={(2,0)}]
\node (v2) at (-0.5,0.4) {$2$};
\node at (-0.45,0.9) {$\su(n_\beta)^{(2)}$};
\end{scope}
\draw  (v1) edge (v2);
\end{tikzpicture}}&$n_\alpha\le 2n_\beta$; $n_\beta\le2n_\alpha+7$
\\ 
 \raisebox{-.4\height}{\begin{tikzpicture}
\node (v1) at (-0.5,0.4) {1};
\node at (-0.45,0.9) {$\sp(n_\alpha)_\theta^{(1)}$};
\begin{scope}[shift={(2,0)}]
\node (v2) at (-0.5,0.4) {$2$};
\node at (-0.45,0.9) {$\su(n_\beta)^{(2)}$};
\end{scope}
\draw  (v1) edge (v2);
\end{tikzpicture}}&$n_\beta=2n_\alpha+8$; $\theta=0,\pi$
\\ 
 \raisebox{-.4\height}{\begin{tikzpicture}
\node (v1) at (-0.5,0.4) {1};
\node at (-0.45,0.9) {$\sp(n_\alpha)^{(1)}$};
\begin{scope}[shift={(2,0)}]
\node (v2) at (-0.5,0.4) {$k$};
\node at (-0.45,0.9) {$\so(10)^{(2)}$};
\end{scope}
\draw  (v1) edge (v2);
\end{tikzpicture}}&$n_\alpha\le 6-k$
\\ 
 \raisebox{-.4\height}{\begin{tikzpicture}
\node (v1) at (-0.5,0.4) {1};
\node at (-0.45,0.9) {$\sp(n_\alpha)^{(1)}$};
\begin{scope}[shift={(2,0)}]
\node (v2) at (-0.5,0.4) {3};
\node at (-0.45,0.9) {$\so(10)^{(2)}$};
\end{scope}
\node (v3) at (0.5,0.4) {\tiny{$2$}};
\draw  (v1) -- (v3);
\draw [<-] (v2) -- (v3);
\end{tikzpicture}}&$n_\alpha\le 2$
\\ 
 \raisebox{-.4\height}{\begin{tikzpicture}
\node (v1) at (-0.5,0.4) {3};
\node at (-0.45,0.9) {$\so(10)^{(2)}$};
\begin{scope}[shift={(2,0)}]
\node (v2) at (-0.5,0.4) {1};
\node at (-0.45,0.9) {$\sp(1)^{(1)}$};
\end{scope}
\node (v3) at (0.5,0.4) {\tiny{$2$}};
\draw  (v1) -- (v3);
\draw [<-] (v2) -- (v3);
\end{tikzpicture}}&
\\ 
 \raisebox{-.4\height}{\begin{tikzpicture}
\node (v1) at (-0.5,0.4) {1};
\node at (-0.45,0.9) {$\su(n_\alpha)^{(2)}$};
\begin{scope}[shift={(2,0)}]
\node (v2) at (-0.5,0.4) {$2$};
\node at (-0.45,0.9) {$\su(n_\beta)^{(2)}$};
\end{scope}
\draw  (v1) edge (v2);
\end{tikzpicture}}&$n_\alpha\le 2n_\beta$; $n_\beta\le n_\alpha+8$
\\
 \raisebox{-.4\height}{\begin{tikzpicture}
\node (v1) at (-0.5,0.4) {1};
\node at (-0.45,0.9) {$\su(\wh{n}_\alpha)^{(2)}$};
\begin{scope}[shift={(2,0)}]
\node (v2) at (-0.5,0.4) {$2$};
\node at (-0.45,0.9) {$\su(n_\beta)^{(2)}$};
\end{scope}
\draw  (v1) edge (v2);
\end{tikzpicture}}&$n_\alpha\le 2n_\beta$; $n_\beta\le n_\alpha-8$
\\
 \raisebox{-.4\height}{\begin{tikzpicture}
\node (v1) at (-0.5,0.4) {1};
\node at (-0.45,0.9) {$\su(\tilde6)^{(2)}$};
\begin{scope}[shift={(2,0)}]
\node (v2) at (-0.5,0.4) {$2$};
\node at (-0.45,0.9) {$\su(n_\beta)^{(2)}$};
\end{scope}
\draw  (v1) edge (v2);
\end{tikzpicture}}&$3\le n_\beta\le15$
\\
 \raisebox{-.4\height}{\begin{tikzpicture}
\node (v1) at (-0.5,0.4) {2};
\node at (-0.45,0.9) {$\su(n_\alpha)^{(2)}$};
\begin{scope}[shift={(2,0)}]
\node (v2) at (-0.5,0.4) {$2$};
\node at (-0.45,0.9) {$\su(n_\beta)^{(2)}$};
\end{scope}
\draw  (v1) edge (v2);
\end{tikzpicture}}&$n_\alpha\le 2n_\beta$; $n_\beta\le 2n_\alpha$
\\
 \raisebox{-.4\height}{\begin{tikzpicture}
\node (v1) at (-0.5,0.4) {2};
\node at (-0.45,0.9) {$\su(n_\alpha)^{(2)}$};
\begin{scope}[shift={(2,0)}]
\node (v2) at (-0.5,0.4) {$4$};
\node at (-0.45,0.9) {$\so(n_\beta)^{(1)}$};
\end{scope}
\node (v3) at (0.5,0.4) {\tiny{2}};
\draw  (v1) -- (v3);
\draw  (v2) -- (v3);
\end{tikzpicture}}&$n_\alpha\le n_\beta-8$; $n_\beta\le2n_\alpha$
\\ 
 \raisebox{-.4\height}{\begin{tikzpicture}
\node (v1) at (-0.5,0.4) {2};
\node at (-0.45,0.9) {$\su(n_\alpha)^{(2)}$};
\begin{scope}[shift={(2,0)}]
\node (v2) at (-0.5,0.4) {$4$};
\node at (-0.45,0.9) {$\so(2n_\beta)^{(2)}$};
\end{scope}
\node (v3) at (0.5,0.4) {\tiny{2}};
\draw  (v1) -- (v3);
\draw  (v2) -- (v3);
\end{tikzpicture}}&$n_\alpha\le 2n_\beta-8$; $n_\beta\le n_\alpha$
\\ 
 \raisebox{-.4\height}{\begin{tikzpicture}
\node (v1) at (-0.5,0.4) {2};
\node at (-0.45,0.9) {$\su(n_\alpha)^{(2)}$};
\begin{scope}[shift={(2,0)}]
\node (v2) at (-0.5,0.4) {$2$};
\node at (-0.45,0.9) {$\su(n_\beta)^{(2)}$};
\end{scope}
\node (v3) at (0.5,0.4) {\tiny{$e$}};
\draw  (v1) -- (v3);
\draw [<-] (v2) -- (v3);
\end{tikzpicture}}&$n_\alpha\le 2n_\beta$; $en_\beta\le2n_\alpha$; $e=2,3$
\\ \hline
\end{tabular}
\end{center}
\caption{List of all the new combinations of KK theory building blocks that can arise by including new KK theory building blocks arising from twists of the first type.}\label{KR2}
\end{table}

Using similar arguments, we can compile a list of possible ways in which the new KK theory building blocks can be combined with other KK theory building blocks. We present this list in Table \ref{KR2}. We note that we only need to study the possible combinations of the new KK building blocks arising from twists of the first type (see Section \ref{I}). The possible combinations for a KK building block arising from a twist of the second type, say
\be
\begin{tikzpicture} [scale=1.9]
\node (v2) at (-4.6,0.8) {$k$};
\node at (-4.6,1.1) {$\fg^{(q)}$};
\node (v1) at (-5.5,0.8) {$\left[\Z^{(2)}_2\right]$};
\draw  (v1) edge (v2);
\end{tikzpicture}
\ee
are the same as the possible combinations for the KK building block
\be
\begin{tikzpicture} [scale=1.9]
\node (v2) at (-4.6,0.8) {$k$};
\node at (-4.6,1.1) {$\fg^{(q)}$};
\end{tikzpicture}
\ee
which does not involve the extra $\Z_2$ twist. This is because the $\Z_2$ only affects hypers transforming in spinor of $\so(11),\so(12)$ and $\F$ of $\fe_7$, but it is not possible to gauge the global symmetries associated to these hypers and obtain a $6d$ SCFT. So, the $\Z_2$ twist acts on parts of the theory which does not affect possible combinations.

Let us notice that the theory
\be\label{TT}
\begin{tikzpicture}
\node (v1) at (-0.5,0.4) {1};
\node at (-0.45,0.9) {$\sp(3)^{(1)}$};
\begin{scope}[shift={(2,0)}]
\node (v2) at (-0.5,0.4) {3};
\node at (-0.45,0.9) {$\so(10)^{(2)}$};
\end{scope}
\node (v3) at (0.5,0.4) {\tiny{$2$}};
\draw  (v1) -- (v3);
\draw [<-] (v2) -- (v3);
\end{tikzpicture}
\ee
does not appear in our list even though the theory
\be
\begin{tikzpicture}
\node (v1) at (-2.3,0.4) {$3$};
\node (v2) at (0.1,0.4) {$1$};
\node at (0.15,0.9) {$\sp(3)$};
\draw  (v1) edge (v2);
\node at (-2.25,0.9) {$\so(10)$};
\begin{scope}[shift={(2.4,0)}]
\node (v3) at (0.1,0.4) {$3$};
\node at (0.15,0.9) {$\so(10)$};
\end{scope}
\draw  (v2) edge (v3);
\end{tikzpicture}
\ee
is a consistent $6d$ SCFT. The reason for the inconsistency of (\ref{TT}) can be found in the itemized list appearing towards the end of Section 3.4 of \cite{Bhardwaj:2019fzv}.

\subsection{Gluing rules for associated CY3}\label{GR}
The CY3 associated to two KK theory building blocks $\alpha$ and $\beta$ connected by an edge is obtained by gluing some curves (comprised of fibers and blowups) in the CY3 associated $\alpha$ with some curves (comprised of fibers and blowups) in the CY3 associated $\beta$. The gluings are independent of the diagonal GS couplings associated to $\alpha$ and $\beta$, and depend only on the type of edge and the associated twisted affine algebras\footnote{They are independent of the matter content as well.} \cite{Bhardwaj:2018vuu,Bhardwaj:2019fzv}. In the following, we will present such gluing rules. Our notation would be to call the building block appearing on the left as $\alpha$ and the building block appearing on the right as $\beta$. We will denote the surfaces coming from CY3 associated to $\alpha$ as $S_{i,\alpha}$ and the surfaces coming from CY3 associated to $\beta$ as $S_{i,\beta}$, where $i$ is the labeling of different surfaces which can be found in Section \ref{PCY} for the new KK theory building blocks and Section 5.2 of \cite{Bhardwaj:2019fzv} for the old KK theory building blocks.

\noindent\ubf{Gluing rules for \raisebox{-.25\height}{\begin{tikzpicture}
\node (w1) at (-0.5,0.9) {$\sp(n_\alpha)_\theta^{(1)}$};
\begin{scope}[shift={(3,0)}]
\node (w2) at (-0.5,0.9) {$\su(2n_\beta)^{(2)}$};
\end{scope}
\draw (w1)--(w2);
\end{tikzpicture}}}:
We can take any geometry with $0\le\nu\le 2n_\alpha+8-2n_\beta$ for $\sp(n_\alpha)_\theta^{(1)}$, and any geometry with $0\le\nu\le 2n_\beta-n_\alpha$ for $\su(2n_\beta)^{(2)}$. The gluing rules below work irrespective of the value of $\theta$. The gluing rules are:
\bit
\item $f-x_1-x_2,x_{2n_\beta-1},x_{2n_\beta}$ in $S_{0,\alpha}$ are glued to $f,f-x_1,y_1$ in $S_{0,\beta}$.
\item $x_i-x_{i+1},x_{2n_\beta-i}-x_{2n_\beta+1-i}$ in $S_{0,\alpha}$ are glued to $f,f$ in $S_{i,\beta}$ for $i=1,\cdots,n_\beta-1$.
\item $x_{n_\beta}-x_{n_\beta+1}$ in $S_{0,\alpha}$ is glued to $f$ in $S_{n_\beta,\beta}$.
\item $x_{i}-x_{i+1},y_{i+1}-y_i$ in $S_{0,\beta}$ are glued to $f,f$ in $S_{i,\alpha}$ for $i=1,\cdots,n_\alpha-1$.
\item $x_{n_\alpha}-y_{n_\alpha}$ in $S_{0,\beta}$ is glued to $f$ in $S_{n_\alpha,\alpha}$.
\eit

\noindent\ubf{Gluing rules for \raisebox{-.25\height}{\begin{tikzpicture}
\node (w1) at (-0.5,0.9) {$\sp(n_\alpha)_\theta^{(1)}$};
\begin{scope}[shift={(3.4,0)}]
\node (w2) at (-0.5,0.9) {$\su(2n_\beta+1)^{(2)}$};
\end{scope}
\draw (w1)--(w2);
\end{tikzpicture}}}:
We can take any geometry with $1\le\nu\le 2n_\alpha+7-2n_\beta$ for $\sp(n_\alpha)_\theta^{(1)}$, and any geometry with $0\le\nu\le 2n_\beta+1-n_\alpha$ for $\su(2n_\beta+1)^{(2)}$. The gluing rules below work irrespective of the value of $\theta$. The gluing rules are:
\bit
\item $f-x_1-x_2,x_1-x_2,x_{2n_\beta+1},x_{2n_\beta+1}$ in $S_{0,\alpha}$ are glued to $f,f,x_1,y_1$ in $S_{0,\beta}$.
\item $x_{i+1}-x_{i+2},x_{2n_\beta+1-i}-x_{2n_\beta+2-i}$ in $S_{0,\alpha}$ are glued to $f,f$ in $S_{i,\beta}$ for $i=1,\cdots,n_\beta-1$.
\item $x_{n_\beta+1}-x_{n_\beta+2}$ in $S_{0,\alpha}$ is glued to $f$ in $S_{n_\beta,\beta}$.
\item $x_{i+1}-x_{i},y_{i+1}-y_i$ in $S_{0,\beta}$ are glued to $f,f$ in $S_{i,\alpha}$ for $i=1,\cdots,n_\alpha-1$.
\item $f-x_{n_\alpha},f-y_{n_\alpha}$ in $S_{0,\beta}$ are glued to $f-x_1,x_1$ in $S_{n_\alpha,\alpha}$.
\eit

\noindent\ubf{Gluing rules for \raisebox{-.25\height}{\begin{tikzpicture}
\node (w1) at (-0.5,0.9) {$\sp(n_\alpha)^{(1)}$};
\begin{scope}[shift={(3,0)}]
\node (w2) at (-0.5,0.9) {$\so(10)^{(2)}$};
\end{scope}
\draw (w1)--(w2);
\end{tikzpicture}},}\\
\ni\ubf{\raisebox{-.25\height}{\begin{tikzpicture}
\node (w1) at (-0.7,0.9) {$\sp(n_\alpha)^{(1)}$};
\begin{scope}[shift={(3.1,0)}]
\node (w2) at (-0.5,0.9) {$\so(10)^{(2)}$};
\end{scope}
\node (w3) at (1,0.9) {\scriptsize{2}};
\draw (w1)--(w3);
\draw[->] (w3)--(w2);
\end{tikzpicture}} and \raisebox{-.25\height}{\begin{tikzpicture}
\node (w1) at (-0.7,0.9) {$\so(10)^{(2)}$};
\begin{scope}[shift={(3.1,0)}]
\node (w2) at (-0.5,0.9) {$\sp(1)^{(1)}$};
\end{scope}
\node (w3) at (1,0.9) {\scriptsize{2}};
\draw (w1)--(w3);
\draw[->] (w3)--(w2);
\end{tikzpicture}}}: Same as the ones provided in \cite{Bhardwaj:2019fzv}.

\noindent\ubf{Gluing rules for \raisebox{-.25\height}{\begin{tikzpicture}
\node (w1) at (-0.5,0.9) {$\su(2n_\alpha)^{(2)}$};
\begin{scope}[shift={(3,0)}]
\node (w2) at (-0.5,0.9) {$\su(2n_\beta)^{(2)}$};
\end{scope}
\draw (w1)--(w2);
\end{tikzpicture}}}:
We can take any geometry with $0\le\nu\le \text{max}(\nu)-n_\beta$ for $\su(2n_\alpha)^{(2)}$, and any geometry with $0\le\nu\le \text{max}(\nu)-n_\alpha$ for $\su(2n_\beta)^{(2)}$, where max$(\nu)$ denotes the maximum value of $\nu$ allowed for that KK building block. The gluing rules are:
\bit
\item $f-x_1,y_1,f-x_2,y_2$ in $S_{0,\alpha}$ are glued to $y_2,y_1,f-x_2,f-x_1$ in $S_{0,\beta}$.
\item $x_i-x_{i+1},y_{i+1}-y_i$ in $S_{0,\alpha}$ are glued to $f,f$ in $S_{i,\beta}$ for $i=1,\cdots,n_\beta-1$.
\item $x_{n_\beta}-y_{n_\beta}$ in $S_{0,\alpha}$ is glued to $f$ in $S_{n_\beta,\beta}$.
\item $x_{i}-x_{i+1},y_{i+1}-y_i$ in $S_{0,\beta}$ are glued to $f,f$ in $S_{i,\alpha}$ for $i=1,\cdots,n_\alpha-1$.
\item $x_{n_\alpha}-y_{n_\alpha}$ in $S_{0,\beta}$ is glued to $f$ in $S_{n_\alpha,\alpha}$.
\eit

\noindent\ubf{Gluing rules for \raisebox{-.25\height}{\begin{tikzpicture}
\node (w1) at (-0.5,0.9) {$\su(2n_\alpha+1)^{(2)}$};
\begin{scope}[shift={(3.4,0)}]
\node (w2) at (-0.5,0.9) {$\su(2n_\beta)^{(2)}$};
\end{scope}
\draw (w1)--(w2);
\end{tikzpicture}}}:
We can take any geometry with $0\le\nu\le \text{max}(\nu)-n_\beta$ for $\su(2n_\alpha+1)^{(2)}$, and any geometry with $1\le\nu\le \text{max}(\nu)-n_\alpha$ for $\su(2n_\beta)^{(2)}$, where max$(\nu)$ denotes the maximum value of $\nu$ allowed for that KK building block. The gluing rules are:
\bit
\item $x_1,y_1,x_2,y_2$ in $S_{0,\alpha}$ are glued to $y_1,y_1,f-x_1,f-x_1$ in $S_{0,\beta}$.
\item $x_{i+1}-x_{i},y_{i+1}-y_i$ in $S_{0,\alpha}$ are glued to $f,f$ in $S_{i,\beta}$ for $i=1,\cdots,n_\beta-1$.
\item $f-x_{n_\beta},f-y_{n_\beta}$ in $S_{0,\alpha}$ is glued to $f-x_1,x_1$ in $S_{n_\beta,\beta}$.
\item $x_{i}-x_{i+1},y_{i+1}-y_i$ in $S_{0,\beta}$ are glued to $f,f$ in $S_{i,\alpha}$ for $i=1,\cdots,n_\alpha-1$.
\item $x_{n_\alpha}-y_{n_\alpha}$ in $S_{0,\beta}$ is glued to $f$ in $S_{n_\alpha,\alpha}$.
\eit

\noindent\ubf{Gluing rules for \raisebox{-.25\height}{\begin{tikzpicture}
\node (w1) at (-0.5,0.9) {$\su(2n_\alpha+1)^{(2)}$};
\begin{scope}[shift={(3.8,0)}]
\node (w2) at (-0.5,0.9) {$\su(2n_\beta+1)^{(2)}$};
\end{scope}
\draw (w1)--(w2);
\end{tikzpicture}}}:
We can take any geometry with $1\le\nu\le \text{max}(\nu)-n_\beta$ for $\su(2n_\alpha+1)^{(2)}$, and any geometry with $1\le\nu\le \text{max}(\nu)-n_\alpha$ for $\su(2n_\beta+1)^{(2)}$, where max$(\nu)$ denotes the maximum value of $\nu$ allowed for that KK building block. The gluing rules are:
\bit
\item $x_1,y_1,x_1,y_1$ in $S_{0,\alpha}$ are glued to $y_1,y_1,x_1,x_1$ in $S_{0,\beta}$.
\item $x_{i+1}-x_{i},y_{i+1}-y_i$ in $S_{0,\alpha}$ are glued to $f,f$ in $S_{i,\beta}$ for $i=1,\cdots,n_\beta-1$.
\item $f-x_{n_\beta},f-y_{n_\beta}$ in $S_{0,\alpha}$ is glued to $f-x_1,x_1$ in $S_{n_\beta,\beta}$.
\item $x_{i+1}-x_{i},y_{i+1}-y_i$ in $S_{0,\beta}$ are glued to $f,f$ in $S_{i,\alpha}$ for $i=1,\cdots,n_\alpha-1$.
\item $f-x_{n_\alpha},f-y_{n_\alpha}$ in $S_{0,\beta}$ is glued to $f-x_1,x_1$ in $S_{n_\alpha,\alpha}$.
\eit

\noindent\ubf{Gluing rules for \raisebox{-.25\height}{\begin{tikzpicture}
\node (w1) at (-0.7,0.9) {$\su(2n_\alpha)^{(2)}$};
\begin{scope}[shift={(3.1,0)}]
\node (w2) at (-0.5,0.9) {$\so(2n_\beta)^{(1)}$};
\end{scope}
\node (w3) at (1,0.9) {\scriptsize{2}};
\draw (w1)--(w3);
\draw (w3)--(w2);
\end{tikzpicture}}}:
We can take any geometry with $n_\beta\le\nu\le 2n_\alpha$ for $\su(2n_\alpha)^{(2)}$, and any geometry with $0\le\nu\le 2n_\beta-8-2n_\alpha$ for $\so(2n_\beta)^{(1)}$. The gluing rules are:
\bit
\item $f-x_1-x_2$ in $S_{n_\alpha,\alpha}$ is glued to $f$ in $S_{0,\beta}$.
\item $x_i-x_{i+1}$ in $S_{n_\alpha,\alpha}$ is glued to $f$ in $S_{i,\beta}$ for $i=1,\cdots,n_\beta-1$.
\item $f,x_{n_\beta-1},x_{n_\beta}$ in $S_{n_\alpha,\alpha}$ are glued to $x_{2n_\alpha}-y_{2n_\alpha},f-x_1,y_1$ in $S_{n_\beta,\beta}$.
\item $x_{i}-x_{i+1},y_{i+1}-y_i,x_{2n_\alpha-i}-x_{2n_\alpha+1-i},y_{2n_\alpha+1-i}-y_{2n_\alpha-i}$ in $S_{n_\beta,\beta}$ are glued to $f,f,f,f$ in $S_{n_\alpha-i,\alpha}$ for $i=1,\cdots,n_\alpha-1$.
\item $x_{n_\alpha-1}-x_{n_\alpha+1},x_{n_\alpha}-x_{n_\alpha+2},y_{n_\alpha+1}-y_{n_\alpha-1},y_{n_\alpha+2}-y_{n_\alpha}$ in $S_{n_\beta,\beta}$ are glued to $f,f,f,f$ in $S_{0,\alpha}$.
\eit

\noindent\ubf{Gluing rules for \raisebox{-.25\height}{\begin{tikzpicture}
\node (w1) at (-0.7,0.9) {$\su(2n_\alpha)^{(2)}$};
\begin{scope}[shift={(3.5,0)}]
\node (w2) at (-0.5,0.9) {$\so(2n_\beta+1)^{(1)}$};
\end{scope}
\node (w3) at (1,0.9) {\scriptsize{2}};
\draw (w1)--(w3);
\draw (w3)--(w2);
\end{tikzpicture}}}:
We can take any geometry with $n_\beta+1\le\nu\le 2n_\alpha$ for $\su(2n_\alpha)^{(2)}$, and any geometry with $0\le\nu\le 2n_\beta-7-2n_\alpha$ for $\so(2n_\beta+1)^{(1)}$. The gluing rules are:
\bit
\item $f-x_1-x_2$ in $S_{n_\alpha,\alpha}$ is glued to $f$ in $S_{0,\beta}$.
\item $x_i-x_{i+1}$ in $S_{n_\alpha,\alpha}$ is glued to $f$ in $S_{i,\beta}$ for $i=1,\cdots,n_\beta-1$.
\item $f-x_{n_\beta+1},x_{n_\beta+1},x_{n_\beta},x_{n_\beta}$ in $S_{n_\alpha,\alpha}$ are glued to $f-x_{2n_\alpha},f-y_{2n_\alpha},x_1,y_1$ in $S_{n_\beta,\beta}$.
\item $x_{i+1}-x_{i},y_{i+1}-y_i,x_{2n_\alpha+1-i}-x_{2n_\alpha-i},y_{2n_\alpha+1-i}-y_{2n_\alpha-i}$ in $S_{n_\beta,\beta}$ are glued to $f,f,f,f$ in $S_{n_\alpha-i,\alpha}$ for $i=1,\cdots,n_\alpha-1$.
\item $x_{n_\alpha+1}-x_{n_\alpha-1},x_{n_\alpha+2}-x_{n_\alpha},y_{n_\alpha+1}-y_{n_\alpha-1},y_{n_\alpha+2}-y_{n_\alpha}$ in $S_{n_\beta,\beta}$ are glued to $f,f,f,f$ in $S_{0,\alpha}$.
\eit

\noindent\ubf{Gluing rules for \raisebox{-.25\height}{\begin{tikzpicture}
\node (w1) at (-0.7,0.9) {$\su(2n_\alpha)^{(2)}$};
\begin{scope}[shift={(3.1,0)}]
\node (w2) at (-0.5,0.9) {$\so(2n_\beta)^{(2)}$};
\end{scope}
\node (w3) at (1,0.9) {\scriptsize{2}};
\draw (w1)--(w3);
\draw (w3)--(w2);
\end{tikzpicture}}}:
We can take any geometry with $n_\beta+1\le\nu\le 2n_\alpha$ for $\su(2n_\alpha)^{(2)}$, and any geometry with $0\le\nu\le 2n_\beta-8-2n_\alpha$ for $\so(2n_\beta)^{(2)}$. The gluing rules are:
\bit
\item $f-x_1-x_2,x_1-x_2$ in $S_{n_\alpha,\alpha}$ are glued to $f,f$ in $S_{0,\beta}$.
\item $x_{i+1}-x_{i+2}$ in $S_{n_\alpha,\alpha}$ is glued to $f$ in $S_{i,\beta}$ for $i=1,\cdots,n_\beta-2$.
\item $f-x_{n_\beta+1},x_{n_\beta+1},x_{n_\beta},x_{n_\beta}$ in $S_{n_\alpha,\alpha}$ are glued to $f-x_{2n_\alpha},f-y_{2n_\alpha},x_1,y_1$ in $S_{n_\beta-1,\beta}$.
\item $x_{i+1}-x_{i},y_{i+1}-y_i,x_{2n_\alpha+1-i}-x_{2n_\alpha-i},y_{2n_\alpha+1-i}-y_{2n_\alpha-i}$ in $S_{n_\beta-1,\beta}$ are glued to $f,f,f,f$ in $S_{n_\alpha-i,\alpha}$ for $i=1,\cdots,n_\alpha-1$.
\item $x_{n_\alpha+1}-x_{n_\alpha-1},x_{n_\alpha+2}-x_{n_\alpha},y_{n_\alpha+1}-y_{n_\alpha-1},y_{n_\alpha+2}-y_{n_\alpha}$ in $S_{n_\beta-1,\beta}$ are glued to $f,f,f,f$ in $S_{0,\alpha}$.
\eit

\noindent\ubf{Gluing rules for \raisebox{-.25\height}{\begin{tikzpicture}
\node (w1) at (-0.7,0.9) {$\su(2n_\alpha+1)^{(2)}$};
\begin{scope}[shift={(3.55,0)}]
\node (w2) at (-0.5,0.9) {$\so(2n_\beta)^{(1)}$};
\end{scope}
\node (w3) at (1.4,0.9) {\scriptsize{2}};
\draw (w1)--(w3);
\draw (w3)--(w2);
\end{tikzpicture}}}:
We can take any geometry with $n_\beta\le\nu\le 2n_\alpha+1$ for $\su(2n_\alpha+1)^{(2)}$, and any geometry with $0\le\nu\le 2n_\beta-9-2n_\alpha$ for $\so(2n_\beta)^{(1)}$. The gluing rules are:
\bit
\item $f-x_1-x_2$ in $S_{n_\alpha,\alpha}$ is glued to $f$ in $S_{0,\beta}$.
\item $x_i-x_{i+1}$ in $S_{n_\alpha,\alpha}$ is glued to $f$ in $S_{i,\beta}$ for $i=1,\cdots,n_\beta-1$.
\item $f,x_{n_\beta-1},x_{n_\beta}$ in $S_{n_\alpha,\alpha}$ are glued to $x_{2n_\alpha+1}-y_{2n_\alpha+1},f-x_1,y_1$ in $S_{n_\beta,\beta}$.
\item $x_{i}-x_{i+1},y_{i+1}-y_i,x_{2n_\alpha+1-i}-x_{2n_\alpha+2-i},y_{2n_\alpha+2-i}-y_{2n_\alpha+1-i}$ in $S_{n_\beta,\beta}$ are glued to $f,f,f,f$ in $S_{n_\alpha-i,\alpha}$ for $i=1,\cdots,n_\alpha-1$.
\item $x_{n_\alpha}-x_{n_\alpha+1},x_{n_\alpha}-x_{n_\alpha+1},x_{n_\alpha+1}-x_{n_\alpha+2},x_{n_\alpha+1}-x_{n_\alpha+2},y_{n_\alpha+1}-y_{n_\alpha},y_{n_\alpha+1}-y_{n_\alpha},y_{n_\alpha+2}-y_{n_\alpha+1},y_{n_\alpha+2}-y_{n_\alpha+1}$ in $S_{n_\beta,\beta}$ are glued to $f,f,f,f,f,f,f,f$ in $S_{0,\alpha}$.
\eit

\noindent\ubf{Gluing rules for \raisebox{-.25\height}{\begin{tikzpicture}
\node (w1) at (-0.7,0.9) {$\su(2n_\alpha+1)^{(2)}$};
\begin{scope}[shift={(3.9,0)}]
\node (w2) at (-0.5,0.9) {$\so(2n_\beta+1)^{(1)}$};
\end{scope}
\node (w3) at (1.4,0.9) {\scriptsize{2}};
\draw (w1)--(w3);
\draw (w3)--(w2);
\end{tikzpicture}}}:
We can take any geometry with $n_\beta+1\le\nu\le 2n_\alpha+1$ for $\su(2n_\alpha+1)^{(2)}$, and any geometry with $0\le\nu\le 2n_\beta-8-2n_\alpha$ for $\so(2n_\beta+1)^{(1)}$. The gluing rules are:
\bit
\item $f-x_1-x_2$ in $S_{n_\alpha,\alpha}$ is glued to $f$ in $S_{0,\beta}$.
\item $x_i-x_{i+1}$ in $S_{n_\alpha,\alpha}$ is glued to $f$ in $S_{i,\beta}$ for $i=1,\cdots,n_\beta-1$.
\item $f-x_{n_\beta+1},x_{n_\beta+1},x_{n_\beta},x_{n_\beta}$ in $S_{n_\alpha,\alpha}$ are glued to $f-x_{2n_\alpha+1},f-y_{2n_\alpha+1},x_1,y_1$ in $S_{n_\beta,\beta}$.
\item $x_{i+1}-x_{i},y_{i+1}-y_i,x_{2n_\alpha+2-i}-x_{2n_\alpha+1-i},y_{2n_\alpha+2-i}-y_{2n_\alpha+1-i}$ in $S_{n_\beta,\beta}$ are glued to $f,f,f,f$ in $S_{n_\alpha-i,\alpha}$ for $i=1,\cdots,n_\alpha-1$.
\item $x_{n_\alpha+1}-x_{n_\alpha},x_{n_\alpha+1}-x_{n_\alpha},x_{n_\alpha+2}-x_{n_\alpha+1},x_{n_\alpha+2}-x_{n_\alpha+1},y_{n_\alpha+1}-y_{n_\alpha},y_{n_\alpha+1}-y_{n_\alpha},y_{n_\alpha+2}-y_{n_\alpha+1},y_{n_\alpha+2}-y_{n_\alpha+1}$ in $S_{n_\beta,\beta}$ are glued to $f,f,f,f,f,f,f,f$ in $S_{0,\alpha}$.
\eit

\noindent\ubf{Gluing rules for \raisebox{-.25\height}{\begin{tikzpicture}
\node (w1) at (-0.7,0.9) {$\su(2n_\alpha+1)^{(2)}$};
\begin{scope}[shift={(3.55,0)}]
\node (w2) at (-0.5,0.9) {$\so(2n_\beta)^{(2)}$};
\end{scope}
\node (w3) at (1.4,0.9) {\scriptsize{2}};
\draw (w1)--(w3);
\draw (w3)--(w2);
\end{tikzpicture}}}:
We can take any geometry with $n_\beta+1\le\nu\le 2n_\alpha+1$ for $\su(2n_\alpha+1)^{(2)}$, and any geometry with $0\le\nu\le 2n_\beta-9-2n_\alpha$ for $\so(2n_\beta)^{(2)}$. The gluing rules are:
\bit
\item $f-x_1-x_2,x_1-x_2$ in $S_{n_\alpha,\alpha}$ are glued to $f,f$ in $S_{0,\beta}$.
\item $x_{i+1}-x_{i+2}$ in $S_{n_\alpha,\alpha}$ is glued to $f$ in $S_{i,\beta}$ for $i=1,\cdots,n_\beta-2$.
\item $f-x_{n_\beta+1},x_{n_\beta+1},x_{n_\beta},x_{n_\beta}$ in $S_{n_\alpha,\alpha}$ are glued to $f-x_{2n_\alpha+1},f-y_{2n_\alpha+1},x_1,y_1$ in $S_{n_\beta-1,\beta}$.
\item $x_{i+1}-x_{i},y_{i+1}-y_i,x_{2n_\alpha+2-i}-x_{2n_\alpha+1-i},y_{2n_\alpha+2-i}-y_{2n_\alpha+1-i}$ in $S_{n_\beta-1,\beta}$ are glued to $f,f,f,f$ in $S_{n_\alpha-i,\alpha}$ for $i=1,\cdots,n_\alpha-1$.
\item $x_{n_\alpha+1}-x_{n_\alpha},x_{n_\alpha+1}-x_{n_\alpha},x_{n_\alpha+2}-x_{n_\alpha+1},x_{n_\alpha+2}-x_{n_\alpha+1},y_{n_\alpha+1}-y_{n_\alpha},y_{n_\alpha+1}-y_{n_\alpha},y_{n_\alpha+2}-y_{n_\alpha+1},y_{n_\alpha+2}-y_{n_\alpha+1}$ in $S_{n_\beta-1,\beta}$ are glued to $f,f,f,f,f,f,f,f$ in $S_{0,\alpha}$.
\eit

\noindent\ubf{Gluing rules for \raisebox{-.25\height}{\begin{tikzpicture}
\node (w1) at (-0.7,0.9) {$\su(2n_\alpha)^{(2)}$};
\begin{scope}[shift={(3.2,0)}]
\node (w2) at (-0.5,0.9) {$\su(2n_\beta)^{(2)}$};
\end{scope}
\node (w3) at (1,0.9) {\scriptsize{2}};
\draw (w1)--(w3);
\draw[->] (w3)--(w2);
\end{tikzpicture}}}:
We can take any geometry with $2n_\beta\le\nu\le 2n_\alpha$ for $\su(2n_\alpha)^{(2)}$, and any geometry with $0\le\nu\le 2n_\beta-n_\alpha$ for $\su(2n_\beta)^{(2)}$. The gluing rules are:
\bit
\item $f-x_1-x_2,x_{2n_\beta-1},x_{2n_\beta}$ in $S_{n_\alpha,\alpha}$ are glued to $f,f-x_1,y_1$ in $S_{0,\beta}$.
\item $x_i-x_{i+1},x_{2n_\beta-i}-x_{2n_\beta+1-i}$ in $S_{n_\alpha,\alpha}$ are glued to $f,f$ in $S_{i,\beta}$ for $i=1,\cdots,n_\beta-1$.
\item $x_{n_\beta}-x_{n_\beta+1}$ in $S_{n_\alpha,\alpha}$ is glued to $f$ in $S_{n_\beta,\beta}$.
\item $x_{i}-x_{i+1},y_{i+1}-y_i$ in $S_{0,\beta}$ are glued to $f,f$ in $S_{n_\alpha-i,\alpha}$ for $i=1,\cdots,n_\alpha-1$.
\item $x_{n_\alpha-1}-y_{n_\alpha},x_{n_\alpha}-y_{n_\alpha-1}$ in $S_{0,\beta}$ are glued to $f,f$ in $S_{0,\alpha}$.
\eit

\noindent\ubf{Gluing rules for \raisebox{-.25\height}{\begin{tikzpicture}
\node (w1) at (-0.7,0.9) {$\su(2n_\alpha)^{(2)}$};
\begin{scope}[shift={(3.6,0)}]
\node (w2) at (-0.5,0.9) {$\su(2n_\beta+1)^{(2)}$};
\end{scope}
\node (w3) at (1,0.9) {\scriptsize{2}};
\draw (w1)--(w3);
\draw[->] (w3)--(w2);
\end{tikzpicture}}}:
We can take any geometry with $2n_\beta+1\le\nu\le 2n_\alpha-1$ for $\su(2n_\alpha)^{(2)}$, and any geometry with $0\le\nu\le 2n_\beta+1-n_\alpha$ for $\su(2n_\beta+1)^{(2)}$. The gluing rules are:
\bit
\item $f-x_1-x_2,x_1-x_2,x_{2n_\beta+1},x_{2n_\beta+1}$ in $S_{n_\alpha,\alpha}$ are glued to $f,f,x_1,y_1$ in $S_{0,\beta}$.
\item $x_{i+1}-x_{i+2},x_{2n_\beta+1-i}-x_{2n_\beta+2-i}$ in $S_{n_\alpha,\alpha}$ are glued to $f,f$ in $S_{i,\beta}$ for $i=1,\cdots,n_\beta-1$.
\item $x_{n_\beta+1}-x_{n_\beta+2}$ in $S_{n_\alpha,\alpha}$ is glued to $f$ in $S_{n_\beta,\beta}$.
\item $x_{i+1}-x_{i},y_{i+1}-y_i$ in $S_{0,\beta}$ are glued to $f,f$ in $S_{n_\alpha-i,\alpha}$ for $i=1,\cdots,n_\alpha-1$.
\item $f-y_{n_\alpha-1},f-y_{n_\alpha},f-x_{n_\alpha-1},f-x_{n_\alpha}$ in $S_{0,\beta}$ are glued to $f-x_1,y_1,f-y_1,x_1$ in $S_{0,\alpha}$.
\eit

\noindent\ubf{Gluing rules for \raisebox{-.25\height}{\begin{tikzpicture}
\node (w1) at (-0.7,0.9) {$\su(2n_\alpha+1)^{(2)}$};
\begin{scope}[shift={(3.6,0)}]
\node (w2) at (-0.5,0.9) {$\su(2n_\beta)^{(2)}$};
\end{scope}
\node (w3) at (1.4,0.9) {\scriptsize{2}};
\draw (w1)--(w3);
\draw[->] (w3)--(w2);
\end{tikzpicture}}}:
We can take any geometry with $2n_\beta\le\nu\le 2n_\alpha+1$ for $\su(2n_\alpha+1)^{(2)}$, and any geometry with $0\le\nu\le 2n_\beta-n_\alpha-1$ for $\su(2n_\beta)^{(2)}$. The gluing rules are:
\bit
\item $f-x_1-x_2,x_{2n_\beta-1},x_{2n_\beta}$ in $S_{n_\alpha,\alpha}$ are glued to $f,f-x_1,y_1$ in $S_{0,\beta}$.
\item $x_i-x_{i+1},x_{2n_\beta-i}-x_{2n_\beta+1-i}$ in $S_{n_\alpha,\alpha}$ are glued to $f,f$ in $S_{i,\beta}$ for $i=1,\cdots,n_\beta-1$.
\item $x_{n_\beta}-x_{n_\beta+1}$ in $S_{n_\alpha,\alpha}$ is glued to $f$ in $S_{n_\beta,\beta}$.
\item $x_{i}-x_{i+1},y_{i+1}-y_i$ in $S_{0,\beta}$ are glued to $f,f$ in $S_{n_\alpha-i,\alpha}$ for $i=1,\cdots,n_\alpha-1$.
\item $x_{n_\alpha}-x_{n_\alpha+1},x_{n_\alpha}-y_{n_\alpha+1},x_{n_\alpha+1}-y_{n_\alpha},y_{n_\alpha+1}-y_{n_\alpha}$ in $S_{0,\beta}$ are glued to $f,f,f,f$ in $S_{0,\alpha}$.
\eit

\noindent\ubf{Gluing rules for \raisebox{-.25\height}{\begin{tikzpicture}
\node (w1) at (-0.7,0.9) {$\su(2n_\alpha+1)^{(2)}$};
\begin{scope}[shift={(4,0)}]
\node (w2) at (-0.5,0.9) {$\su(2n_\beta+1)^{(2)}$};
\end{scope}
\node (w3) at (1.4,0.9) {\scriptsize{2}};
\draw (w1)--(w3);
\draw[->] (w3)--(w2);
\end{tikzpicture}}}:
We can take any geometry with $2n_\beta+1\le\nu\le 2n_\alpha$ for $\su(2n_\alpha+1)^{(2)}$, and any geometry with $0\le\nu\le 2n_\beta-n_\alpha$ for $\su(2n_\beta+1)^{(2)}$. The gluing rules are:
\bit
\item $f-x_1-x_2,x_1-x_2,x_{2n_\beta+1},x_{2n_\beta+1}$ in $S_{n_\alpha,\alpha}$ are glued to $f,f,x_1,y_1$ in $S_{0,\beta}$.
\item $x_{i+1}-x_{i+2},x_{2n_\beta+1-i}-x_{2n_\beta+2-i}$ in $S_{n_\alpha,\alpha}$ are glued to $f,f$ in $S_{i,\beta}$ for $i=1,\cdots,n_\beta-1$.
\item $x_{n_\beta+1}-x_{n_\beta+2}$ in $S_{n_\alpha,\alpha}$ is glued to $f$ in $S_{n_\beta,\beta}$.
\item $x_{i+1}-x_{i},y_{i+1}-y_i$ in $S_{0,\beta}$ are glued to $f,f$ in $S_{n_\alpha-i,\alpha}$ for $i=1,\cdots,n_\alpha-1$.
\item $f-y_{n_\alpha+1},f-y_{n_\alpha},f-x_{n_\alpha+1},f-x_{n_\alpha},x_{n_\alpha+1}-x_{n_\alpha},y_{n_\alpha+1}-y_{n_\alpha}$ in $S_{0,\beta}$ are glued to $f-x_1,y_1,f-y_1,x_1,f,f$ in $S_{0,\alpha}$.
\eit

\noindent\ubf{Gluing rules for \raisebox{-.25\height}{\begin{tikzpicture}
\node (w1) at (-0.7,0.9) {$\su(2n_\alpha)^{(2)}$};
\begin{scope}[shift={(3.2,0)}]
\node (w2) at (-0.5,0.9) {$\su(2n_\beta)^{(2)}$};
\end{scope}
\node (w3) at (1,0.9) {\scriptsize{3}};
\draw (w1)--(w3);
\draw[->] (w3)--(w2);
\end{tikzpicture}}}:
We can take any geometry with $3n_\beta\le\nu\le 2n_\alpha$ for $\su(2n_\alpha)^{(2)}$, and any geometry with $n_\alpha\le\nu\le 2n_\beta$ for $\su(2n_\beta)^{(2)}$. The gluing rules are:
\bit
\item $f-x_1-x_2,x_{2n_\beta-1}-x_{2n_\beta+1},x_{2n_\beta}-x_{2n_\beta+2}$ in $S_{n_\alpha,\alpha}$ are glued to $f,f,f$ in $S_{0,\beta}$.
\item $x_i-x_{i+1},x_{2n_\beta-i}-x_{2n_\beta+1-i},x_{2n_\beta+i}-x_{2n_\beta+1+i}$ in $S_{n_\alpha,\alpha}$ are glued to $f,f,f$ in $S_{i,\beta}$ for $i=1,\cdots,n_\beta-1$.
\item $x_{n_\beta}-x_{n_\beta+1},x_{3n_\beta}$ in $S_{n_\alpha,\alpha}$ are glued to $f,x_{n_\alpha}$ in $S_{n_\beta,\beta}$.
\item $x_{n_\alpha-i}-x_{n_\alpha-i+1}$ in $S_{n_\beta,\beta}$ are glued to $f,f$ in $S_{n_\alpha-i,\alpha}$ for $i=1,\cdots,n_\alpha-1$.
\item $f-x_1-x_2$ in $S_{n_\beta,\beta}$ is glued to $f$ in $S_{0,\alpha}$.
\eit

\noindent\ubf{Gluing rules for \raisebox{-.25\height}{\begin{tikzpicture}
\node (w1) at (-0.7,0.9) {$\su(2n_\alpha)^{(2)}$};
\begin{scope}[shift={(3.6,0)}]
\node (w2) at (-0.5,0.9) {$\su(2n_\beta+1)^{(2)}$};
\end{scope}
\node (w3) at (1,0.9) {\scriptsize{3}};
\draw (w1)--(w3);
\draw[->] (w3)--(w2);
\end{tikzpicture}}}:
We can take any geometry with $3n_\beta+2\le\nu\le 2n_\alpha$ for $\su(2n_\alpha)^{(2)}$, and any geometry with $n_\alpha\le\nu\le 2n_\beta+1$ for $\su(2n_\beta+1)^{(2)}$. The gluing rules are:
\bit
\item $f-x_1-x_2,x_1-x_2,x_{2n_\beta+1}-x_{2n_\beta+2},x_{2n_\beta+1}-x_{2n_\beta+2},x_{2n_\beta+2}-x_{2n_\beta+3},x_{2n_\beta+2}-x_{2n_\beta+3}$ in $S_{n_\alpha,\alpha}$ are glued to $f,f,f,f,f,f$ in $S_{0,\beta}$.
\item $x_{i+1}-x_{i+2},x_{2n_\beta+1-i}-x_{2n_\beta+2-i},x_{2n_\beta+2+i}-x_{2n_\beta+3+i}$ in $S_{n_\alpha,\alpha}$ are glued to $f,f,f$ in $S_{i,\beta}$ for $i=1,\cdots,n_\beta-1$.
\item $x_{n_\beta+1}-x_{n_\beta+2},x_{3n_\beta+2}$ in $S_{n_\alpha,\alpha}$ are glued to $f,x_{n_\alpha}$ in $S_{n_\beta,\beta}$.
\item $x_{n_\alpha-i}-x_{n_\alpha-i+1}$ in $S_{n_\beta,\beta}$ are glued to $f,f$ in $S_{n_\alpha-i,\alpha}$ for $i=1,\cdots,n_\alpha-1$.
\item $f-x_1-x_2$ in $S_{n_\beta,\beta}$ is glued to $f$ in $S_{0,\alpha}$.
\eit

\noindent\ubf{Gluing rules for \raisebox{-.25\height}{\begin{tikzpicture}
\node (w1) at (-0.7,0.9) {$\su(2n_\alpha+1)^{(2)}$};
\begin{scope}[shift={(3.6,0)}]
\node (w2) at (-0.5,0.9) {$\su(2n_\beta)^{(2)}$};
\end{scope}
\node (w3) at (1.4,0.9) {\scriptsize{3}};
\draw (w1)--(w3);
\draw[->] (w3)--(w2);
\end{tikzpicture}}}:
We can take any geometry with $3n_\beta\le\nu\le 2n_\alpha+1$ for $\su(2n_\alpha+1)^{(2)}$, and any geometry with $n_\alpha+1\le\nu\le 2n_\beta$ for $\su(2n_\beta)^{(2)}$. The gluing rules are:
\bit
\item $f-x_1-x_2,x_{2n_\beta-1}-x_{2n_\beta+1},x_{2n_\beta}-x_{2n_\beta+2}$ in $S_{n_\alpha,\alpha}$ are glued to $f,f,f$ in $S_{0,\beta}$.
\item $x_i-x_{i+1},x_{2n_\beta-i}-x_{2n_\beta+1-i},x_{2n_\beta+i}-x_{2n_\beta+1+i}$ in $S_{n_\alpha,\alpha}$ are glued to $f,f,f$ in $S_{i,\beta}$ for $i=1,\cdots,n_\beta-1$.
\item $x_{n_\beta}-x_{n_\beta+1},x_{3n_\beta}$ in $S_{n_\alpha,\alpha}$ are glued to $f,x_{n_\alpha+1}$ in $S_{n_\beta,\beta}$.
\item $x_{n_\alpha+1-i}-x_{n_\alpha+2-i}$ in $S_{n_\beta,\beta}$ are glued to $f,f$ in $S_{n_\alpha-i,\alpha}$ for $i=1,\cdots,n_\alpha-1$.
\item $f-x_1-x_2,x_1-x_2$ in $S_{n_\beta,\beta}$ are glued to $f,f$ in $S_{0,\alpha}$.
\eit

\noindent\ubf{Gluing rules for \raisebox{-.25\height}{\begin{tikzpicture}
\node (w1) at (-0.7,0.9) {$\su(2n_\alpha+1)^{(2)}$};
\begin{scope}[shift={(4,0)}]
\node (w2) at (-0.5,0.9) {$\su(2n_\beta+1)^{(2)}$};
\end{scope}
\node (w3) at (1.4,0.9) {\scriptsize{3}};
\draw (w1)--(w3);
\draw[->] (w3)--(w2);
\end{tikzpicture}}}:
We can take any geometry with $3n_\beta+2\le\nu\le 2n_\alpha+1$ for $\su(2n_\alpha+1)^{(2)}$, and any geometry with $n_\alpha+1\le\nu\le 2n_\beta+1$ for $\su(2n_\beta+1)^{(2)}$. The gluing rules are:
\bit
\item $f-x_1-x_2,x_1-x_2,x_{2n_\beta+1}-x_{2n_\beta+2},x_{2n_\beta+1}-x_{2n_\beta+2},x_{2n_\beta+2}-x_{2n_\beta+3},x_{2n_\beta+2}-x_{2n_\beta+3}$ in $S_{n_\alpha,\alpha}$ are glued to $f,f,f,f,f,f$ in $S_{0,\beta}$.
\item $x_{i+1}-x_{i+2},x_{2n_\beta+1-i}-x_{2n_\beta+2-i},x_{2n_\beta+2+i}-x_{2n_\beta+3+i}$ in $S_{n_\alpha,\alpha}$ are glued to $f,f,f$ in $S_{i,\beta}$ for $i=1,\cdots,n_\beta-1$.
\item $x_{n_\beta+1}-x_{n_\beta+2},x_{3n_\beta+2}$ in $S_{n_\alpha,\alpha}$ are glued to $f,x_{n_\alpha+1}$ in $S_{n_\beta,\beta}$.
\item $x_{n_\alpha+1-i}-x_{n_\alpha+2-i}$ in $S_{n_\beta,\beta}$ are glued to $f,f$ in $S_{n_\alpha-i,\alpha}$ for $i=1,\cdots,n_\alpha-1$.
\item $f-x_1-x_2,x_1-x_2$ in $S_{n_\beta,\beta}$ are glued to $f,f$ in $S_{0,\alpha}$.
\eit

\subsection{Reading the data of permutation twist from the gluing rules}\label{RD}
As we have seen above, there are sometimes multiple ways to combine two KK building blocks. For example,
\be
\begin{tikzpicture}
\node at (-0.5,0.4) {2};
\node at (-0.45,0.9) {$\su(m)^{(2)}$};
\end{tikzpicture}
\ee
and
\be
\begin{tikzpicture}
\node at (-0.5,0.4) {2};
\node at (-0.45,0.9) {$\su(n)^{(2)}$};
\end{tikzpicture}
\ee
can be combined to produce either
\be\label{D1}
\begin{tikzpicture}
\node (v1) at (-0.5,0.4) {2};
\node at (-0.45,0.9) {$\su(m)^{(2)}$};
\begin{scope}[shift={(2,0)}]
\node (v2) at (-0.5,0.4) {$2$};
\node at (-0.45,0.9) {$\su(n)^{(2)}$};
\end{scope}
\draw  (v1) edge (v2);
\end{tikzpicture}
\ee
or
\be\label{D2}
\begin{tikzpicture}
\node (v1) at (-0.5,0.4) {2};
\node at (-0.45,0.9) {$\su(m)^{(2)}$};
\begin{scope}[shift={(2,0)}]
\node (v2) at (-0.5,0.4) {$2$};
\node at (-0.45,0.9) {$\su(n)^{(2)}$};
\end{scope}
\node (v3) at (0.5,0.4) {\tiny{$2$}};
\draw  (v1) -- (v3);
\draw [<-] (v2) -- (v3);
\end{tikzpicture}
\ee
or
\be\label{D3}
\begin{tikzpicture}
\node (v1) at (-0.5,0.4) {2};
\node at (-0.45,0.9) {$\su(m)^{(2)}$};
\begin{scope}[shift={(2,0)}]
\node (v2) at (-0.5,0.4) {$2$};
\node at (-0.45,0.9) {$\su(n)^{(2)}$};
\end{scope}
\node (v3) at (0.5,0.4) {\tiny{$3$}};
\draw  (v1) -- (v3);
\draw [<-] (v2) -- (v3);
\end{tikzpicture}
\ee
The difference between these three combinations is captured in the corresponding gluing rules. In this subsection, our aim is to review how one can recover the type of edge joining two building blocks from the gluing rules associated to the edge \cite{Bhardwaj:2020gyu}.

The difference between (\ref{D1}), (\ref{D2}) and (\ref{D3}) can be captured by a matrix of Chern-Simons couplings $\left[\Omega_{\alpha\beta}\right]$ descending from the GS coupling of the underlying $6d$ SCFT (see \cite{Bhardwaj:2019fzv}). For a general KK theory, $\left[\Omega_{\alpha\beta}\right]$ is an $r\times r$ matrix if the KK theory is produced by gluing $r$ building blocks. The type of edge between two building blocks $\alpha$ and $\beta$ is captured by the two off-diagonal $\Omega_{\alpha\beta}$ and $\Omega_{\beta\alpha}$ of the matrix. This matrix is encoded in the CY3 associated to the KK theory as follows \cite{Bhardwaj:2020gyu}:\\
Let $\tilde e_\alpha$ be the $e$ curve of a specific Hirzebruch surface out of the surfaces $S_{i,\alpha}$ for various $i$. This specific surface can be taken to be the affine surface $S_{0,\alpha}$ if the associated affine algebra $\fg_\alpha^{(q)}$ is untwisted, that is has $q=1$. It can be taken to be the surface $S_{n,\alpha}$ for $\su(2n)^{(2)}$, $\su(2n+1)^{(2)}$; the surface $S_{n-2}$ for $\so(2n)^{(2)}$; the surface $S_{2}$ for $\so(8)^{(3)}$; and the surface $S_3$ for $\fe_6^{(2)}$ (the labeling of surfaces can be found in this paper and \cite{Bhardwaj:2019fzv}).\\
Let us define a surface $S_\alpha$ as
\be
S_\alpha:=\sum_id^\vee_{i,\alpha}S_{i,\alpha}
\ee
where $d^\vee_{i,\alpha}$ are the dual Coxeter numbers associated to $\fg_\alpha^{(q)}$.\\
Then, we have the relationship
\be\label{r}
\Omega_{\alpha\beta}=-S_\alpha\cdot\tilde e_\beta
\ee

Using (\ref{r}) on the gluing rules proposed for (\ref{D1}), (\ref{D2}) and (\ref{D3}), we find that the associated off-diagonal entries are respectively
\be
\Omega_{\alpha\beta}=\Omega_{\beta\alpha}=-1
\ee
\be
\Omega_{\alpha\beta}=-2,~\Omega_{\beta\alpha}=-1
\ee
and
\be
\Omega_{\alpha\beta}=-3,~\Omega_{\beta\alpha}=-1
\ee
These are precisely the off-diagonal entries in the matrix of CS couplings associated to (\ref{D1}), (\ref{D2}) and (\ref{D3}) respectively \cite{Bhardwaj:2019fzv}. 

Notice that the correspondence between these off-diagonal entries and the types of edges shown in graphs (\ref{D1}), (\ref{D2}), (\ref{D3}) is precisely the correspondence between the off-diagonal entries for a Cartan matrix and the type of edge in the associated Dynkin graph. $p$ number of directed edges from a node $\alpha$ to another node $\beta$ in a Dynkin graph translate to the off-diagonal entries $\Omega_{\alpha\beta}=-p$ and $\Omega_{\beta\alpha}=-1$ in the associated Cartan matrix, and $p$ number of undirected edges from $\alpha$ to $\beta$ translate to the off-diagonal entries $\Omega_{\alpha\beta}=\Omega_{\beta\alpha}=-p$. The reader can similarly check that this correspondence between type of edge and off-diagonal entries $\Omega_{\alpha\beta},\Omega_{\beta\alpha}$ holds true for other combinations of building blocks shown in Table \ref{KR2}. This provides a non-trivial consistency check on the proposed gluing rules.

\subsection{Low energy effective gauge theory}\label{LE2}
Using the techniques of Section \ref{LE}, it is also possible to study the impact of gluing KK theory building blocks upon the associated low-energy theories, which is the topic of discussion in this subsection.

Consider the KK theory
\be\label{E1}
\begin{tikzpicture}
\node (v1) at (-2.3,0.4) {$k$};
\node (v2) at (0.1,0.4) {$l$};
\node at (0.15,0.9) {$\su(2n_\beta)^{(2)}$};
\draw  (v1) edge (v2);
\node at (-2.25,0.9) {$\su(2n_\alpha)^{(2)}$};
\end{tikzpicture}
\ee
According to the arguments at the beginning of this section, the gluing rules suggested in Section \ref{GR} above should imply that, if the low-energy gauge algebras chosen for the two building blocks are $\so(2n_\alpha)$ and $\sp(n_\beta)$ respectively, then contracting a maximal set of blowups should lead to a half-hyper charged in bifundamental of $\so(2n_\alpha)\oplus\sp(n_\beta)$.

This can be easily verified. To obtain the above low-energy limit, we can contract all fibers to zero size except the fibers for $S_{n_\alpha,\alpha}$ and $S_{0,\beta}$. This forces all blowups $x_i$ for $i=1,\cdots,n_\alpha$ living in $S_{0,\beta}$ (and participating in the gluing rules) to remain at non-zero volume, while the blowups $y_i$ for $i=1,\cdots,n_\alpha$ living in $S_{0,\beta}$ can be contracted to zero volume. On the other hand, all the blowups living in $S_{0,\alpha}$ (and participating in the gluing rules) can be consistently contracted to zero size. To read the matter content in the low-energy theory, we restrict our attention only to those surfaces whose corresponding fibers are contracted to zero volume. The gluing rules are then reduced to:
\bit
\item $x_i-x_{i+1},y_{i+1}-y_i$ in $S_{0,\alpha}$ are glued to $f,f$ in $S_{i,\beta}$ for $i=1,\cdots,n_\beta-1$.
\item $x_{n_\beta}-y_{n_\beta}$ in $S_{0,\alpha}$ is glued to $f$ in $S_{n_\beta,\beta}$.
\eit
In other words, the above gluing rules are telling us that a total of $n_\beta$ fundamental hypers of $\so(2n_\alpha)$ are gauged by an $\sp(n_\beta)$. Thus, these blowups (living in $S_{0,\alpha}$) must give rise to a half-hyper in bifundamental of $\so(2n_\alpha)\oplus\sp(n_\beta)$, as expected.

What happens if instead we choose the low-energy gauge algebra to be $\sp(n_\alpha)\oplus\sp(n_\beta)$? In this case, since a \emph{half-hyper} in bifundamental is not possible, we would expect to obtain no matter degrees of freedom charged under a mixed representation of $\sp(n_\alpha)\oplus\sp(n_\beta)$. This can again be verified using the gluing rules presented in Section \ref{GR}. To obtain the above low-energy limit, we can contract all fibers to zero size except the fibers for $S_{0,\alpha}$ and $S_{0,\beta}$. The reader can verify that this limit forces all the blowups living in $S_{0,\alpha}$ and $S_{0,\beta}$ (and participating in the gluing rules) to remain at positive, non-zero volume. Since none of the blowups participating in the gluing rules give rise to massless particles, the low-energy $\sp(n_\alpha)\oplus\sp(n_\beta)$ has no hypers charged in a mixed representation of $\sp(n_\alpha)\oplus\sp(n_\beta)$, as expected.

In a similar way, one can check that the low-energy mixed hyper content for other combinations of KK building blocks, as expected from the arguments presented at the beginning of this section, is concretely reproduced by the gluing rules proposed in Section \ref{GR}, thus providing strong consistency checks between the arguments and proposals presented in this paper.

\section*{Acknowledgements}
The author thanks Gabi Zafrir for many useful discussions.\\
This work is supported by NSF grant PHY-1719924.

\bibliographystyle{ytphys}
\let\bbb\bibitem\def\bibitem{\itemsep4pt\bbb}
\bibliography{ref}

\providecommand{\href}[2]{#2}\begingroup\raggedright\begin{thebibliography}{10}

\bibitem{DelZotto:2017pti}
M.~Del~Zotto, J.~J. Heckman, and D.~R. Morrison, ``{6D SCFTs and Phases of 5D
  Theories},'' \href{http://dx.doi.org/10.1007/JHEP09(2017)147}{{\em JHEP}
  {\bfseries 09} (2017) 147},
\href{http://arxiv.org/abs/1703.02981}{{\ttfamily arXiv:1703.02981 [hep-th]}}.

\bibitem{Jefferson:2018irk}
P.~Jefferson, S.~Katz, H.-C. Kim, and C.~Vafa, ``{On Geometric Classification
  of 5d SCFTs},'' \href{http://dx.doi.org/10.1007/JHEP04(2018)103}{{\em JHEP}
  {\bfseries 04} (2018) 103},
\href{http://arxiv.org/abs/1801.04036}{{\ttfamily arXiv:1801.04036 [hep-th]}}.

\bibitem{Bhardwaj:2018yhy}
L.~Bhardwaj and P.~Jefferson, ``{Classifying 5d SCFTs via 6d SCFTs: Rank
  one},''
\href{http://arxiv.org/abs/1809.01650}{{\ttfamily arXiv:1809.01650 [hep-th]}}.

\bibitem{Bhardwaj:2018vuu}
L.~Bhardwaj and P.~Jefferson, ``{Classifying 5d SCFTs via 6d SCFTs: Arbitrary
  rank},''
\href{http://arxiv.org/abs/1811.10616}{{\ttfamily arXiv:1811.10616 [hep-th]}}.

\bibitem{Apruzzi:2018nre}
F.~Apruzzi, L.~Lin, and C.~Mayrhofer, ``{Phases of 5d SCFTs from M-/F-theory on
  Non-Flat Fibrations},'' \href{http://dx.doi.org/10.1007/JHEP05(2019)187}{{\em
  JHEP} {\bfseries 05} (2019) 187},
\href{http://arxiv.org/abs/1811.12400}{{\ttfamily arXiv:1811.12400 [hep-th]}}.

\bibitem{Apruzzi:2019vpe}
F.~Apruzzi, C.~Lawrie, L.~Lin, S.~Schafer-Nameki, and Y.-N. Wang, ``{5d
  Superconformal Field Theories and Graphs},''
  \href{http://dx.doi.org/10.1016/j.physletb.2019.135077}{{\em Phys.Lett.}
  {\bfseries B800} (2019) 135077},
\href{http://arxiv.org/abs/1906.11820}{{\ttfamily arXiv:1906.11820 [hep-th]}}.

\bibitem{Apruzzi:2019opn}
F.~Apruzzi, C.~Lawrie, L.~Lin, S.~Schafer-Nameki, and Y.-N. Wang, ``{Fibers add
  Flavor, Part I: Classification of 5d SCFTs, Flavor Symmetries and BPS
  States},'' \href{http://dx.doi.org/10.1007/JHEP11(2019)068}{{\em JHEP}
  {\bfseries 11} (2019) 068},
\href{http://arxiv.org/abs/1907.05404}{{\ttfamily arXiv:1907.05404 [hep-th]}}.

\bibitem{Bhardwaj:2019ngx}
L.~Bhardwaj, ``{Dualities of $5d$ gauge theories from S-duality},''
\href{http://arxiv.org/abs/1909.05250}{{\ttfamily arXiv:1909.05250 [hep-th]}}.

\bibitem{Apruzzi:2019enx}
F.~Apruzzi, C.~Lawrie, L.~Lin, S.~Schafer-Nameki, and Y.-N. Wang, ``{Fibers add
  Flavor, Part II: 5d SCFTs, Gauge Theories, and Dualities},''
\href{http://arxiv.org/abs/1909.09128}{{\ttfamily arXiv:1909.09128 [hep-th]}}.

\bibitem{Bhardwaj:2019jtr}
L.~Bhardwaj, ``{On the classification of $5d$ SCFTs},''
\href{http://arxiv.org/abs/1909.09635}{{\ttfamily arXiv:1909.09635 [hep-th]}}.

\bibitem{Bhardwaj:2019fzv}
L.~Bhardwaj, P.~Jefferson, H.-C. Kim, H.-C. Tarazi, and C.~Vafa, ``{Twisted
  Circle Compactification of 6d SCFTs},''
\href{http://arxiv.org/abs/1909.11666}{{\ttfamily arXiv:1909.11666 [hep-th]}}.

\bibitem{Bhardwaj:2019xeg}
L.~Bhardwaj, ``{Do all 5d SCFTs descend from 6d SCFTs?},''
\href{http://arxiv.org/abs/1912.00025}{{\ttfamily arXiv:1912.00025 [hep-th]}}.

\bibitem{Apruzzi:2019syw}
F.~Apruzzi, S.~Schafer-Nameki, and Y.-N. Wang, ``{5d SCFTs from Decoupling and
  Gluing},'' \href{http://arxiv.org/abs/1912.04264}{{\ttfamily arXiv:1912.04264
  [hep-th]}}.

\bibitem{Bhardwaj:2020gyu}
L.~Bhardwaj and G.~Zafrir, ``{Classification of 5d N=1 gauge theories},''
  \href{http://arxiv.org/abs/2003.04333}{{\ttfamily arXiv:2003.04333
  [hep-th]}}.

\bibitem{Jefferson:2017ahm}
P.~Jefferson, H.-C. Kim, C.~Vafa, and G.~Zafrir, ``{Towards Classification of
  5d SCFTs: Single Gauge Node},''
\href{http://arxiv.org/abs/1705.05836}{{\ttfamily arXiv:1705.05836 [hep-th]}}.

\bibitem{Hayashi:2016abm}
H.~Hayashi, S.-S. Kim, K.~Lee, and F.~Yagi, ``{Equivalence of several
  descriptions for 6d SCFT},''
  \href{http://dx.doi.org/10.1007/JHEP01(2017)093}{{\em JHEP} {\bfseries 01}
  (2017) 093},
\href{http://arxiv.org/abs/1607.07786}{{\ttfamily arXiv:1607.07786 [hep-th]}}.

\bibitem{Hayashi:2018bkd}
H.~Hayashi, S.-S. Kim, K.~Lee, and F.~Yagi, ``{5-brane webs for 5d $
  \mathcal{N} $ = 1 G$_{2}$ gauge theories},''
  \href{http://dx.doi.org/10.1007/JHEP03(2018)125}{{\em JHEP} {\bfseries 03}
  (2018) 125},
\href{http://arxiv.org/abs/1801.03916}{{\ttfamily arXiv:1801.03916 [hep-th]}}.

\bibitem{Hayashi:2018lyv}
H.~Hayashi, S.-S. Kim, K.~Lee, and F.~Yagi, ``{Dualities and 5-brane webs for
  5d rank 2 SCFTs},'' \href{http://dx.doi.org/10.1007/JHEP12(2018)016}{{\em
  JHEP} {\bfseries 12} (2018) 016},
\href{http://arxiv.org/abs/1806.10569}{{\ttfamily arXiv:1806.10569 [hep-th]}}.

\bibitem{Closset:2019mdz}
C.~Closset and M.~Del~Zotto, ``{On 5d SCFTs and their BPS quivers. Part I:
  B-branes and brane tilings},''
  \href{http://arxiv.org/abs/1912.13502}{{\ttfamily arXiv:1912.13502
  [hep-th]}}.

\bibitem{Cabrera:2019hya}
S.~Cabrera, A.~Hanany, and F.~Yagi, ``{Tropical Geometry and Five Dimensional
  Higgs Branches at Infinite Coupling},''
  \href{http://dx.doi.org/10.1007/JHEP01(2019)068}{{\em JHEP} {\bfseries 01}
  (2019) 068}, \href{http://arxiv.org/abs/1810.01379}{{\ttfamily
  arXiv:1810.01379 [hep-th]}}.

\bibitem{Kim:2019uqw}
J.~Kim, S.-S. Kim, K.-H. Lee, K.~Lee, and J.~Song, ``{Instantons from
  Blow-up},''
\href{http://arxiv.org/abs/1908.11276}{{\ttfamily arXiv:1908.11276 [hep-th]}}.

\bibitem{Kim:2019dqn}
H.-C. Kim, S.-S. Kim, and K.~Lee, ``{Higgsing and Twisting of 6d $D_N$ gauge
  theories},''
\href{http://arxiv.org/abs/1908.04704}{{\ttfamily arXiv:1908.04704 [hep-th]}}.

\bibitem{Fluder:2019szh}
M.~Fluder, S.~M. Hosseini, and C.~F. Uhlemann, ``{Black hole microstate
  counting in Type IIB from 5d SCFTs},''
  \href{http://dx.doi.org/10.1007/JHEP05(2019)134}{{\em JHEP} {\bfseries 05}
  (2019) 134},
\href{http://arxiv.org/abs/1902.05074}{{\ttfamily arXiv:1902.05074 [hep-th]}}.

\bibitem{Uhlemann:2019ypp}
C.~F. Uhlemann, ``{Exact results for 5d SCFTs of long quiver type},''
\href{http://arxiv.org/abs/1909.01369}{{\ttfamily arXiv:1909.01369 [hep-th]}}.

\bibitem{Saxena:2019wuy}
V.~Saxena, ``{Rank-two 5d SCFTs from M-theory at isolated toric singularities:
  a systematic study},''
\href{http://arxiv.org/abs/1911.09574}{{\ttfamily arXiv:1911.09574 [hep-th]}}.

\bibitem{Hayashi:2019yxj}
H.~Hayashi, S.-S. Kim, K.~Lee, and F.~Yagi, ``{Rank-3 antisymmetric matter on
  5-brane webs},'' \href{http://dx.doi.org/10.1007/JHEP05(2019)133}{{\em JHEP}
  {\bfseries 05} (2019) 133},
\href{http://arxiv.org/abs/1902.04754}{{\ttfamily arXiv:1902.04754 [hep-th]}}.

\bibitem{Closset:2018bjz}
C.~Closset, M.~Del~Zotto, and V.~Saxena, ``{Five-dimensional SCFTs and gauge
  theory phases: an M-theory/type IIA perspective},''
  \href{http://dx.doi.org/10.21468/SciPostPhys.6.5.052}{{\em SciPost Phys.}
  {\bfseries 6} no.~5, (2019) 052},
\href{http://arxiv.org/abs/1812.10451}{{\ttfamily arXiv:1812.10451 [hep-th]}}.

\bibitem{Hayashi:2015zka}
H.~Hayashi, S.-S. Kim, K.~Lee, and F.~Yagi, ``{6d SCFTs, 5d Dualities and Tao
  Web Diagrams},'' \href{http://dx.doi.org/10.1007/JHEP05(2019)203}{{\em JHEP}
  {\bfseries 05} (2019) 203},
\href{http://arxiv.org/abs/1509.03300}{{\ttfamily arXiv:1509.03300 [hep-th]}}.

\bibitem{Hayashi:2020sly}
H.~Hayashi, S.-S. Kim, K.~Lee, and F.~Yagi, ``{Complete prepotential for 5d
  ${\cal N} = 1$ superconformal field theories},''
  \href{http://dx.doi.org/10.1007/JHEP02(2020)074}{{\em JHEP} {\bfseries 02}
  (2020) 074}, \href{http://arxiv.org/abs/1912.10301}{{\ttfamily
  arXiv:1912.10301 [hep-th]}}.

\bibitem{Eckhard:2020jyr}
J.~Eckhard, S.~Schafer-Nameki, and Y.-N. Wang, ``{Trifectas for $T_N$ in 5d},''
  \href{http://arxiv.org/abs/2004.15007}{{\ttfamily arXiv:2004.15007
  [hep-th]}}.

\bibitem{Bourget:2020gzi}
A.~Bourget, J.~F. Grimminger, A.~Hanany, M.~Sperling, and Z.~Zhong, ``{Magnetic
  Quivers from Brane Webs with O5 Planes},''
  \href{http://arxiv.org/abs/2004.04082}{{\ttfamily arXiv:2004.04082
  [hep-th]}}.

\bibitem{Uhlemann:2019ors}
C.~F. Uhlemann, ``{$AdS_6$/$CFT_5$ with O7-planes},''
\href{http://arxiv.org/abs/1912.09716}{{\ttfamily arXiv:1912.09716 [hep-th]}}.

\bibitem{Gu:2019pqj}
J.~Gu, B.~Haghighat, A.~Klemm, K.~Sun, and X.~Wang, ``{Elliptic Blowup
  Equations for 6d SCFTs. III: E-strings, M-strings and Chains},''
  \href{http://arxiv.org/abs/1911.11724}{{\ttfamily arXiv:1911.11724
  [hep-th]}}.

\bibitem{Cota:2019cjx}
C.~F. Cota, A.~Klemm, and T.~Schimannek, ``{Topological strings on genus one
  fibered Calabi-Yau 3-folds and string dualities},''
  \href{http://dx.doi.org/10.1007/JHEP11(2019)170}{{\em JHEP} {\bfseries 11}
  (2019) 170}, \href{http://arxiv.org/abs/1910.01988}{{\ttfamily
  arXiv:1910.01988 [hep-th]}}.

\bibitem{Choi:2019miv}
S.~Choi and S.~Kim, ``{Large AdS$_6$ black holes from CFT$_5$},''
  \href{http://arxiv.org/abs/1904.01164}{{\ttfamily arXiv:1904.01164
  [hep-th]}}.

\bibitem{Chaney:2018gjc}
A.~Chaney and C.~F. Uhlemann, ``{On minimal Type IIB AdS$_{6}$ solutions with
  commuting 7-branes},'' \href{http://dx.doi.org/10.1007/JHEP12(2018)110}{{\em
  JHEP} {\bfseries 12} (2018) 110},
  \href{http://arxiv.org/abs/1810.10592}{{\ttfamily arXiv:1810.10592
  [hep-th]}}.

\bibitem{Bastian:2018fba}
B.~Bastian, S.~Hohenegger, A.~Iqbal, and S.-J. Rey, ``{Five-Dimensional Gauge
  Theories from Shifted Web Diagrams},''
  \href{http://dx.doi.org/10.1103/PhysRevD.99.046012}{{\em Phys. Rev. D}
  {\bfseries 99} no.~4, (2019) 046012},
  \href{http://arxiv.org/abs/1810.05109}{{\ttfamily arXiv:1810.05109
  [hep-th]}}.

\bibitem{Cheng:2018wll}
S.~Cheng and S.-S. Kim, ``{Refined topological vertex for 5d $Sp(N)$ gauge
  theories with antisymmetric matter},''
  \href{http://arxiv.org/abs/1809.00629}{{\ttfamily arXiv:1809.00629
  [hep-th]}}.

\bibitem{Bah:2018lyv}
I.~Bah, A.~Passias, and P.~Weck, ``{Holographic duals of five-dimensional SCFTs
  on a Riemann surface},''
  \href{http://dx.doi.org/10.1007/JHEP01(2019)058}{{\em JHEP} {\bfseries 01}
  (2019) 058}, \href{http://arxiv.org/abs/1807.06031}{{\ttfamily
  arXiv:1807.06031 [hep-th]}}.

\bibitem{Assel:2018rcw}
B.~Assel and A.~Sciarappa, ``{Wilson loops in 5d $\mathcal{N}=1$ theories and
  S-duality},'' \href{http://dx.doi.org/10.1007/JHEP10(2018)082}{{\em JHEP}
  {\bfseries 10} (2018) 082}, \href{http://arxiv.org/abs/1806.09636}{{\ttfamily
  arXiv:1806.09636 [hep-th]}}.

\bibitem{Ashok:2017bld}
S.~Ashok, M.~Billo, E.~Dell'Aquila, M.~Frau, V.~Gupta, R.~John, and A.~Lerda,
  ``{Surface operators in 5d gauge theories and duality relations},''
  \href{http://dx.doi.org/10.1007/JHEP05(2018)046}{{\em JHEP} {\bfseries 05}
  (2018) 046}, \href{http://arxiv.org/abs/1712.06946}{{\ttfamily
  arXiv:1712.06946 [hep-th]}}.

\bibitem{Garozzo:2020pmz}
I.~Garozzo, N.~Mekareeya, M.~Sacchi, and G.~Zafrir, ``{Symmetry enhancement and
  duality walls in 5d gauge theories},''
  \href{http://arxiv.org/abs/2003.07373}{{\ttfamily arXiv:2003.07373
  [hep-th]}}.

\end{thebibliography}\endgroup

\end{document}